\documentclass[sigconf, nonacm]{acmart}

\newcommand\vldbdoi{XX.XX/XXX.XX}

\newcommand\vldbavailabilityurl{URL_TO_YOUR_ARTIFACTS}
\newcommand\vldbpagestyle{plain} 

\usepackage{balance}
\usepackage{subcaption}
\usepackage[linesnumbered, ruled, vlined]{algorithm2e}
\usepackage{xcolor}
\usepackage{colortbl}
\usepackage{xspace}
\usepackage{listings}

\usepackage{url}
\usepackage{soul}
\sethlcolor{lightgray}
\usepackage[export]{adjustbox}
\usepackage{enumitem}
\usepackage{tabularx}
\usepackage{graphicx}
\graphicspath{{figures/}}
\usepackage{multirow}
\usepackage{makecell}

\usepackage[noabbrev,capitalize,nameinlink]{cleveref}

\usepackage[inline]{aplcomments}
\newcommenter{todo}{1.0, 0, 0.0}
\newcommenter{xinyu}{1.0, 0.49, 0}
\newcommenter{yulong}{0.9, 0.5, 0}
\newcommand{\edit}[1]{{\textcolor{black} {#1}}}

\renewcommand{\marginpar}[1]{}

\newcommand{\weakpt}[2]{\noindent \underline{\textbf{R#1.W#2}}}
\newcommand{\detailedev}[2]{\noindent \underline{\textbf{R#1.D#2}}}
\newcommand{\revision}[2]{\noindent \underline{\textbf{R#1.R#2}}}

\newcommand*\ttvar[1]{\texttt{\expandafter\dottvar\detokenize{#1}\relax}}
\newcommand*\dottvar[1]{\ifx\relax#1\else
  \expandafter\ifx\string-#1\string-\allowbreak\else#1\fi
  \expandafter\dottvar\fi}
\let\oldnl\nl
\definecolor{comment-color}{rgb}{0.5,0.1,0.1}
\definecolor{lightgray}{gray}{0.85}
\definecolor{darkblue}{HTML}{2b8cbe}
\definecolor{mediumblue}{HTML}{a6bddb}
\definecolor{lightblue}{HTML}{f1eef6}
\newcommand{\nonl}{\renewcommand{\nl}{\let\nl\oldnl}}

\newcommand{\ignore}[1]{}

\newcommand{\formatSequenceFile}{SequenceFile\xspace}
\newcommand{\formatRCFile}{RCFile\xspace}
\newcommand{\formatCarbonData}{CarbonData\xspace}
\newcommand{\formatArtus}{Artus\xspace}
\newcommand{\formatCapacitor}{Capacitor\xspace}
\newcommand{\formatArrow}{Arrow\xspace}
\newcommand{\formatDWRF}{DWRF\xspace}
\newcommand{\formatAlpha}{Alpha\xspace}
\newcommand{\pq}{Parquet\xspace}
\newcommand{\orc}{ORC\xspace}
\newcommand{\zm}{%
  \ifvmode%
    Z%
  \else%
    z%
  \fi%
  one map\xspace%
}
\newcommand{\zms}{%
  \ifvmode%
    Z%
  \else%
    z%
  \fi%
  one maps\xspace%
}
\newcommand{\bloomf}{Bloom Filter\xspace}
\newcommand{\bloomfs}{Bloom Filters\xspace}
\newcommand{\bitpack}{Bitpacking\xspace}

\newcolumntype{C}[1]{>{\centering\arraybackslash}m{#1}}

\usepackage{refcount}

\newcommand{\myfootnotetext}[1]{\footnotetext{#1\label{fn:text}%
        \edef\fnmark{\getpagerefnumber{fn:mark}}%
        \edef\fntext{\getpagerefnumber{fn:text}}%
        \ifx\fnmark\fntext\else\ClassWarning{}{footnote mark and text on different pages!}\fi}}

\begin{document}

\title{An Empirical Evaluation of Columnar Storage\\ Formats (Extended Version)}

\newcommand{\footremember}[2]{%
  \footnote{#2}
  \newcounter{#1}
  \setcounter{#1}{\value{footnote}}%
}
\newcommand{\footrecall}[1]{%
  \footnotemark[\value{#1}]%
}
\newcommand\vldbauthors{Xinyu Zeng, Yulong Hui, Jiahong Shen, Andrew Pavlo, Wes McKinney, Huanchen Zhang}
\author{
  Xinyu Zeng, Yulong Hui, Jiahong Shen, Andrew Pavlo$^{\dagger}$, Wes McKinney$^{\ddagger}$, Huanchen Zhang
}

\affiliation{%
  \institution{Tsinghua University,\ \ \  $^{\dagger}$Carnegie Mellon University
    , \ \ \  $^{\ddagger}$Voltron Data}
}
\email{{zeng-xy21, huiyl22, shen-jh20}@mails.tsinghua.edu.cn}
\email{pavlo@cs.cmu.edu,wes@voltrondata.com,
  huanchen@tsinghua.edu.cn}

\renewcommand{\shortauthors}{Zeng, Hui, Shen, Pavlo, McKinney, Zhang et al.}

\begin{abstract}
  Columnar storage is a core component of a modern data analytics system. Although many
  database management systems (DBMSs) have proprietary storage formats, most provide extensive
  support to open-source storage formats such as \pq and \orc to facilitate cross-platform data
  sharing. But these formats were developed over a decade ago, in the early 2010s, for the Hadoop
  ecosystem. Since then, both the hardware and workload landscapes have changed.

  In this paper, we revisit the most widely adopted open-source columnar storage formats (\pq
  and \orc) with a deep dive into their internals. We designed a benchmark to stress-test the
  formats' performance and space efficiency under different workload configurations. From our
  comprehensive evaluation of \pq and \orc, we identify design decisions advantageous with modern
  hardware and real-world data distributions. These include using dictionary encoding by default,
  favoring decoding speed over compression ratio for integer encoding algorithms, making block
  compression optional, and embedding finer-grained auxiliary data structures.
  \edit{We also point out the inefficiencies in the format designs when handling common
    machine learning workloads and using GPUs for decoding.}
  Our analysis identified important considerations that may guide future formats to better fit
  modern technology trends.

\end{abstract}

\maketitle

\pagestyle{\vldbpagestyle}


\section{Introduction} \label{sec:intro}
Columnar storage has been widely adopted for data analytics because of its advantages such as
irrelevant attribute skipping, efficient data compression, and vectorized query
processing~\cite{abadi2013design, ailamaki2001weaving-pax, copeland1985decomposition-dsm}. In the
early 2010s, organizations developed data processing engines for the open-source big data
ecosystem~\cite{hadoop}, including Hive~\cite{hive,thusoo2009hive}, Impala~\cite{impala},
Spark~\cite{spark,zaharia2012resilient}, and Presto~\cite{presto,sethi2019presto}, to
respond to the petabytes of data generated per day and the growing demand for large-scale data
analytics. To facilitate data sharing across the various Hadoop-based query engines, vendors
proposed open-source columnar storage formats~\cite{he2011rcfile, parquet, orc,
      carbondata}, represented by \pq and \orc, that have become the \textit{de facto}
standard for data storage in today's data warehouses and data lakes~\cite{armbrust2020delta, hudi,
      iceberg, presto, spark, dremio, influxdata}.

These formats, however, were developed more than a decade ago. The hardware landscape has changed
since then: persistent storage performance has improved by orders of magnitude, achieving
gigabytes per second~\cite{samsungssd}.  Meanwhile, the rise of data lakes
means more column-oriented files reside in cheap cloud storage (e.g., AWS
S3~\cite{s3}, Azure Blob Storage~\cite{azure-blob}, Google Cloud Storage~\cite{gcs}), which exhibits
both high bandwidth and high latency. On the software side, a number of new lightweight compression
schemes~\cite{boncz2020fsst,liakos2022chimp,zhang2020hope,afroozeh2023fastlanes}, as well as indexing and filtering
techniques~\cite{zhang2018surf, hentschel2018column, sidirourgos2013column,li2013bitweaving}, have
been proposed in academia, while existing open columnar formats are based on DBMS
methods from the 2000s~\cite{abadi2006integrating}.

Prior studies on storage formats focus on measuring the end-to-end performance of Hadoop-based
query engines~\cite{ivanov2020impact, floratou2014sql}. They fail to analyze the design
decisions and their trade-offs. Moreover, they use synthetic workloads
that do not consider skewed data distributions observed in the real
world~\cite{tableau}. Such data sets are less suitable for storage format benchmarking.

The goal of this paper is to analyze state-of-the-art columnar file formats and to identify
important design considerations to provide insights for developing next-generation
column-oriented storage formats. To achieve this, we created a benchmark with predefined
workloads whose configurations were extracted from a collection of real-world data sets.
We then performed a comprehensive analysis for the major components in \pq and \orc,
including encoding algorithms, block compression, metadata organization,
indexing and filtering, and nested data modeling.
\edit{In particular, we investigated how efficiently the columnar formats support common machine learning workloads and whether their designs are friendly to GPUs.}
We detail the lessons learned in \cref{sec:lessons} and summarize our main findings below.

First, there is no clear winner between \pq and \orc in format efficiency.
\pq has a slight file size advantage because of its aggressive dictionary encoding.
\pq also has faster column decoding due to its simpler integer encoding algorithms,
while \orc is more effective in selection pruning due to
\edit{the finer granularity of its \zms (a type of sparse index).}

Second, most columns in real-world data sets
\edit{\marginpar{\detailedev{1}{1(a)}}have a small number of distinct values (or low ``NDV ratios'' defined in \cref{sec:properties})},
which is ideal for dictionary encoding.
As a result, the efficiency of integer-encoding algorithms (i.e., to compress dictionary codes)
is critical to the format's size and decoding speed.
Third, faster and cheaper storage devices means that it is better to use faster decoding schemes
to reduce computation costs than to pursue more aggressive compression
to save I/O bandwidth. Formats should not apply general-purpose block compression
(e.g., Snappy~\cite{snappy}, zstd~\cite{zstd}) by default because the
bandwidth savings do not justify the decompression overhead.

Fourth, \pq and \orc provide simplistic support for auxiliary data structures (e.g., \zms, \bloomfs).
As bottlenecks shift from storage to computation, there are opportunities to embed more sophisticated structures
and precomputed results into the format to trade inexpensive space for less computation.

\edit{Fifth, existing columnar formats are inefficient in serving common machine learning (ML) workloads.
      Current designs are suboptimal in handling projections of thousands of features during ML training
      and low-selectivity selection during top-k similarity search in the vector embeddings.
      Finally, the current formats do not provide enough parallel units to fully utilize the computing power of GPUs.
      Also, unlike the CPUs, more aggressive compression is preferred in the formats with GPU processing
      because the I/O overhead (including PCIe transfer) dominates the file scan time.}

We make the following contributions in this paper. First, we created a feature taxonomy for
columnar storage formats like \pq and \orc. Second, we designed
a benchmark to stress-test the formats and identify their performance vs. space trade-offs
under different workloads. Lastly, we conducted a comprehensive set of experiments on
\pq and \orc using our benchmark and summarized the lessons learned for the future format design.

\section{Background and Related Work}
\label{sec:background}

The Big Data ecosystem in the early 2010s gave rise to open-source
file formats. 
Apache Hadoop first introduced two row-oriented
formats, \textbf{\formatSequenceFile}~\cite{sequencefile}  organized as key-value pairs, and \textbf{Avro}~\cite{avro} based on JSON.
At the same time, column-oriented DBMSs, such as C-Store~\cite{C-Store},
MonetDB~\cite{idreos2012monetdb}, and VectorWise~\cite{zukowski2012vectorwise},
developed the fundamental methods for efficient analytical query
processing~\cite{abadi2013design}: columnar compression, vectorized processing, and
late materialization.
The Hadoop community then adopted these ideas from columnar systems and developed more efficient
formats.

In 2011, Facebook/Meta released a column-oriented format
for Hadoop called \textbf{\formatRCFile}~\cite{he2011rcfile}.
Two years later,
Meta refined \formatRCFile and announced the PAX
(Partition Attribute Across)-based~\cite{ailamaki2001weaving-pax} \textbf{\orc} (Optimized Record
Columnar File) format~\cite{orc,huai2014major}.
A month after \orc's release, Twitter and Cloudera released the first version of
\textbf{\pq}~\cite{parquet}. Their format borrowed insights from earlier columnar storage
research, such as the PAX model and the record-shredding
and assembly algorithm from Google's Dremel~\cite{melnik2010dremel}.

Since then, both \pq and \orc have become top-level Apache Foundation projects. They are also
supported by most data processing platforms, including
Hive~\cite{hive}, Presto/Trino~\cite{presto,sethi2019presto}, and Spark~\cite{spark,zaharia2012resilient}.
Even database products with proprietary storage formats (e.g.,
Redshift~\cite{redshift}, Snowflake~\cite{snowflake}, ClickHouse~\cite{clickhouse}, and
BigQuery~\cite{bigquery}) support \pq and \orc through external tables.

Huawei's \textbf{\formatCarbonData}~\cite{carbondata} is another open-source columnar format
that provides built-in inverted indexing and column groups.
Because of its closer relationship with Spark, previous work failed to evaluate the format
in isolation~\cite{trivedi2018albis}.
Recent work concludes that \formatCarbonData has a worse performance compared with \pq and \orc
and has a less active community~\cite{currerirdf}.

A number of large companies have developed their own proprietary
columnar formats in the last decade.
Google's \textbf{\formatCapacitor} format is used by many of their systems~\cite{capacitor},
including BigQuery~\cite{melnik2020dremel} and Napa~\cite{agiwal2021napa}.
It is based on the techniques from Dremel~\cite{melnik2010dremel} and
\citeauthor{abadi2006integrating}~\cite{abadi2006integrating} that optimize layout based on
workload behavior.
YouTube developed the \textbf{\formatArtus} format in 2019 for the Procella DBMS that supports
adaptive encoding without block compression and
$O(1)$ seek time for nested schemas~\cite{chattopadhyay2019procella}.
Meta's \textbf{\formatDWRF} is a variant of \orc with better support for reading and encrypting
nested data~\cite{dwrf}.
Meta recently developed \textbf{\formatAlpha} to improve the training workloads of machine
learning (ML) applications~\cite{meta-alpha}.

\textbf{\formatArrow} is an in-memory columnar format designed for
efficient exchange of data with limited or no serialization between different application processes or at library API boundaries~\cite{arrow}.
Unlike \pq or \orc, \formatArrow supports random access 
and thus does not require block-based decoding on reads.
Because \formatArrow is not meant for long-term disk storage~\cite{arrow-response-blog},
we do not evaluate it in this paper.

The recent lakehouse~\cite{armbrust2021lakehouse} trend led to
an expansion of formats to support better metadata management
(e.g., ACID transactions).
Representative projects include Delta Lake~\cite{armbrust2020delta},
Apache Iceberg~\cite{iceberg}, and Apache Hudi~\cite{hudi}.
They add an auxiliary metadata layer and do not directly modify the underlying columnar file
formats.

\begin{table*}[t]
    \small{
\begin{tabularx}{\textwidth}{p{0.2in}>{\raggedleft\arraybackslash}p{1.3in}p{2.5in}X}
                                                              &                                              &
  \textbf{\pq}                                                &
  \textbf{\orc}
  \\
  \Xhline{1pt}
  \multirow{7}{*}{\rotatebox{90}{\textsc{\textbf{Features}}}} &
  Internal Layout (\S \ref{ssec:layout})
                                                              & PAX
                                                              & PAX
  \\
                                                              & Encoding Variants (\S \ref{ssec:encoding})
                                                              & plain, RLE\_DICT, RLE, Delta, \bitpack         
                                                              & plain, RLE\_DICT,  RLE, Delta, \bitpack, FOR
  \\
                                                              & Compression (\S \ref{ssec:compression})
                                                              & Snappy, gzip, LZO, zstd, LZ4, Brotli
                                                              & Snappy, zlib, LZO, zstd, LZ4
  \\
                                                              & Type System (\S \ref{ssec:types})
                                                              & Separate logical and physical type system
                                                              & One unified type system
  \\
                                                              & Zone Map / Index (\S \ref{ssec:index})
                                                              & Min-max per smallest \zm/row group/file
                                                              & Min-max per smallest \zm/row group/file
  \\
                                                              & \bloomf (\S \ref{ssec:index})
                                                              & Supported per column chunk
                                                              & Supported per smallest \zm
  \\
                                                              & Nested Data Encoding (\S \ref{ssec:nested})
                                                              & Dremel Model
                                                              & Length and presence
  \\
\end{tabularx}
}
    \caption{
        \textbf{Feature Taxonomy} --
        An overview of the features of columnar storage formats.
    }
    \label{tab:taxonomy}
\end{table*}

\begin{table}[t]
    \small{

\begin{tabularx}{\linewidth}{p{0.01in}>{\raggedleft\arraybackslash}p{0.9in}p{0.9in}X}
                 & \textbf{This Paper}  &
    \textbf{\pq} &
    \textbf{\orc}
    \\
    \Xhline{1pt}
                 &
    Row Group
                 & Row Group
                 & Stripe                 \\
                 & Smallest Zone Map
                 & Page Index (a Page)
                 & Row Index (10k rows)   \\
                 & Compression Unit
                 & Page
                 & Compression Chunk      \\
\end{tabularx}}
    \caption{
        \textbf{Concepts Mapping} --
        Terms used in this paper and the corresponding ones in the formats.
    }
    \label{tab:concepts}
\end{table}

\edit{There are also scientific data storage formats for HPC workloads, including
    HDF5~\cite{folk2011overview, hdf5}, BP5~\cite{bp5}, NetCDF~\cite{netcdf}, and Zarr~\cite{zarr}.
    \marginpar{\weakpt{1}{1}}They target heterogeneous data that has complex file structures, types,
    and organizations. Their data is typically multi-dimensional arrays and not support column-wise encoding.
    Although they expose several language APIs (e.g., Python API to interoperate with Pandas
    and Numpy), few DBMSs support these formats because of their lack of columnar storage
    features.}

Most of the previous investigations on columnar formats target entire query processing systems
without analyzing the format internals in
isolation~\cite{ivanov2020impact, floratou2014sql,pirzadeh2017performance}.
\citeauthor{trivedi2018albis} compared the read performance of \pq, \orc, \formatArrow, and JSON
on the NVMe SSDs~\cite{trivedi2018albis},
but they only measured sequential scans with synthetic data sets (i.e., TPC-DS~\cite{tpc-ds}).
There are also older industry articles that compare popular columnar
formats, but they do not provide an in-depth analysis of the internal design
details~\cite{abadi-arrow-blog,format-bench-blog,format-bench-blog-2}.

Other research proposes ways to optimize these existing columnar formats under specific
workloads or hardware
configurations~\cite{bian2017wide,madden1self,bian2022pixels}.
For example, \citeauthor{jiang2021good} use ML to select the
best encoding algorithms for \pq according to the query history~\cite{jiang2021good}.
\edit{BtrBlocks integrates a sampling-based encoding selection algorithm to achieve the optimal
    decompression speed with network-optimized instances~\cite{btrblocks}.
    \citeauthor{li-selection}
    proposed using BMI
    instructions to improve selection performance on Parquet~\cite{li-selection}.}
None of these techniques, however, have been incorporated in the most popular formats.

\section{Feature Taxonomy} \label{sec:taxonomy}



In this section, we present a taxonomy of columnar formats features (see \cref{tab:taxonomy}).
For each feature category,
we first describe the common designs between \pq and \orc and then highlight
their differences as well as the rationale behind the divergence.

\subsection{Format Layout} \label{ssec:layout}

\begin{figure*}[t!]
  \begin{subfigure}[t]{.49\linewidth}%
    \center
    \includegraphics[width=\linewidth]{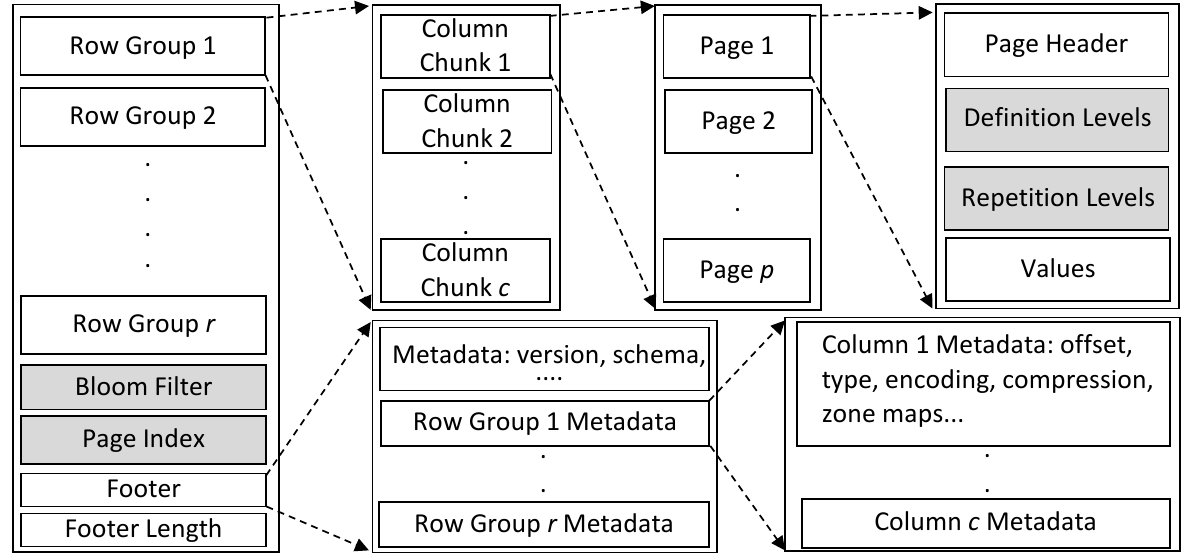}
    \caption{\pq layout.}
    \label{fig:parquet_layout}
  \end{subfigure}
  \hfill
  \begin{subfigure}[t]{.49\linewidth}%
    \center
    \includegraphics[width=\linewidth]{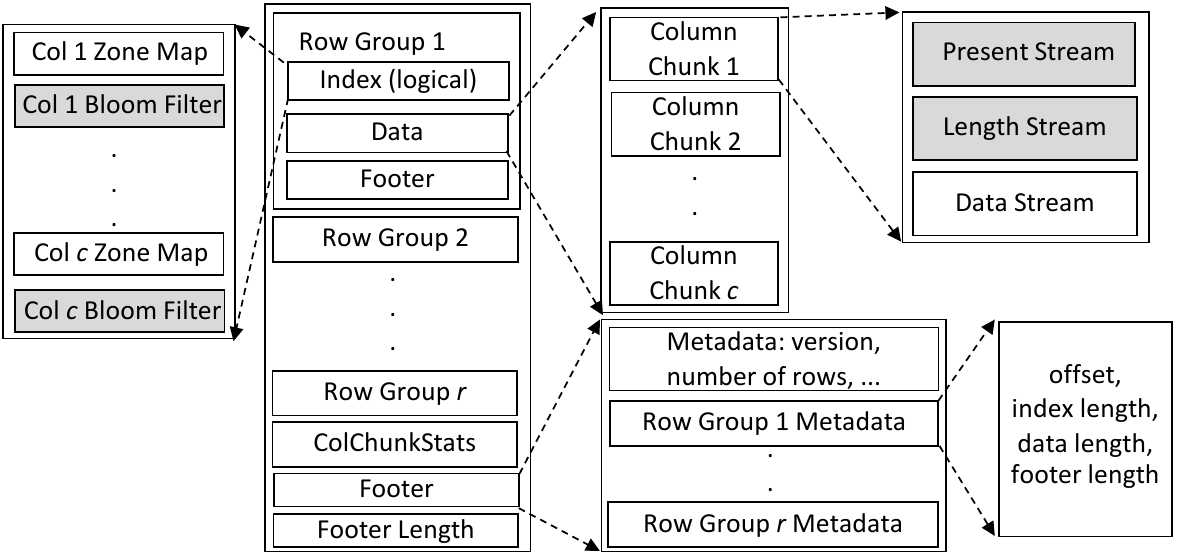}
    \caption{\orc layout. }
    \label{fig:orc_layout}
  \end{subfigure}
  \caption{
    \textbf{Format Layout} --
    Blocks in \colorbox{lightgray}{gray} are optional depending on configurations/data.
  }
  \label{fig:layout}
\end{figure*}

As shown in \cref{fig:layout}, both \pq and \orc employ the PAX format.
The DBMS first partitions a table horizontally into row groups. It then stores tuples
column-by-column within each row group, with each attribute forming a column chunk. The hybrid
columnar layout enables the DBMS to use vectorized query processing and mitigates the tuple
reconstruction overhead in a row group.
\edit{\marginpar{\weakpt{2}{1}}
  Many systems and libraries, such as DuckDB~\cite{duckdb} and Arrow~\cite{arrow-multithreaded-reads},
  leverage the PAX layout to perform parallel reads: each column chunk can be decoded by a separate thread.}

Both formats first apply lightweight encoding schemes (\cref{ssec:encoding}) to
the values for each column chunk. The formats then use general-purpose block compression
algorithms (e.g., Snappy~\cite{snappy}, zstd~\cite{zstd}) to reduce the column chunk's
size. The entry point of a \pq/\orc file is called a footer. Besides file-level metadata
such as table schema and tuple count, the footer keeps the metadata for each row group, including
its offset in the file and \zms for each column chunk. For clarity in our exposition,
in \cref{tab:concepts} we also summarize the mapping between the terminologies used in this
paper and those used in \pq/\orc.

Although the layouts of \pq and \orc are similar, they differ in how they map logical
blocks to physical storage. For example, (non-Java) \pq uses a row-group size based on the
number of rows (e.g., 1M~rows) whereas
\orc uses fixed physical storage size (e.g., 64~MB).
\pq seeks to guarantee that there are enough entries within a row group to leverage
vectorized query processing, but it may suffer from large memory footprints, especially with
wide tables. On the other hand, \orc limits the physical size of a row group to
better control memory usage, but it may lead to insufficient entries with large attributes.
\edit{\marginpar{\weakpt{2}{1}}
  Different systems configure the row group sizes differently to trade-off between
  compression ratio, metadata overhead, and scan parallelism.
  For example, DuckDB sets a relatively small row-group size for \pq to facilitate parallel reads
  even with moderate file sizes.}

Another difference is \pq maps its the compression unit to the smallest zone map.
\orc provides flexibility in tuning the
performance-space trade-off of a block compression algorithm. However, misalignment between the
smallest zone map and compression units imposes extra complexity during query processing (e.g., a value may
be split across unit boundaries).



\subsection{Encoding} \label{ssec:encoding}

Applying lightweight compression schemes to the columns can reduce both storage and network
costs~\cite{abadi2006integrating}. \pq and \orc support standard OLAP compression techniques,
such as Dictionary Encoding, Run-Length Encoding (RLE), and \bitpack.

\pq applies Dictionary Encoding aggressively to every column regardless of the data type
by default,
while \orc only uses it for strings. They both apply another layer of integer encoding on the
dictionary codes. The advantage of applying Dictionary Encoding to an integer
column, as in \pq, is that it might achieve additional compression for
large-value integers.
\edit{\marginpar{\detailedev{1}{1(b)}}
  However, the dictionary codes are assigned based on the values' first appearances in the column chunk
  and thus might destroy local serial patterns that could be compressed well by
  Delta Encoding or Frame-of-Reference (FOR)~\cite{goldstein1998compressing,zukowski2006super,lemire2015decoding}.
  Therefore, \pq only uses \bitpack and RLE to further compress the dictionary codes.}

\pq imposes a limit (1~MB by default) to the dictionary size for each column chunk. When the
dictionary is full, later values fall back to ``plain'' (i.e., no encoding) because a full
dictionary indicates that the number of distinct values (NDVs) is too large
On the other hand, \orc computes the \textit{NDV-ratio}
(i.e., NDV / row count) of the
column to determine whether to apply Dictionary Encoding to it. If a column's NDV-ratio is greater
than a
predefined threshold (e.g., $0.8$), then \orc disables encoding. Compared
to \pq's dictionary size physical limit, \orc's approach is more intuitive, and the tuning
of the NDV-ratio threshold is independent of the row-group size.

For integer columns, \pq first dictionary encodes and then applies a hybrid of RLE and \bitpack
to the dictionary codes. If the
same value repeats $\ge 8$ times consecutively, it uses RLE; otherwise, it uses bitpacking.
Interestingly, we found that the RLE-threshold $8$ is a non-configurable parameter
hard-coded in every implementation of \pq. Although it saves \pq a tuning knob, such
inflexibility could lead to suboptimal compression ratios for specific data sets (e.g., when the
common repetition length is $7$).

\begin{figure}[t]
  \includegraphics[width=\linewidth]{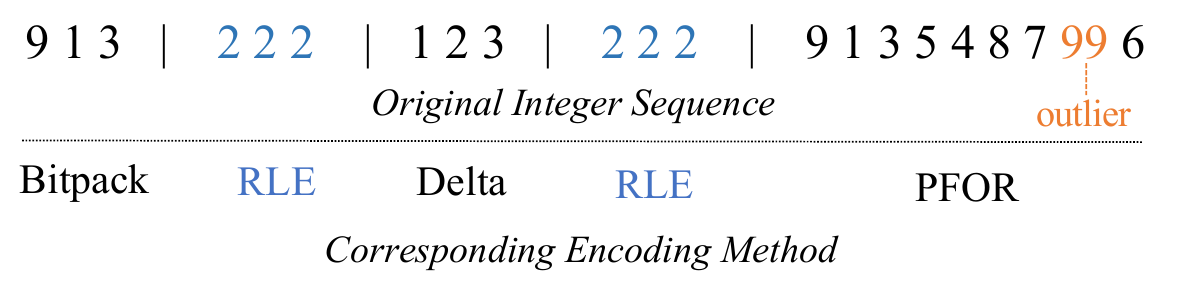}
  \caption{
    \textbf{\orc's Hybrid Integer Encoding} --
    Each encoding subsequence has a header for decoder to decide which algorithm to use at run
    time.
  }
  \label{fig:orc_rle}
\end{figure}

Unlike \pq's RLE + \bitpack scheme, \orc includes four schemes to encode both dictionary codes (for
string columns) and integer columns.
\orc's integer encoder uses a rule-based greedy algorithm to select the best scheme for each
subsequence of values. Starting from the beginning of the sequence, the algorithm keeps
a look-ahead buffer (with a maximum size of 512 values) and tries to detect particular patterns. First, if
there are subsequences of identical values with lengths between 3 and 10, \orc uses RLE to encode
them. If the length of the identical values is greater than 10, or the values of a subsequence
are monotonically increasing or decreasing, \orc applies Delta Encoding to the values. Lastly, for
the remaining subsequences, the algorithm encodes them using either \bitpack or a variant of
\edit{PFOR~\cite{zukowski2006super}}, depending on whether there exist ``outliers'' in a
subsequence. \cref{fig:orc_rle}
is an example of \orc's integer encoding schemes.

The sophistication (compared to \pq) of \orc's integer encoding algorithm allows \orc to seize
more opportunities for compression. However, switching between four encoding schemes slows down the
decoding process and creates more fragmented subsequences that require more metadata to keep track.
\edit{\marginpar{\weakpt{2}{1}}
  All the open-source DBMSs and libraries that we surveyed follow \pq and \orc's default
  encoding schemes without implementing their own tools for selecting encoding algorithms in the files.}

\subsection{Compression} \label{ssec:compression}

Both \pq and \orc enable block compression by default.
The algorithms supported by each format are summarized in \cref{tab:taxonomy}.
Because a general-purpose block compression algorithm is type-agnostic
(i.e., it treats any data as a byte stream), it is mostly orthogonal to
the underlying format layout.
Most block compression algorithms contain parameters to configure the
``compression level'' to make trade-offs between the compression ratio and
the compression/decompression speed.
\pq exposes these tuning knobs directly to the users,
while \orc provides a wrapper with two pre-configured options,
``optimize for speed'' and ``optimize for compression'', for each algorithm.

One of our key observations is that applying block compression to
columnar storage formats is unhelpful (or even detrimental) to the
end-to-end query speed on modern hardware.
\cref{sec:eval} further discusses this issue with experimental evidence.

\subsection{Type System} \label{ssec:types}

\pq provides a minimal set of primitive types (e.g., INT32, FLOAT, BYTE\_ARRAY).
All the other supported types (e.g., INT8, date, timestamp) in \pq
are implemented using those primitives.
For example, INT8 in \pq is encoded as INT32 internally.
Because small integers may be dictionary compressed well,
such a ``type expansion'' has minimal impact on storage efficiency.
On the other hand, every type in \orc has a separate implementation
with a dedicated reader and writer.
Although this could bring more type-specific optimizations,
it makes the implementation bloated.

As for complex types, \pq and \orc both support Struct, List and Map, but \pq does not provide the Union type like \orc. Union allows data values to have different types for the same column name.
Recent work shows that
a Union type can help optimize \pq's Dremel model with schema
changes~\cite{alkowaileet2022columnar}.

\subsection{Index and Filter} \label{ssec:index}

\begin{figure*}[t]
  \begin{subfigure}[b]{.33\linewidth}%
    \center
    \includegraphics[width=\linewidth]{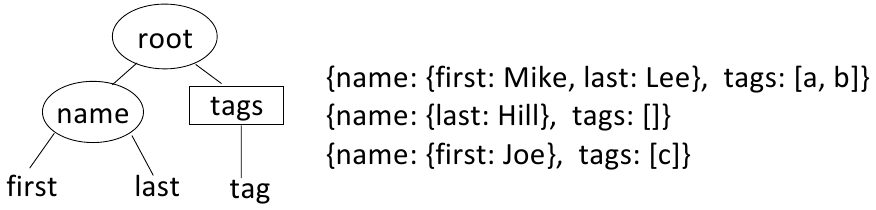}
    \caption{Example schema and three sample records.}
    \label{fig:nested_schema_data}
  \end{subfigure}
  \begin{subfigure}[b]{.32\linewidth}%
    \center
    \includegraphics[width=\linewidth]{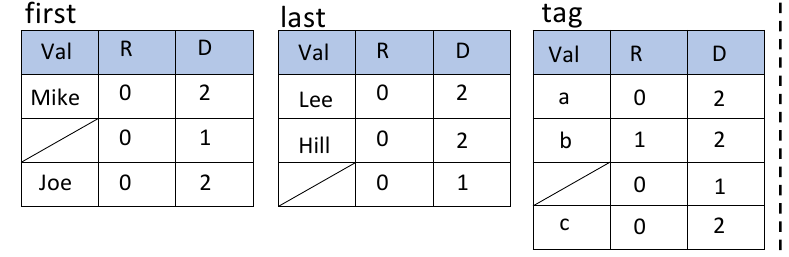}
    \caption{\pq's. R/D=Repetition/Definition Level.}
    \label{fig:nested_parquet}
  \end{subfigure}
  \begin{subfigure}[b]{.34\linewidth}%
    \center
    \includegraphics[width=\linewidth]{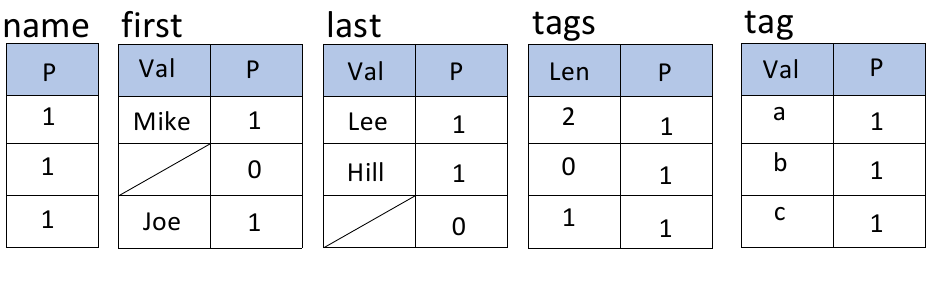}
    \caption{\orc's. Len=length, P=presence.}
    \label{fig:nested_orc}
  \end{subfigure}
  \caption{
    \textbf{Nested Data Example} --
    Assume all nodes except the root can be null.
  }
  \label{fig:nested_example}
\end{figure*}

\pq and \orc include \zms and
optional \bloomfs to enable selection pruning.
A \zm contains the min value, the max value, and the row count
within a predefined range in the file.
If the range of the values of the zone does not satisfy a predicate,
the entire zone can be skipped during the table scan.
Both \pq and \orc contain \zms at the file level and
the row group level.
The smallest \zm granularity in \pq is a physical page
(i.e., the compression unit), while that in \orc is a configurable
value representing the number of rows (10000 rows by default).
Whether to build the smallest \zms is optional in \pq.

In earlier versions of \pq, the smallest \zms are stored
in the page headers.
Because the page headers are co-located with each page and are thus
discontinuous in storage, (only) checking the \zms requires
a number of expensive random I/Os.
In \pq's newest version (2.9.0),
this is fixed by having an optional component called the PageIndex,
stored before the file footer to keep all the smallest \zms.
Similarly, \orc stores its smallest \zms at the beginning of each row group,
as shown in \cref{fig:layout}.

\bloomfs are optional in \pq and \orc.
The \bloomfs in \orc have the same granularity as the smallest \zms,
and they are co-located with each other.
\bloomfs in \pq, however, are created only at the column chunk level
partly because the PageIndex (i.e., the smallest \zms) in \pq is optional.
In terms of the \bloomf implementation,
\pq adopts the Split Block \bloomf (SBBF)~\cite{sbbf},
which is designed to have better cache performance and SIMD support~\cite{parquet-bf-jira}.

\edit{\marginpar{\weakpt{2}{1}}
  According to our survey, Arrow and DuckDB only adopt \zms at the row group level for \pq,
  while InfluxDB and Spark enable PageIndex and \bloomfs to trade space for better selection performance~\cite{influx-parquet-ms}.
  When writing \orc files, Arrow, Spark, and Presto enable row indexes but disable \bloomfs by default.
}

\zms are only effective when the values are clustered (e.g., mostly sorted).
As data processing bottlenecks shift from
storage to computation, whether adding more types of auxiliary data
structures~\cite{hentschel2018column,sidirourgos2013column,li2013bitweaving,zhang2018surf}
to the format will be beneficial to the overall query performance remains an
interesting open question.

\subsection{Nested Data Model} \label{ssec:nested}

As semi-structured data sets such as those in JSON~\cite{json} and Google's Protocol Buffers~\cite{protobuf}
have become prevalent, an open format must support nested data.
\cref{fig:nested_schema_data} shows an example.
The nested data model in \pq is based on Google's Dremel format~\cite{melnik2010dremel}.
As shown in \cref{fig:nested_parquet}, \pq stores the values of each atomic field (the leaf nodes in the hierarchical schema)
as a separate column.
Each column is associated with two integer sequences of the same length, called
the repetition level and the definition level, to encode the nested structure.
The repetition levels link the values to their corresponding ``repeated fields'',
while the definition levels keep track of the NULLs in the ``non-required fields''.

On the other hand, \orc adopts a more intuitive model based on length and presence
to encode nested data~\cite{melnik2020dremel}.
As shown in \cref{fig:nested_orc},
\orc associates a boolean column to each optional field to indicate value presence.
For each repeated field, \orc includes an additional integer column to record the repeated lengths.

For comparison, \orc creates separate presence and length columns for
non-atomic fields (e.g., ``name'' and ``tags'' in \cref{fig:nested_orc}),
while \pq embeds this structural information in the atomic fields via the
repetition and definition levels.
The advantage of \pq's approach is that reads fewer columns
(i.e., atomic fields only) during query processing.
However, \pq often produces a larger file size because the information
about the non-atomic fields could be duplicated in multiple atomic fields
(e.g., ``first'' and ``last'' both contains the information about the presence of ``name''
in \cref{fig:nested_parquet}).

\section{Columnar Storage Benchmark} \label{sec:microbench}
\begin{figure}[t]
  \includegraphics[width=\linewidth]{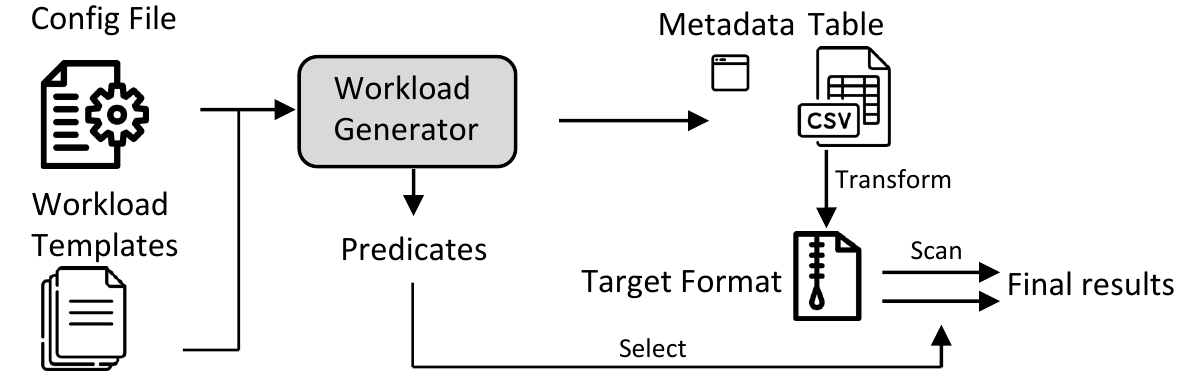}
  \caption{
    \textbf{Benchmark Procedure Overview}
  }
  \label{fig:bench}
\end{figure}

The next step is to stress-test the performance and space efficiency of the storage formats using
data sets using varying value distributions. Standard OLAP benchmarks such as SSB~\cite{ssb},
TPC-H~\cite{tpc-h} and TPC-DS~\cite{tpc-ds} generate data sets with uniform distributions.
Second, although some benchmarks, such as YCSB~\cite{cooper10},
DSB~\cite{ding2021dsb}, and BigDataBench~\cite{bigdatabench} allow users to set data skewness, the
configuration space is often too small to generate distributions that are close to real-world data
sets. Lastly, using real-world data is ideal, but the number of high-quality resources
available is insufficient to cover a comprehensive analysis.

Given this, we designed a benchmark framework based on real-world data
to evaluate multiple aspects of columnar formats. We first define several salient properties of the value
distribution of a column (e.g., sortedness, skew pattern). We then extract these properties from
real-world data sets to form predefined workloads representing applications ranging
from BI to ML. To use our benchmark, as shown in
\cref{fig:bench}, a user first provides a configuration file (or an existing workload template) that
specifies the parameter values of the properties. The workload generator then produces the data
using this configuration and then
generates point and range predicates to evaluate the format's
(filtered) scan performance.

\begin{figure*}[t]
  \begin{subfigure}[b]{0.17\textwidth}%
    \center
    \includegraphics[width=\linewidth]{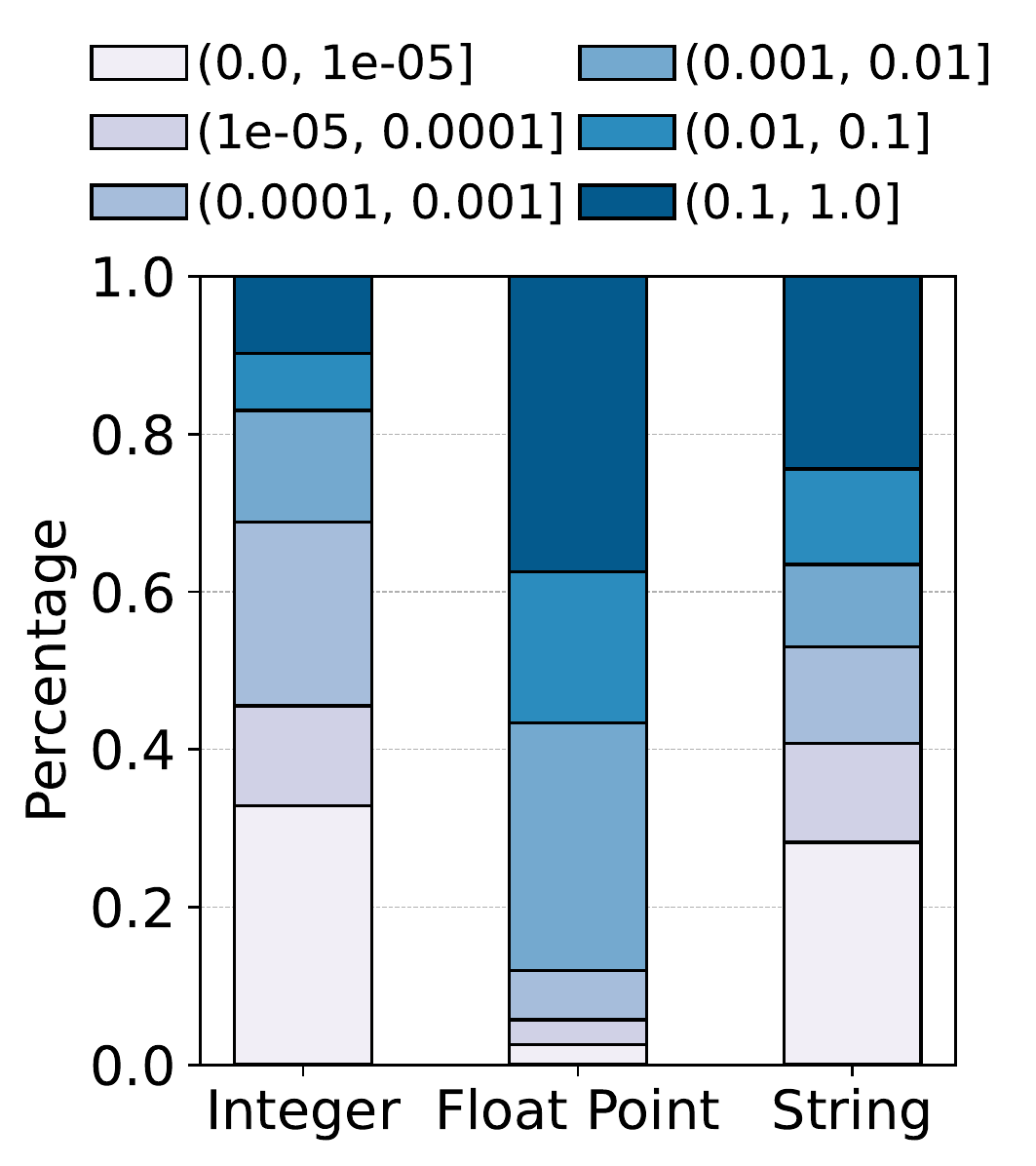}
    \caption{NDV Ratio}\label{fig:car_dist}
  \end{subfigure}%
  \begin{subfigure}[b]{0.161\textwidth}%
    \center
    \includegraphics[width=\linewidth]{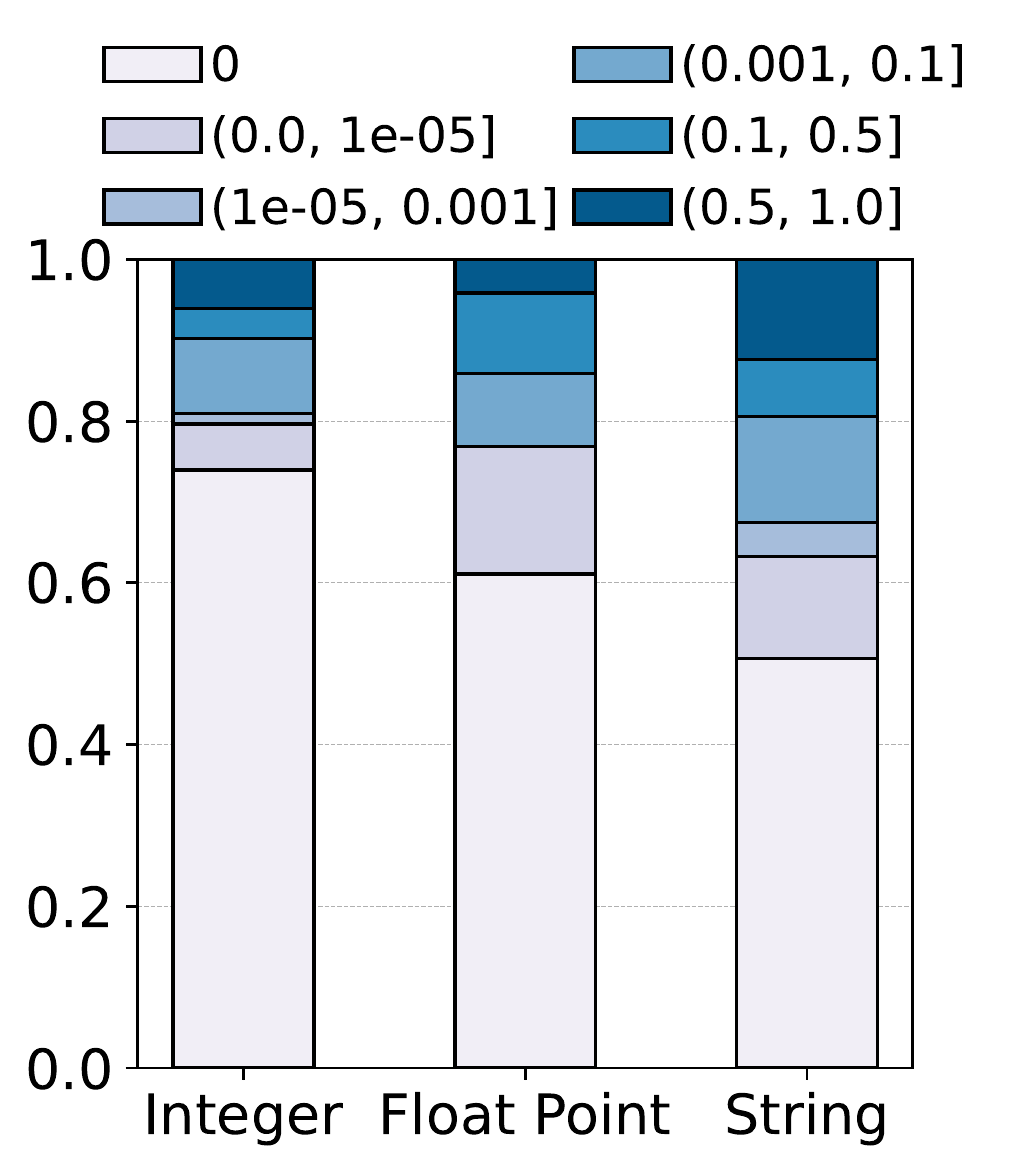}
    \caption{Null Ratio}\label{fig:null_dist}
  \end{subfigure}%
  \begin{subfigure}[b]{0.15\textwidth}%
    \center
    \includegraphics[width=\linewidth]{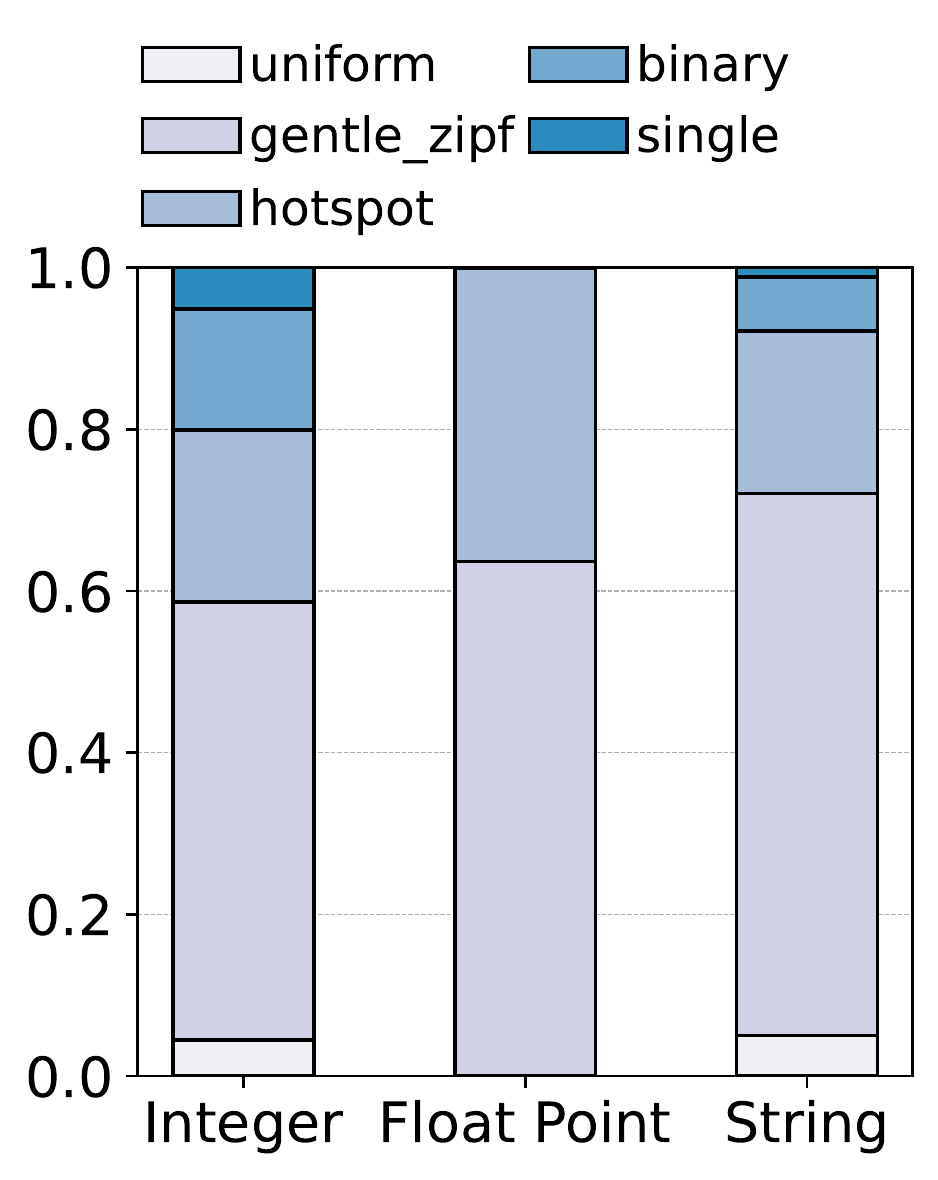}
    \caption{Skew Pattern}\label{fig:skew_dist}
  \end{subfigure}%
  \begin{subfigure}[b]{0.15\textwidth}%
    \center
    \includegraphics[width=\linewidth]{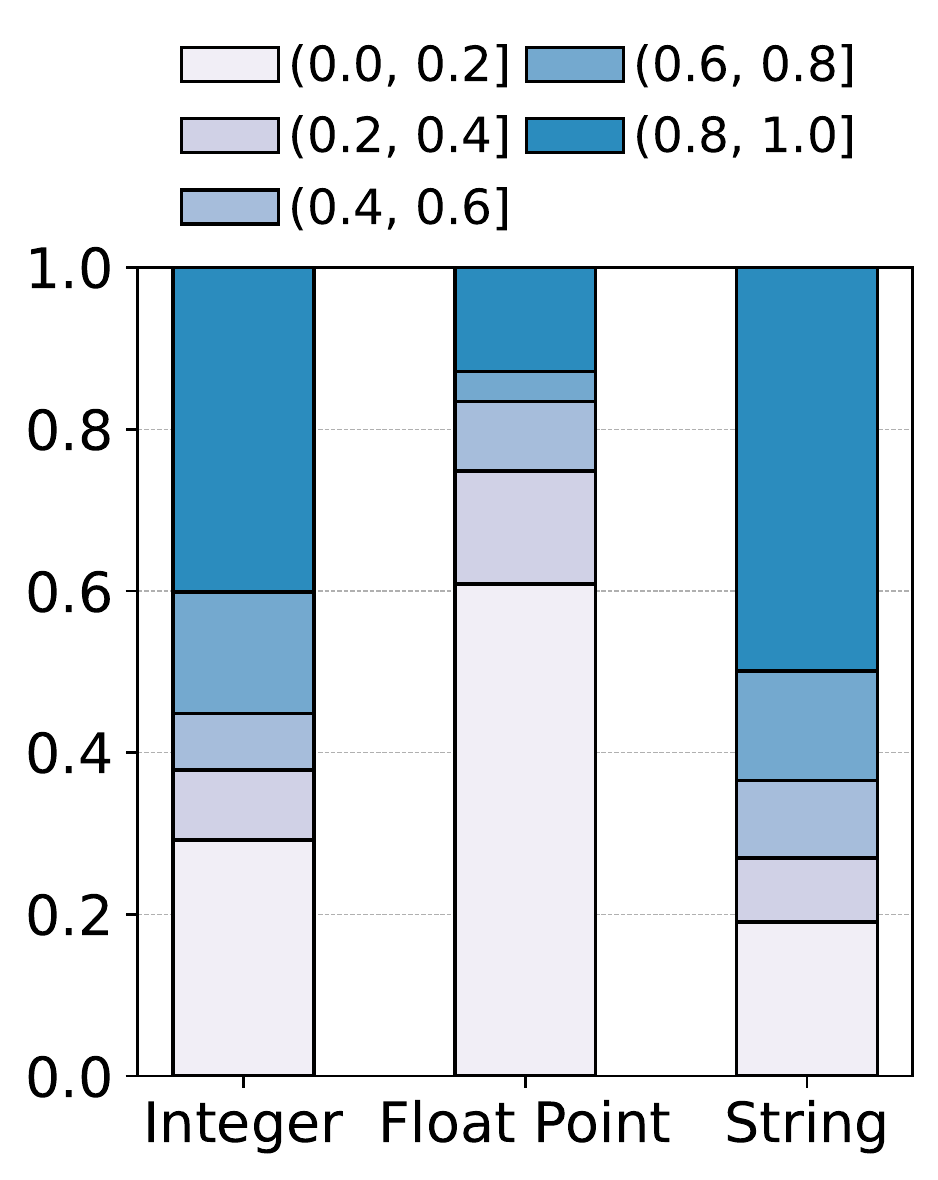}
    \caption{Sortedness}\label{fig:sort_dist}
  \end{subfigure}%
  \begin{subfigure}[b]{0.127\textwidth}%
    \center
    \includegraphics[width=\linewidth]{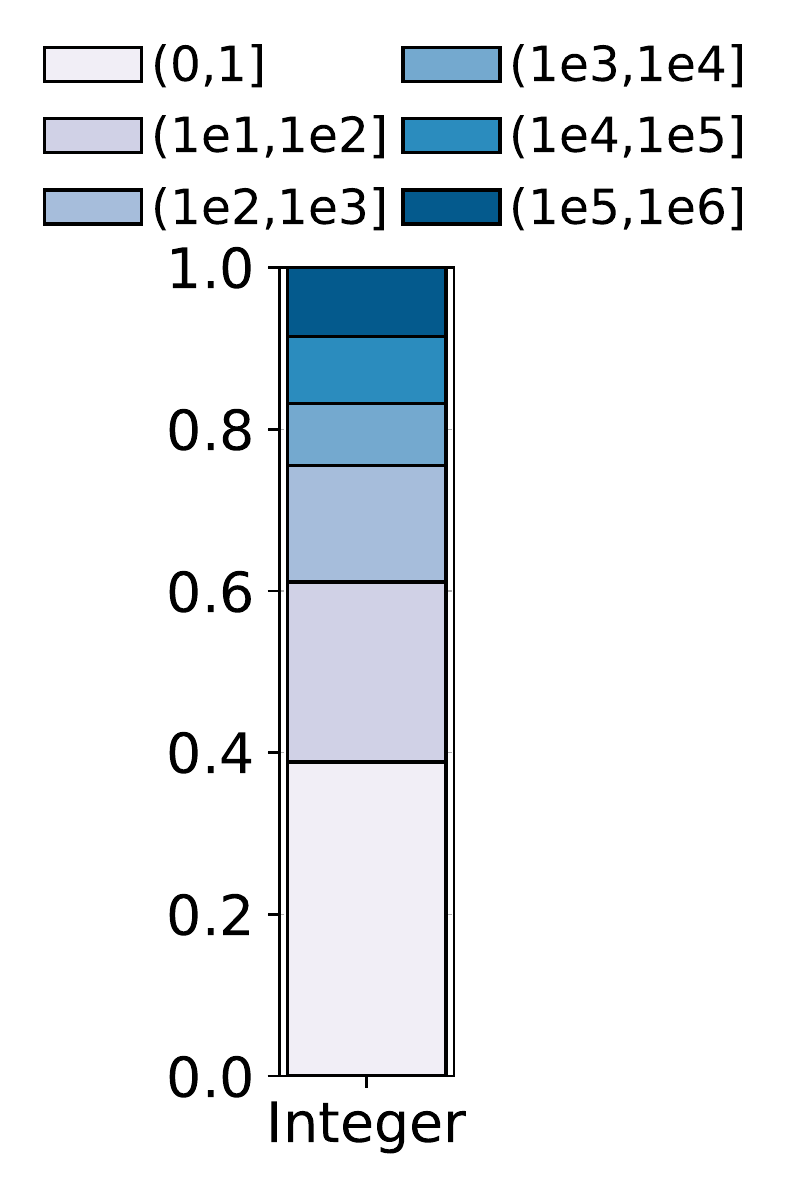}
    \caption{Int Value Range}\label{fig:width_int}
  \end{subfigure}%
  \begin{subfigure}[b]{0.129\textwidth}%
    \center
    \includegraphics[width=\linewidth]{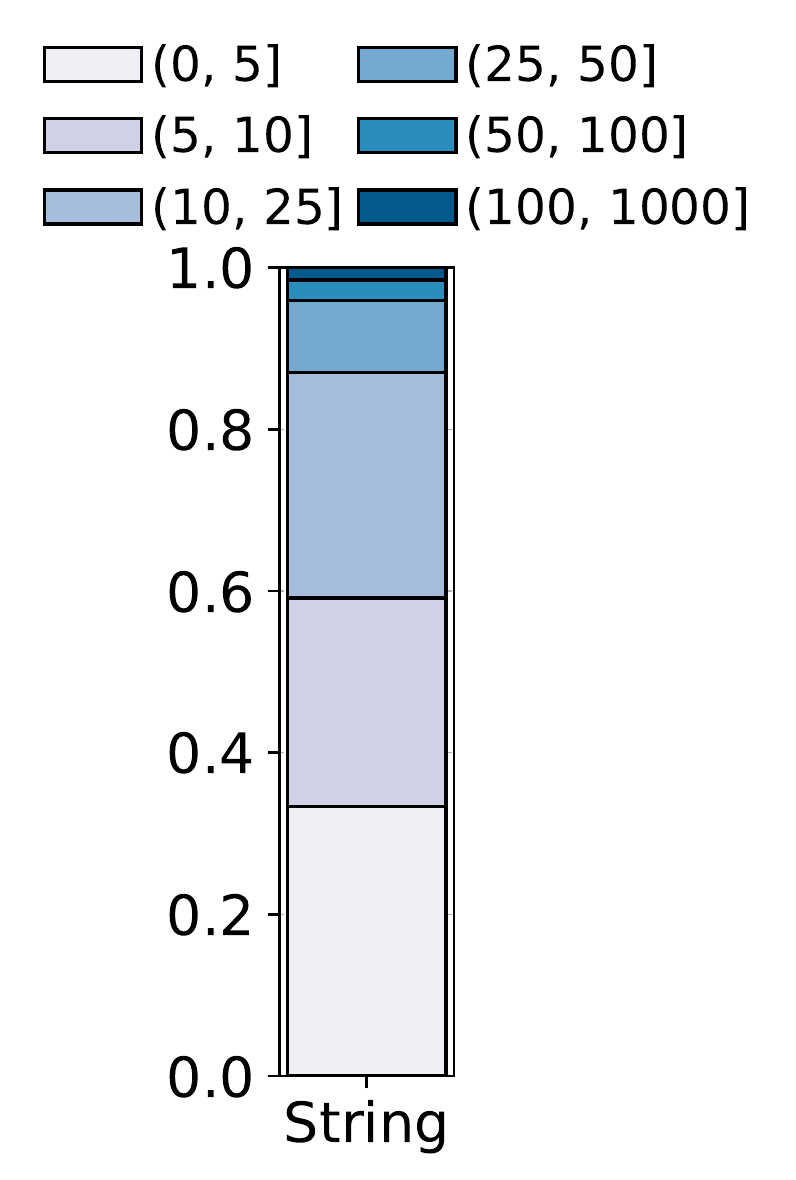}
    \caption{String Length}\label{fig:width_str}
  \end{subfigure}%
  \caption{
    \textbf{Parameter Distribution}
    --
    Percentage of total columns from diverse data sets of different parameter values.
  }

  \label{fig:feature_dist}
\end{figure*}

\subsection{Column Properties} \label{sec:properties}

We first introduce the core properties that define the value distribution of a column.
We use [$a_1, a_2,..., a_N$] to represent the values in a particular column, where $N$ denotes
the number of records.

\paragraph{NDV Ratio:}
Defined as the number of distinct values (NDV) divided by the total number of records in a column:
$f_{cr}=\frac{NDV}{N} $.
A numeric column typically has a higher NDV ratio than a categorical column.
A column with a lower NDV ratio is usually more compressible via Dictionary Encoding and RLE, for
example.

\paragraph{Null Ratio:}
Defined as the number of NULLs divided by the total number of records in a column:
$f_{nr}=\frac{|\{i|a_i \text{ is null}\}|}{N}$.
It is important for a columnar storage format to
handle NULL values efficiently both in terms of space and query processing.

\paragraph{Value Range:}
This property defines the range of the absolute values in a column.
Users pass two parameters: the average value (e.g., $1000$ for an integer column) and the variance
of the value distribution. The value range directly impacts the
compressed file size because most columnar formats apply \bitpack to the values. For string, this is
defined as byte length.

\paragraph{Sortedness:}
The degree of sortedness of a column affects not only the efficiency of
encoding algorithms such as RLE and Delta Encoding, but also the effectiveness of \zms.
Prior work has proposed ways to measure the sortedness of a
sequence~\cite{mannila1985measures},
including \textit{inversions} that counts the number of value pairs with an inversed order,
\textit{runs} that counts the number of ascending subsequences,
and \textit{exchanges} that counts the least number of swaps needed to bring the sequence in
order.
Since these metrics do not correlate strongly with encoding efficiency, we developed a
simple metric that puts more emphasis on local sortedness.
We divide the column into fixed-sized blocks ($512$ entries by default).
Within each block, we compute a sortedness score to reflect its ascending or descending tendency:
\newline $asc\! =\! |\{i | 1\!\leq\! i\! <\! n \text{ and } a_{i}\! <\! a_{i+1} \}|$;
$desc\!=\! |\{i | 1\!\leq\! i\! <\! n \text{ and } a_{i} \!>\! a_{i+1} \}|$
\newline $eq=  |\{i | 1\!\leq \!i \!< \!n \text{ and } a_{i}\! =\! a_{i+1} \}|$;
$f_{sortness}= \frac{\max(asc,desc)+eq-\lfloor\frac{N}{2}\rfloor}{\lceil\frac{N}{2}\rceil-1}$
We then take the average of the per-block scores to represent the column's overall sortedness.
A score of $1$ means that the column is fully sorted, while a score close to $0$ indicates a
high probability that the column's values are randomly distributed.
Although this metric is susceptible to adversarial patterns (e.g., $1, 2, 3, 4, 3, 2, 1$), it is
sufficient for our generator to produce columns with different sortedness levels.
Given a score (e.g., $0.8$), we first sort the values in a block in ascending or descending order
and then swap value pairs randomly until its sortedness degrades to the target score.

\paragraph{Skew Pattern:}
We use the following pseudo-zipfian distribution to
model the value skewness in a column:
$ p(k)={\frac{1}{k^s}}/({\sum_{n=1}^{C}\frac{1}{n^s}})$.
$C$ denotes the total number of distinct values,
and $k$ refers to the frequency rank (e.g., $p(1)$ represents the portion occupied by
the most frequent value).
The Zipf-parameter $s$ determines the column skewness: a larger $s$ leads to
a more skewed distribution.
Based on the range of $s$, we classified the skew patterns into four categories:

\begin{itemize}[leftmargin=.2in]
  \item \textit{Uniform:}
        When $s\leq0.01$. Each value appears in the column with a similar probability.

  \item \textit{Gentle Zipf:}
        When $0.01< s\leq2$. The data is skewed to some extent. The long
        tail of the values still occupies a significant portion of the column.

  \item \textit{Hotspot:}
        When $s>2$. The data is highly skewed. A few hot values cover almost the entire column.

  \item \textit{Single/Binary:}
        This represents extreme cases observed from real-world data where a column
        contains one/two distinct values.
\end{itemize}

The skew pattern is a key factor that determines the performance of
both lightweight encodings
and block compression algorithms.

\subsection{Parameter Distribution in Real-World Data}
\label{ssec:datasets}

We study the following real-world data sets to depict a parameter distribution
of each of the core properties introduced in \cref{sec:properties}.
\begin{itemize}[leftmargin=.1in]
  \item[--]
    Public BI Benchmark~\cite{tableau,cwi-bench}: real-world data and queries from Tableau
    with 206 tables (uncompressed 386GB).
  \item[--] ClickHouse~\cite{ch-data}: sample data sets from the ClickHouse tutorials,
    which represent typical OLAP workloads.
  \item[--] UCI-ML~\cite{uci-ml}: a collection of 622 data sets for ML training.
    We select nine data sets that are larger than 100~MB. All are numerical data excluding unstructured images and embeddings.
  \item[--] Yelp~\cite{yelp}: Yelp's businesses, reviews, and user information.
  \item[--] LOG~\cite{edgar}: log information on internet search traffic for EDGAR filings through SEC.gov.
  \item[--] Geonames~\cite{geonames}: geographical information covering all countries.
  \item[--] IMDb~\cite{imdb}: data sets that describe the basic information, ratings, and reviews of a collection of movies.
\end{itemize}




We extracted the core properties from each of the above data sets
and plotted their parameter distributions in \cref{fig:feature_dist}.
As shown in \cref{fig:car_dist}, over $80\%$ of the integer columns and
$60\%$ of the string columns have an NDV ratio smaller than $0.01$.
Surprisingly, even for floating-point columns, $60\%$ of them have significant
value repetitions with an NDV ratio smaller than $0.1$.
This implies that Dictionary Encoding would be beneficial to most of the real-world columns.
\cref{fig:null_dist} shows that the NULL ratio is low, and string columns
tend to have more NULLs than the other data types.

Most columns in the real world exhibit a skewed value distribution,
as shown in \cref{fig:skew_dist}.
Less than $5\%$ of the columns can be classified as \textit{Uniform}.
Regardless of the data type, a majority ($60 - 70\%$) of the columns fall into
the category of \textit{Gentle Zipf}.
The remaining $\approx30\%$ of the columns contain ``heavy hitters''.
The distribution of the skew patterns indicates that an open columnar format
must handle both the ``heavy hitters'' and the ``long tails'' (from \textit{Gentle Zipf})
efficiently at the same time.

\cref{fig:sort_dist} shows that the distribution of the sortedness scores
is polarized: most columns are either well-sorted or unsorted at all.
This implies that encoding algorithms that excel only at sorted columns
(e.g., Delta Encoding and FOR) could still play an important role.
Lastly, as shown in \cref{fig:width_int}, most integer columns have small values
that are ideal for \bitpack compression.
Long string values are also rare in our data set collection (see \cref{fig:width_str}).
\edit{We also analyzed real-world Parquet files sampled from public available object store buckets and find that they mostly corroborate
  \cref{fig:feature_dist}.
}

\subsection{Predefined Workloads} \label{ssec:predefined}

We extracted the column properties from the real-world data sets introduced
in \cref{sec:properties} and categorized them into five predefined workloads:
\texttt{bi} (based on the Public BI Benchmark),
\texttt{classic} (based on IMDb, Yelp, and a subset of the Clickhouse sample data sets),
\texttt{geo} (based on Geonames and the Cell Towers and Air Traffic data sets from Clickhouse),
\texttt{log} (based on LOG and the machine-generated log data sets from Clickhouse),
and \texttt{ml} (based on UCL-ML).
\cref{tab:workload_dtype} presents the proportion of each data type for each workload,
while \cref{tab:workload_fea} summarizes the parameter settings of the column properties.
Each value in \cref{tab:workload_fea} represents
a weighted average across the
data types
(e.g., if there are $6$ integer columns with an NDV ratio
of $0.1$, $3$ string columns with an NDV ratio of $0.2$, and $1$ float columns with
an NDV ratio of $0.4$, the value reported in \cref{tab:workload_fea} would be $0.16$).
The \texttt{classic} workload has a higher Zipf parameter and
a higher NDV ratio at the same time, indicating a long-tail distribution.
On the other hand, the NDV ratio in \texttt{log} is relatively low, but the
columns are better sorted.
In terms of data types, \texttt{classic} and \texttt{geo} are string-heavy,
while \texttt{log} and \texttt{ml} are float-heavy.

We then created the \texttt{core} workload which is a mix of the five predefined workloads.
It contains 50\% of \texttt{bi} columns, 21\% of \texttt{classic}, 7\% of \texttt{geo},
7\% of \texttt{log}, and 15\% of \texttt{ml}.
We will use \texttt{core} as the default workload in \cref{sec:eval}.
For each workload, we also specify a selectivity for our benchmark to generate predicates to evaluate the filtered scan performance of a columnar storage format.
As shown in \cref{tab:workload_fea}, \texttt{bi} and \texttt{classic} have high selectivities
because these scenarios typically involve large scans.
On the contrary, we use a low selectivity in \texttt{geo} and \texttt{log} because their queries
request data from small geographic areas or
specific time windows.

\begin{table}[t]
  \small{\begin{tabularx}{0.48\textwidth}{C{0.36in}C{0.37in}C{0.4in}C{0.34in}C{0.34in}C{0.34in}>{\centering\arraybackslash}X}
    \textbf{Type} & \textbf{core} & \textbf{bi} & \textbf{classic} & \textbf{geo} & \textbf{log} & \textbf{ml} \\\hline
    Integer       & 0.37          & 0.46        & 0.33             & 0.31         & 0.22         & 0.24        \\
    Float         & 0.21          & 0.20        & 0.06             & 0.08         & 0.46         & 0.39        \\
    String        & 0.41          & 0.34        & 0.61             & 0.61         & 0.32         & 0.37        \\
    Bool          & 0.003         & 0.002       & 0.00             & 0.00         & 0.00         & 0.01        \\
\end{tabularx}}
  \caption{
    \textbf{Data type distribution of different workloads}
    --
    Number in the table indicating the proportion of columns.
  }
  \label{tab:workload_dtype}
\end{table}

\vspace{.1in}

\begin{table}[t]
  \small{\begin{tabularx}{0.48\textwidth}{C{0.72in}C{0.29in}C{0.29in}C{0.29in}C{0.29in}C{0.29in}>{\centering\arraybackslash}X}
    \textbf{Properties} & \textbf{core}                & \textbf{bi}                 & \textbf{classic}           & \textbf{geo}                & \textbf{log}               & \textbf{ml}                \\\hline
    NDV Ratio         & \cellcolor{mediumblue}0.12   & \cellcolor{lightblue}0.08   & \cellcolor{darkblue}0.25   & \cellcolor{darkblue}0.18    & \cellcolor{lightblue}0.08  & \cellcolor{mediumblue}0.12 \\
    Null Ratio        & \cellcolor{mediumblue}0.09   & \cellcolor{darkblue}0.11    & \cellcolor{mediumblue}0.09 & \cellcolor{darkblue}0.13    & \cellcolor{mediumblue}0.02 & \cellcolor{lightblue}0.00  \\
    Value Range       & \cellcolor{mediumblue}medium & \cellcolor{lightblue} small & \cellcolor{darkblue} large & \cellcolor{lightblue} small & \cellcolor{lightblue}small & \cellcolor{darkblue}large  \\
    Sortedness        & \cellcolor{mediumblue}0.54   & \cellcolor{mediumblue}0.57  & \cellcolor{mediumblue}0.49 & \cellcolor{lightblue}0.45   & \cellcolor{darkblue}0.75   & \cellcolor{lightblue}0.30 \\
    Zipf $s$          & \cellcolor{mediumblue}1.12   & \cellcolor{mediumblue}1.10  & \cellcolor{darkblue}1.42   & \cellcolor{lightblue}0.89   & \cellcolor{darkblue}1.26   & \cellcolor{mediumblue}1.00 \\
    Pred. Selectivity & \cellcolor{mediumblue}mid    & \cellcolor{darkblue}high    & \cellcolor{darkblue}high   & \cellcolor{lightblue}low    & \cellcolor{lightblue}low   & \cellcolor{mediumblue}mid  \\
\end{tabularx}
}
  \caption{
    \textbf{Summarized Workload Properties}
    --
    We categorize each property into three levels. The darker the color the
    higher the number.
  }
  \label{tab:workload_fea}
\end{table}

\section{Experimental Evaluation}\label{sec:eval}

In this section, we analyze \pq and \orc's features presented in \cref{sec:taxonomy}.
The purpose is to provide experiment-backed lessons to guide the design
of the next-generation columnar storage formats.
\cref{ssec:exp_setup} describes the experimental setup.
\cref{ssec:general} presents the performance and space results of Parquet and ORC
under default configurations using the predefined workloads in our benchmark.
We then examine the formats' key components with controlled experiments in
\cref{ssec:encoding_exp,ssec:compression_exp,ssec:projection,ssec:index_exp,ssec:nested_exp}.
\edit{Lastly, we test the formats' ability to support ML workloads (\cref{ssec:ml_exp})
  and GPUs (\cref{ssec:gpu_exp})}.

\subsection{Experiment Setup}\label{ssec:exp_setup}

We run the experiments on an AWS \texttt{i3.2xlarge} instance with
8 vCPUs of Intel Xeon CPU E5-2686 v4, 61GB memory, and 1.7TB NVMe SSD.
The operating system is Ubuntu 20.04 LTS.
We use Arrow v9.0.0 to generate the \pq and \orc files.
For all experiments, we use the following configurations
of the formats (unless specified otherwise).
\pq has a row group size of 1m rows, and it sets the dictionary page size limit to 1~MB.
The row group size in \orc is 64~MB, and its NDV-ratio threshold for dictionary encoding is v0.8
(Hive's default).
Snappy compression is enabled (by default ) for both formats.
We use the C++ implementation of \pq (integrated with Arrow C++)~\cite{arrow-cpp-github} and \orc
(v1.8)~\cite{orc-cpp-github}
compiled with g++ v9.4.
To evaluate page-level \zms,
we use the Rust implementation (v32) of \pq in \cref{ssec:index_exp}.
We generate the workloads for the experiments using the benchmark
introduced in \cref{sec:microbench}.
We measure the file sizes and the scan performance (with filters)
in these experiments.
Each reported measurement is the average of three
runs per experiment.

One approach to measuring the (filtered) scan performance of \pq and \orc
is to decode both formats into Arrow tables.
But this approach is unfair
because \pq is tightly coupled with Arrow with native support for
format transformation (e.g., Arrow can decode \pq's dictionary page directly into its dictionary array),
while we must convert \orc into an intermediate in-memory representation
(\texttt{ColumnVectorBatch}) before transforming it into Arrow tables.
Given this, we focus on the raw scan performance of each storage format.
We preallocate a fixed-sized memory buffer.
After decoding the fixed-size unit of data, the system writes the result to the same buffer,
assuming that the previous one has already been consumed by upstream operators.

\begin{figure}[t]
  \begin{minipage}{0.28\linewidth}
    \centering
    \begin{subfigure}[b]{\linewidth}%
      \center
      \includegraphics[width=\linewidth]{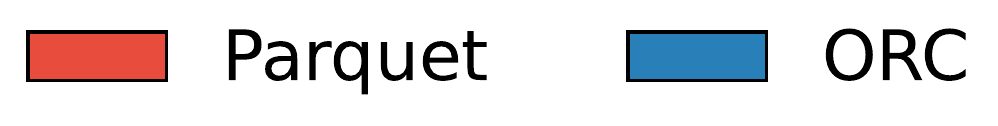}
      \vspace{-.16in}
    \end{subfigure}%
  \end{minipage}
  \begin{subfigure}[b]{0.33\linewidth}%
    \center
    \includegraphics[width=\linewidth]{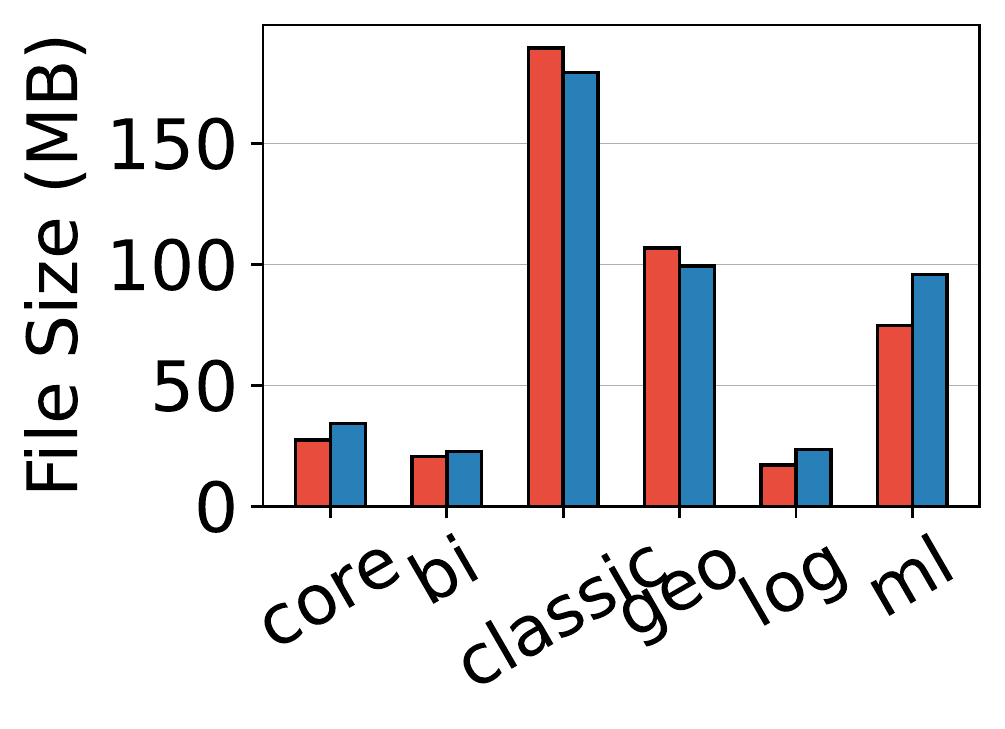}
    \caption{File Size}\label{fig:general_size}
  \end{subfigure}%
  \begin{subfigure}[b]{0.33\linewidth}%
    \center
    \includegraphics[width=\linewidth]{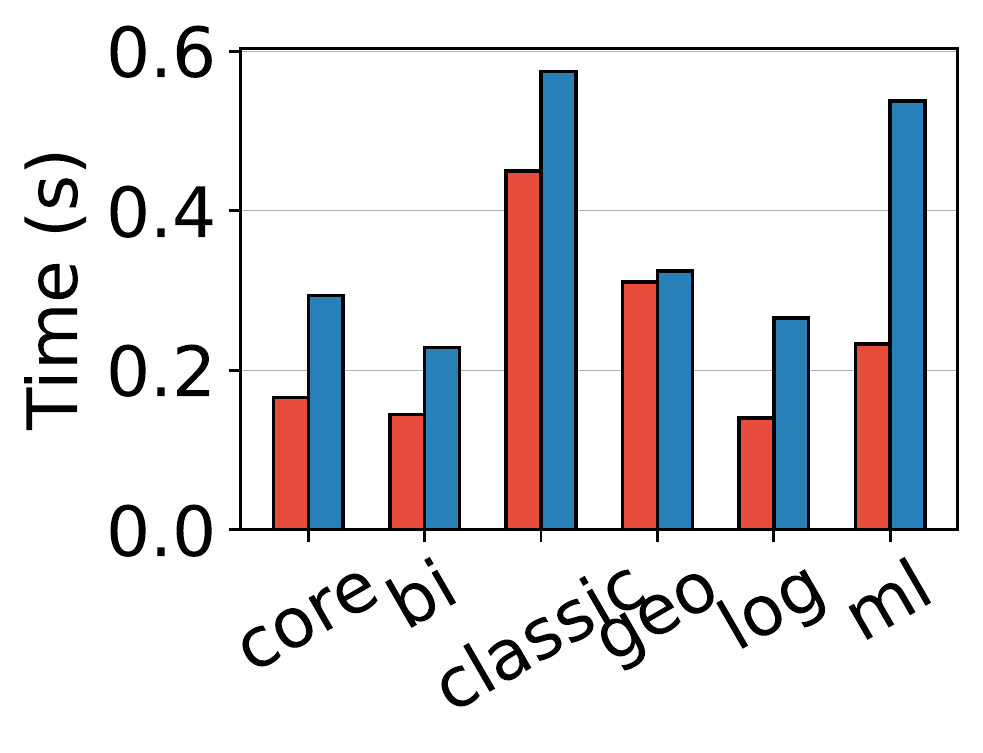}
    \caption{Scan Time}
    \label{fig:general_time}
  \end{subfigure}%
  \begin{subfigure}[b]{0.33\linewidth}%
    \center
    \includegraphics[width=\linewidth]{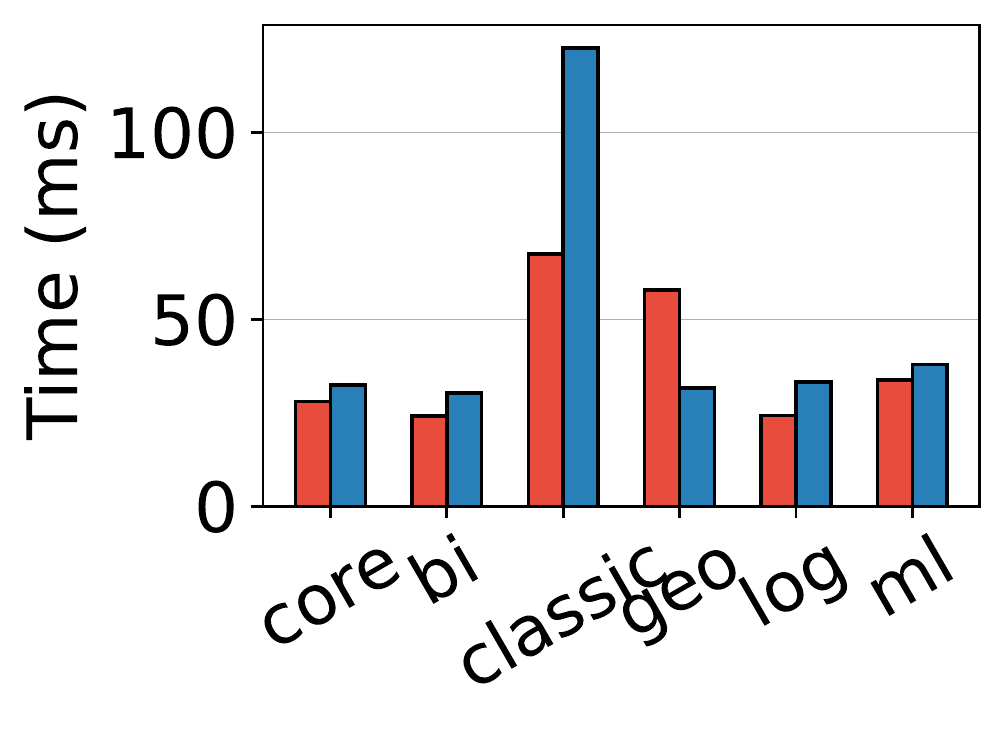}
    \caption{Select Time}
    \label{fig:general_time_filter}
  \end{subfigure}%
  \caption{
    \textbf{Benchmark results with predefined workloads}
  }
  \label{fig:general}
\end{figure}

\subsection{Benchmark Result Overview} \label{ssec:general}

We first present the results of benchmarking \pq and \orc with their default configurations
using the predefined workloads (\cref{ssec:predefined}).
We generate a 20-column table with 1m rows for each workload
and store the data in a single \pq/\orc file.
We then perform a sequential scan of file and report the execution time.
Lastly, we clear the buffer cache and perform 30 select queries.
The selectivities of the range predicates are defined in \cref{tab:workload_fea},
and we report the average latency of the select queries for each workload.


As shown in \cref{fig:general_size}, there is no clear winner between
\pq and \orc in terms of file sizes.
\pq's file size is smaller than \orc's in \texttt{log} and \texttt{ml}
because \pq applies dictionary encoding on float columns where their NDV ratios
are low in real-world data sets (\cref{fig:car_dist}).
However, \orc generates smaller files for \texttt{classic} and
\texttt{geo} because they mostly contain  string data.
We provide further analysis of the encoding schemes in \cref{ssec:encoding_exp}.

The results in \cref{fig:general_time} indicate that \pq is faster than \orc for scans.
The main reason is that \pq's integer / dictionary code encoding scheme
is lightweight: it mostly uses \bitpack and only applies RLE when value repetition is
$\ge 8$ (\cref{ssec:encoding}).
Because RLE decoding is hard to accelerate using SIMD, it has an inferior performance compared to
\bitpack when the repetition count is small.
In contrast, \orc applies RLE more aggressively (when value repetition is $\ge 3$,
and its integer encoding scheme switches between four algorithms,
thus slowing down the decoding process.

\cref{fig:general_time_filter} shows the average latencies of the select queries.
The results generally follow those in sequential scans.
The only exception is \texttt{geo} where \orc outperforms \pq.
The reason is that \orc's smallest \zm has a finer granularity than \pq's
(\cref{ssec:index}).
Compared to other workloads, \texttt{geo} has a relatively high NDV ratio but
a low predicate selectivity, which makes fine-grained \zms more effective.

\begin{table*}[h!]
  \begin{minipage}{0.15\linewidth}
    \centering
    \begin{subfigure}[b]{.8\linewidth}%
      \center
      \includegraphics[width=\linewidth]{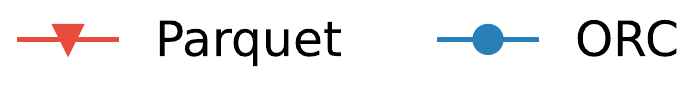}
      \vspace{-.171in}
    \end{subfigure}%
  \end{minipage}
  \begin{tabular}{c c c c c   }
    \rotatebox[origin=c]{90}{\textbf{Integer}} & \raisebox{-0.5\height}{\includegraphics[width=0.22\textwidth]{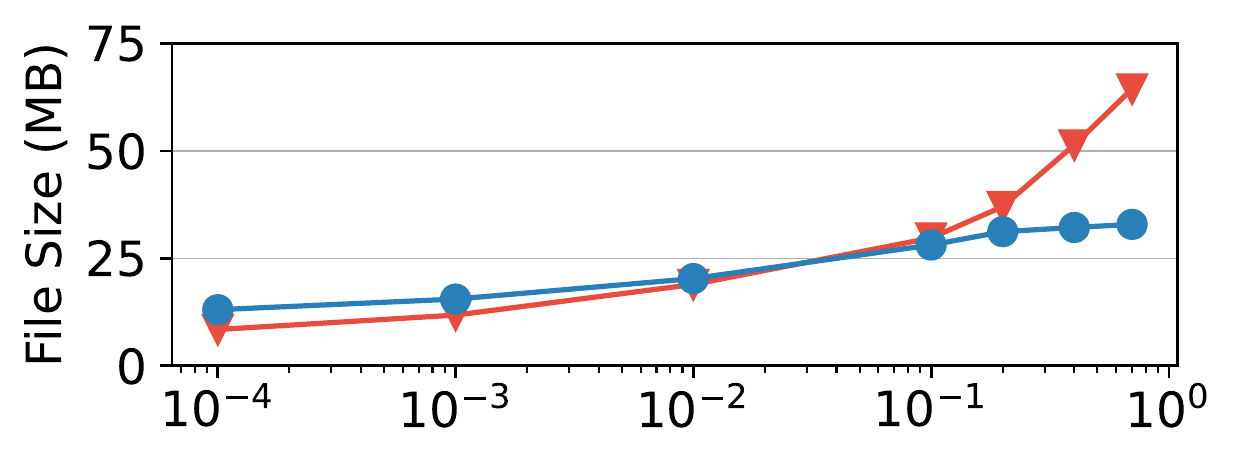}}
                                               & \raisebox{-0.5\height}{\includegraphics[width=0.22\textwidth]{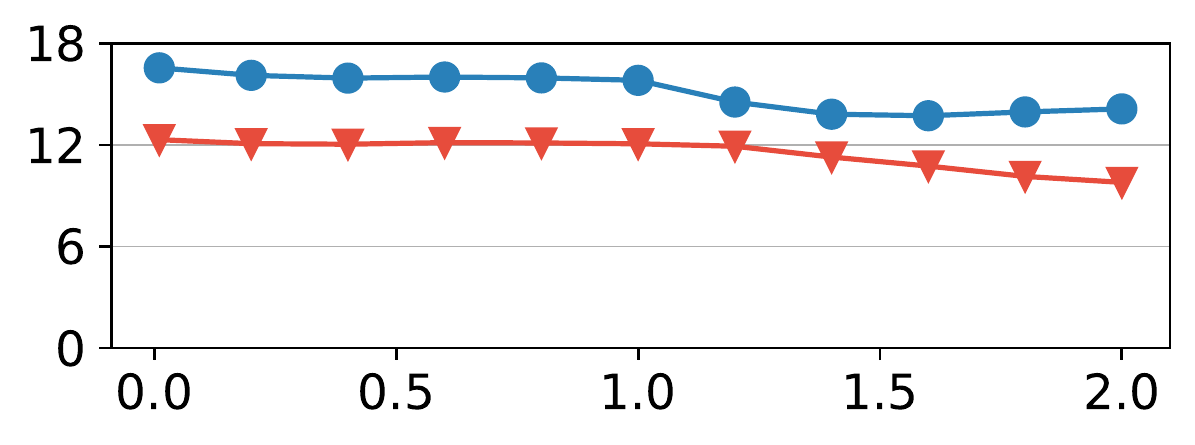}}
                                               &
    \raisebox{-0.5\height}{\includegraphics[width=0.22\textwidth]{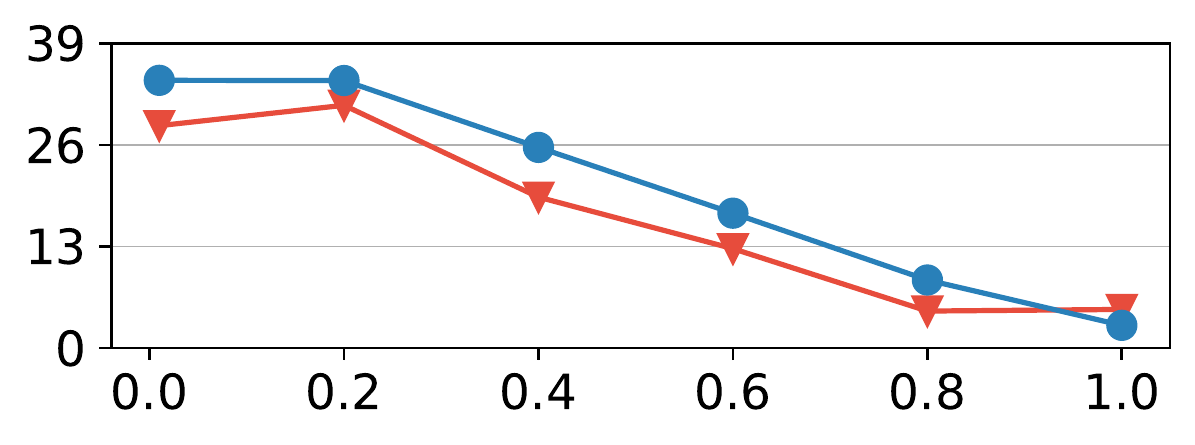}}
                                               &
    \raisebox{-0.5\height}{\includegraphics[width=0.22\textwidth]{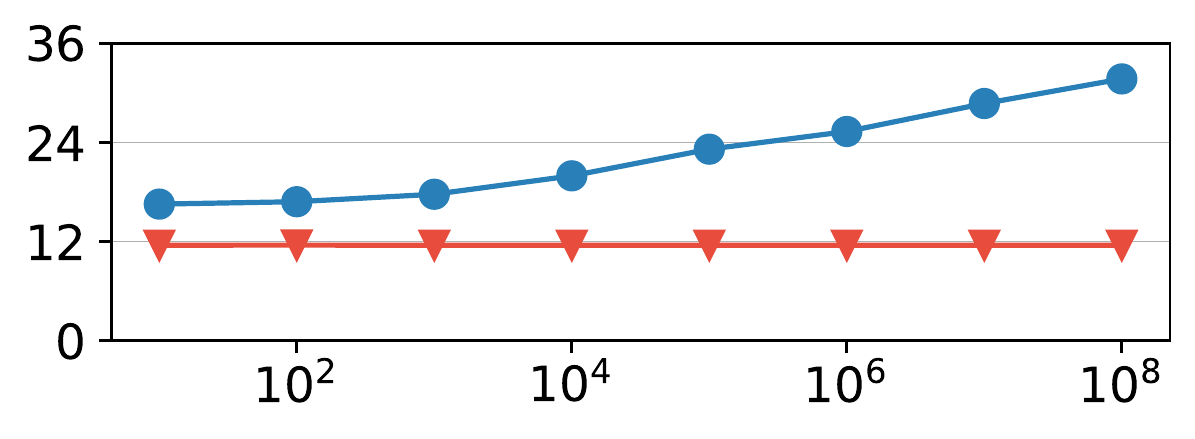}}
    \\
    \rotatebox[origin=c]{90}{\textbf{String}}  & \raisebox{-0.5\height}{\includegraphics[width=0.22\textwidth]{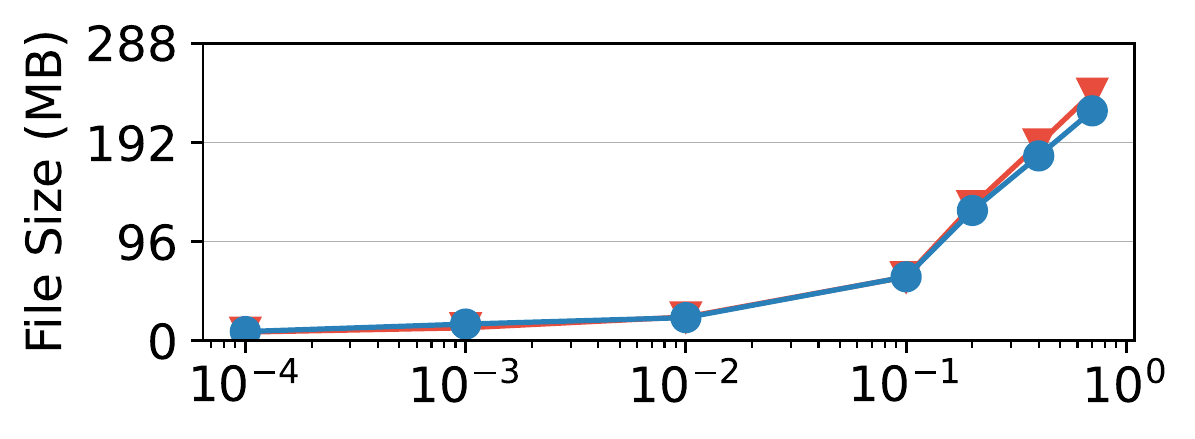}}
                                               & \raisebox{-0.5\height}{\includegraphics[width=0.22\textwidth]{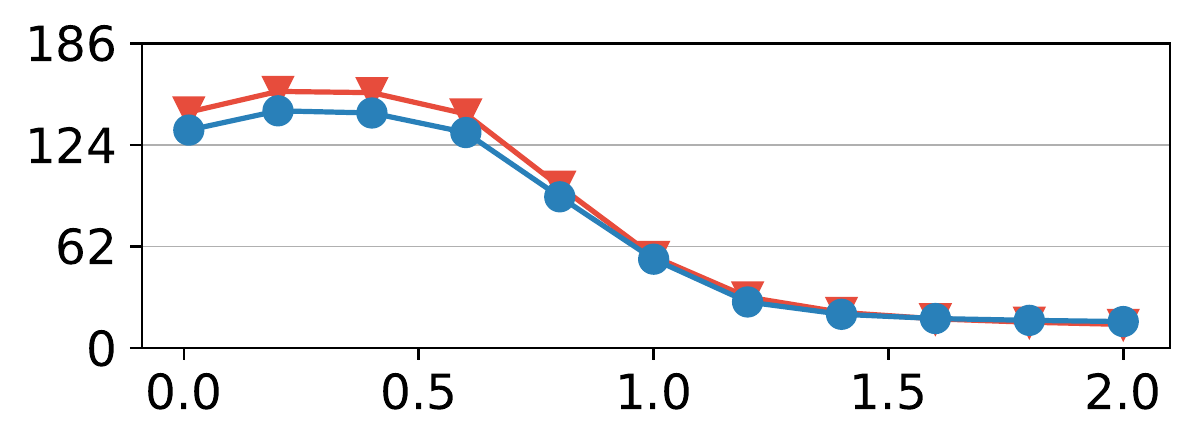}}
                                               &
    \raisebox{-0.5\height}{\includegraphics[width=0.22\textwidth]{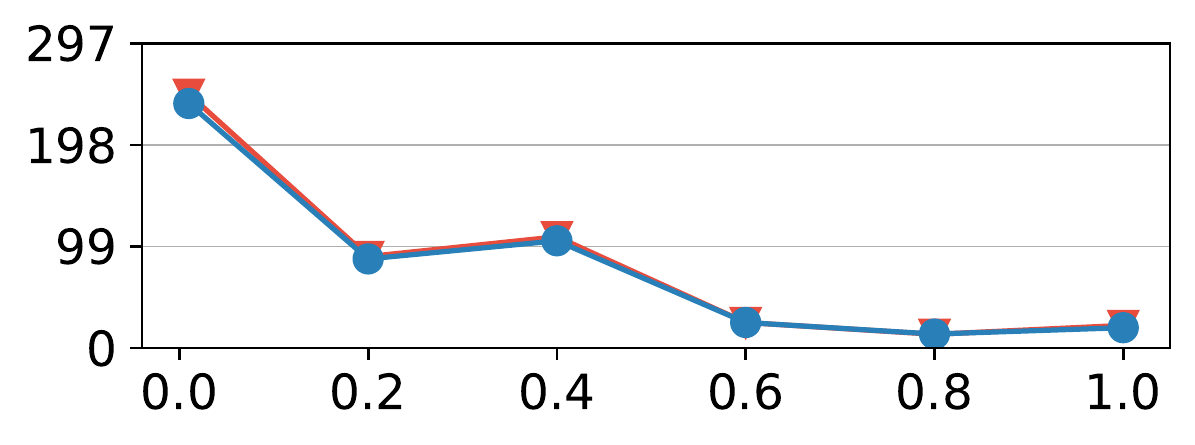}}
                                               &
    \raisebox{-0.5\height}{\includegraphics[width=0.22\textwidth]{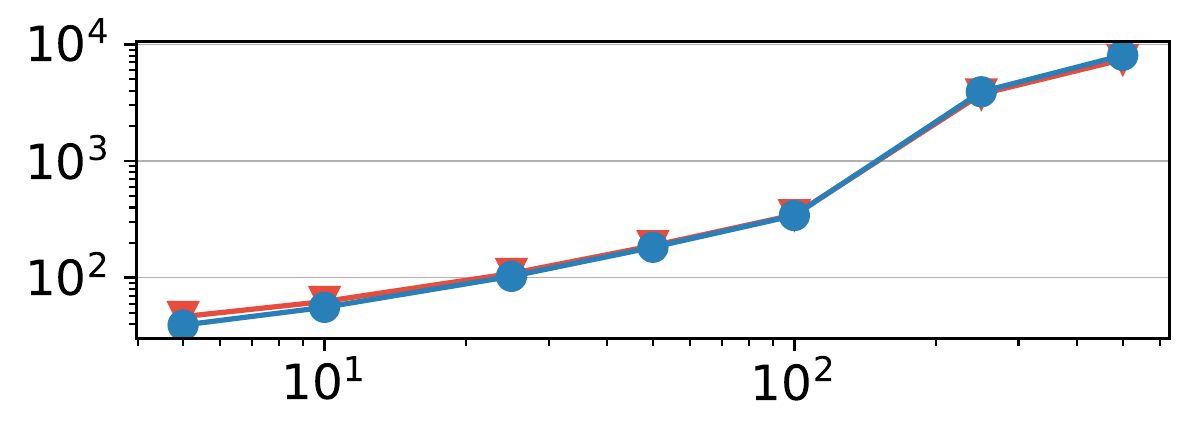}}
    \\
    \rotatebox[origin=c]{90}{\textbf{Float}}   & \raisebox{-0.5\height}{\includegraphics[width=0.22\textwidth]{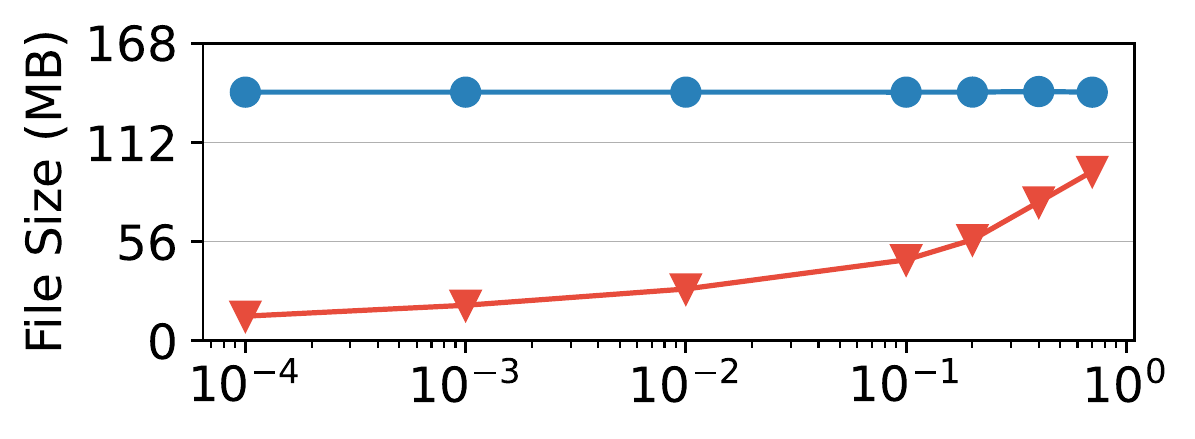}}
                                               & \raisebox{-0.5\height}{\includegraphics[width=0.22\textwidth]{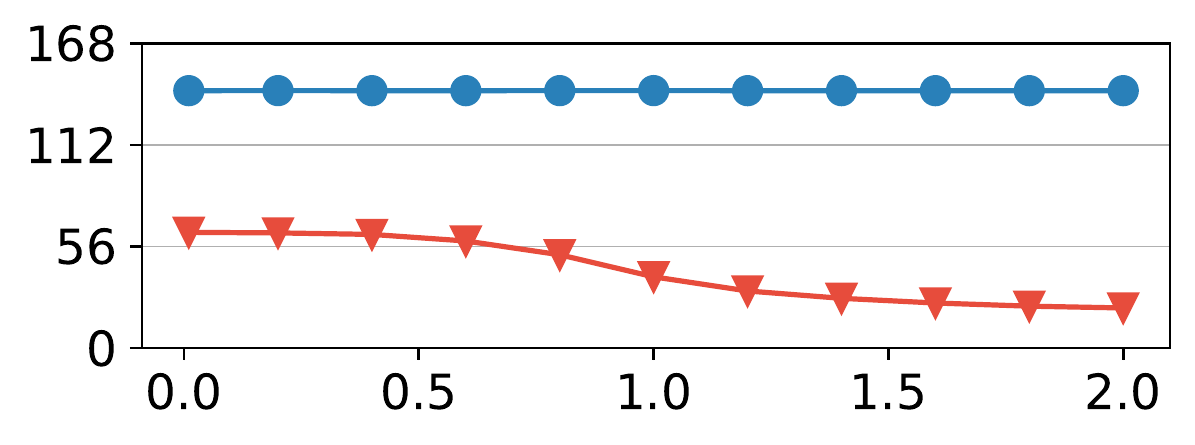}}
                                               &
    \raisebox{-0.5\height}{\includegraphics[width=0.22\textwidth]{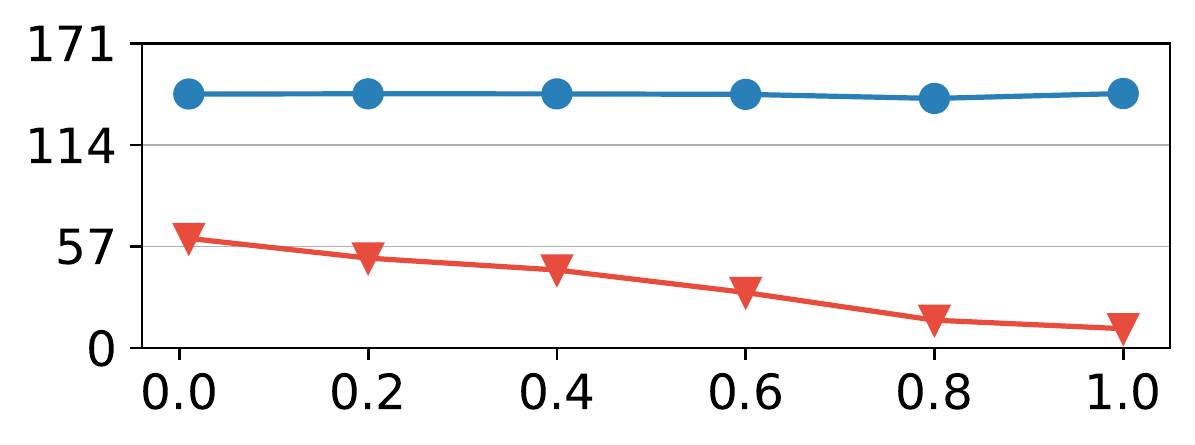}}
                                               &
    \raisebox{-0.5\height}{\includegraphics[width=0.22\textwidth]{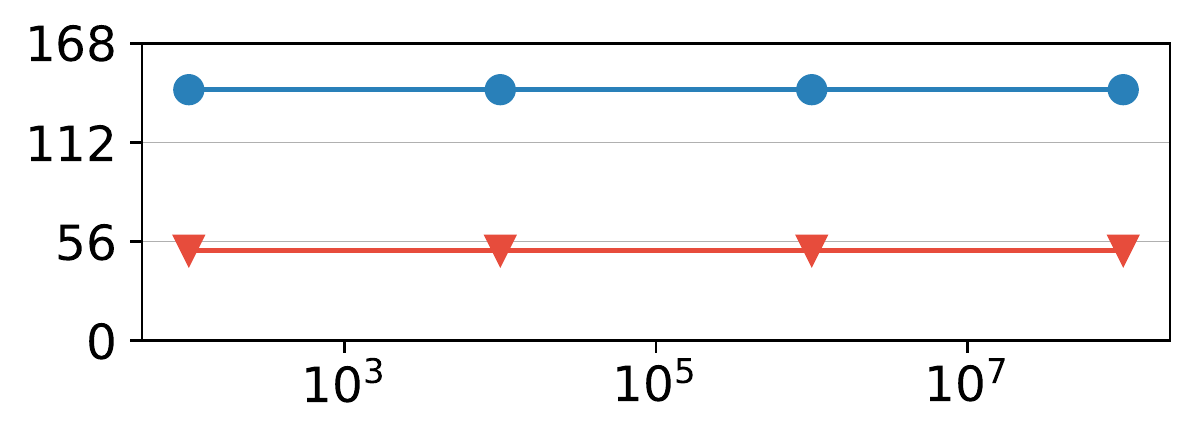}}
    \\
                                               & \textbf{NDV Ratio}                                                                       & \textbf{Zipf $s$} & \textbf{Sortedness} & \textbf{Value Range} \\
  \end{tabular}
  \captionof{figure}{
    \textbf{Encoding size differences} --
    Varying parameters on \texttt{core} workload w/o block compression.
  }
  \label{fig:encoding_size}
\end{table*}

\subsection{Encoding Analysis} \label{ssec:encoding_exp}

We next examine the performance and space efficiency
of the encoding schemes in \pq and \orc in this section.

\subsubsection{Compression Ratio} \label{sssec:encoding_size}


To investigate how \pq and \orc's compression ratios change based on column properties,
we generate a series of tables,
each having 1m rows with 20 columns of the same data type.
For each table, we use the \texttt{core} workload's parameter settings
but modify one of the four column properties: NDV ratio, Value Range, Sortedness,
and the Zipf parameter.
\cref{fig:encoding_size} shows how the file size changes when we sweep the parameter
of different column properties.
We disabled block compression in both \pq and \orc temporarily in these experiments.

As shown in the first row of \cref{fig:encoding_size},
\pq achieves a better compression ratio than \orc for integer columns
with a low to medium NDV ratio (which is common in real-world data sets)
because \pq applies Dictionary Encoding on integers before using \bitpack + RLE.
When the NDV ratio grows larger (e.g., $>0.1$),
this additional layer of Dictionary Encoding becomes less effective
than \orc's more sophisticated integer encoding algorithms.

As the Zipf parameter $s$ becomes larger, the compression ratios on integer columns
improve for both \pq and \orc (row 1, column 2 in \cref{fig:encoding_size}).
The file size reduction happens earlier for \orc ($s = 1$)
than \pq ($s = 1.4$).
This is because RLE kicks in to replace \bitpack earlier in \orc
(with the run length $\ge 3$) than \pq (with the run length $\ge 8$).
We also observe that when the integer column is highly sorted,
\orc compresses those integers better than \pq (row 1, column 3 in \cref{fig:encoding_size})
because of the adoption of Delta Encoding and FOR in its integer encoding.

\pq's file size is stable as the value range of the integers varies
(row 1, column 4 in \cref{fig:encoding_size}).
\pq applies Dictionary Encoding on the integers
and uses \bitpack + RLE on the dictionary codes only.
Because these codes do not change as we vary the value range,
the file size of \pq stays the same in these experiments.
On the other hand, the file size of \orc increases as the value range
gets larger because \orc encodes the original integers directly.

For string columns, as shown in the second row of \cref{fig:encoding_size},
\pq and \orc have almost identical file sizes because they both use Dictionary Encoding on strings.
\orc has a slight size advantage over \pq, especially when the dictionary is large
because \orc applies encoding on the string lengths of the dictionary entries.

The third row of \cref{fig:encoding_size} shows the results for float columns.
\pq dominates \orc in file sizes because Dictionary Encoding is surprisingly
effective on float-point numbers in the real world.

\textbf{Discussion:}
Because of the low NDV ratio of real-world columns (as shown in \cref{fig:feature_dist}),
\pq's strategy of applying Dictionary Encoding on every column
seems to be a reasonable default for future formats.
However, an encoding selection algorithm such as the one described in~\cite{jiang2021good}
is needed to handle the situation when Dictionary Encoding fails.
Also, the format should expose certain encoding parameters, such as the minimum
run length for RLE
\footnote{The minimum run length for RLE is 8 in \pq, and it is hardcoded.}
for tuning so that users can make the
trade-off more smoothly.

\begin{figure}[t]
  \begin{minipage}{0.7\linewidth}
    \centering
    \begin{subfigure}[b]{.9\linewidth}%
      \includegraphics[width=\linewidth]{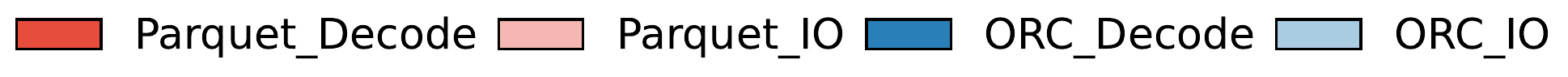}
      \vspace{-.2in}
    \end{subfigure}%
  \end{minipage}
  \begin{subfigure}[t]{0.365\linewidth}%
    \center
    \includegraphics[width=\linewidth]{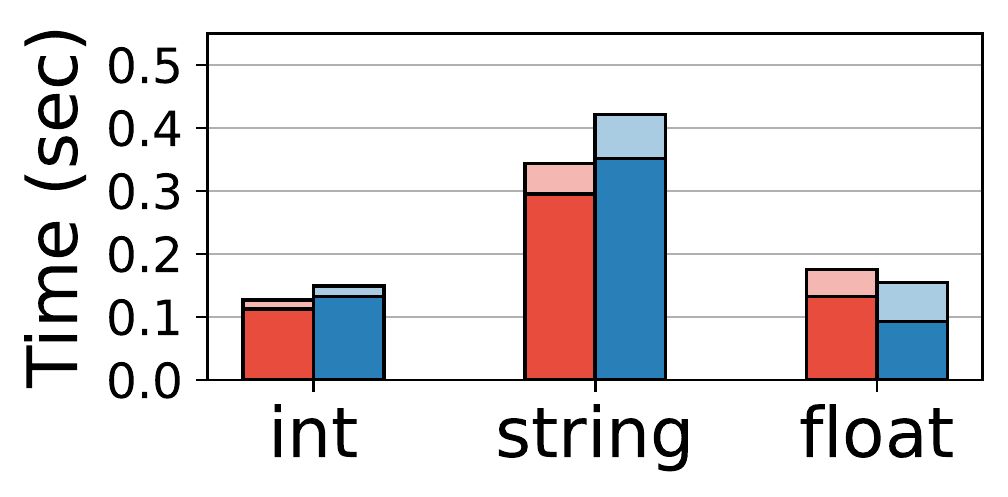}
    \caption{Time [Uncompressed]}\label{fig:scan_none}
  \end{subfigure}%
  \begin{subfigure}[t]{0.34\linewidth}%
    \center
    \includegraphics[width=\linewidth]{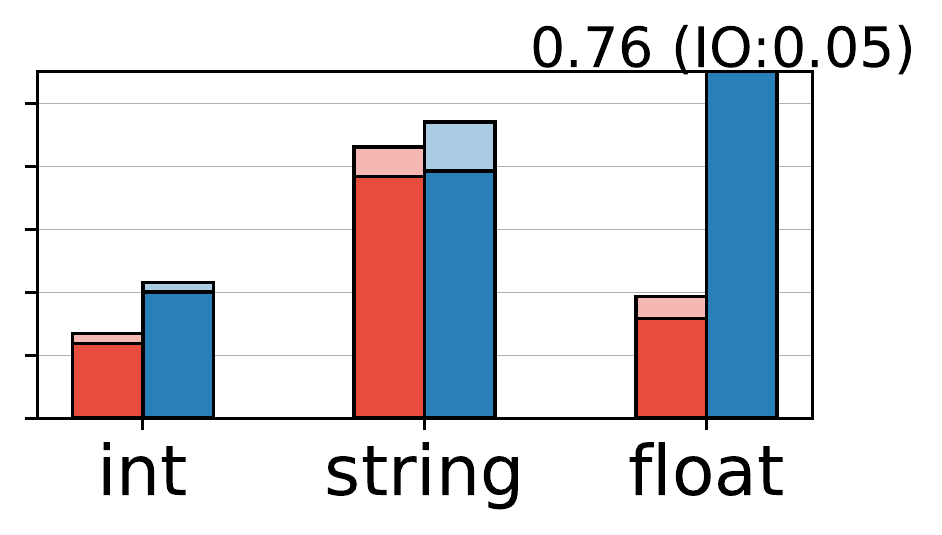}
    \caption{Time [Snappy]}\label{fig:scan_snappy}
  \end{subfigure}%
  \begin{subfigure}[t]{0.3\linewidth}%
    \center
    \includegraphics[width=\linewidth]{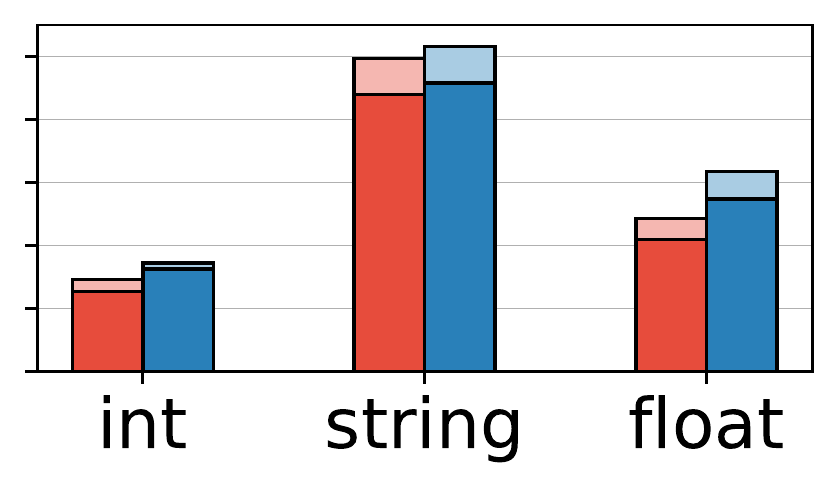}
    \caption{Time [zstd]}\label{fig:scan_zstd}
  \end{subfigure}%
  \\
  \begin{minipage}{0.7\linewidth}
    \centering
    \begin{subfigure}[b]{.33\linewidth}%
      \includegraphics[width=\linewidth]{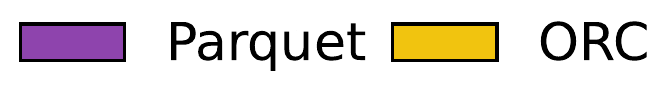}
    \end{subfigure}%
  \end{minipage}
  \begin{subfigure}[t]{0.38\linewidth}%
    \center
    \includegraphics[width=\linewidth]{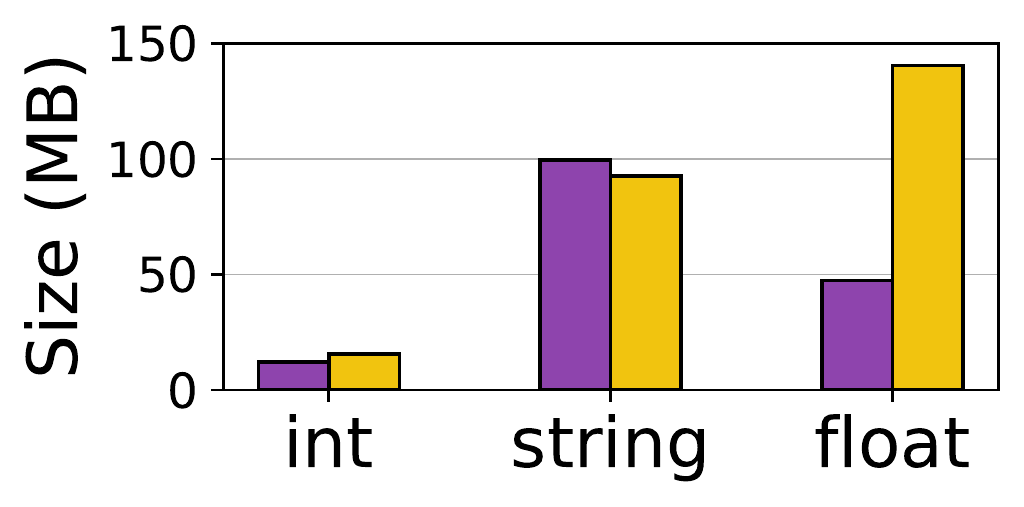}
    \caption{Size [Uncompressed]}\label{fig:size_none}
  \end{subfigure}%
  \begin{subfigure}[t]{0.31\linewidth}%
    \center
    \includegraphics[width=\linewidth]{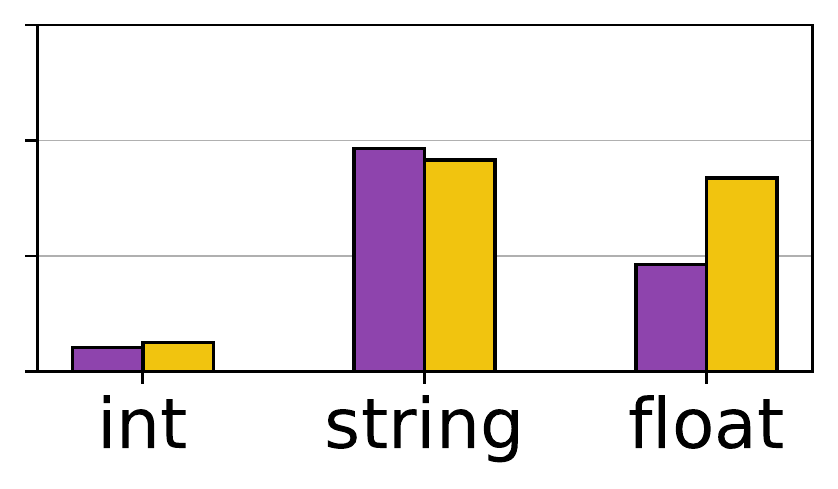}
    \caption{Size [Snappy]}\label{fig:size_snappy}
  \end{subfigure}%
  \begin{subfigure}[t]{0.31\linewidth}%
    \center
    \includegraphics[width=\linewidth]{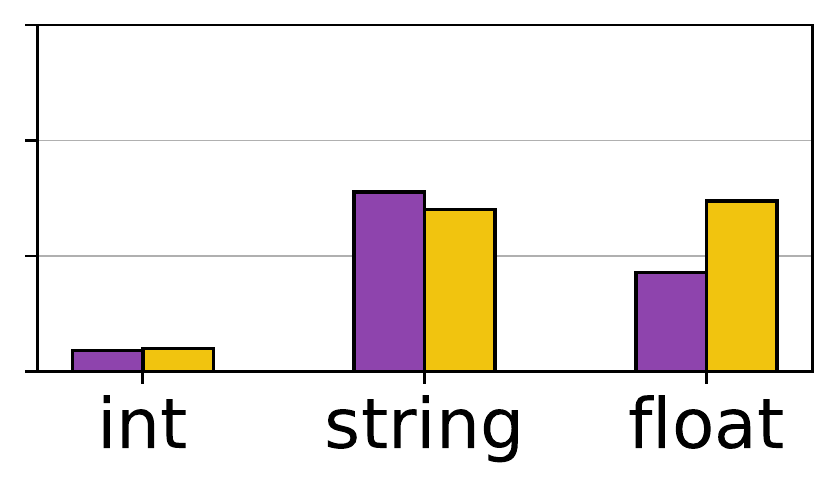}
    \caption{Size [zstd]}\label{fig:size_zstd}
  \end{subfigure}%
  \caption{\textbf{Varying compression on \texttt{core} workload.}
  }\label{fig:compression_eval}
\end{figure}

\begin{figure}[t]
  \begin{subfigure}[t]{0.50\linewidth}%
    \center
    \includegraphics[width=\linewidth]{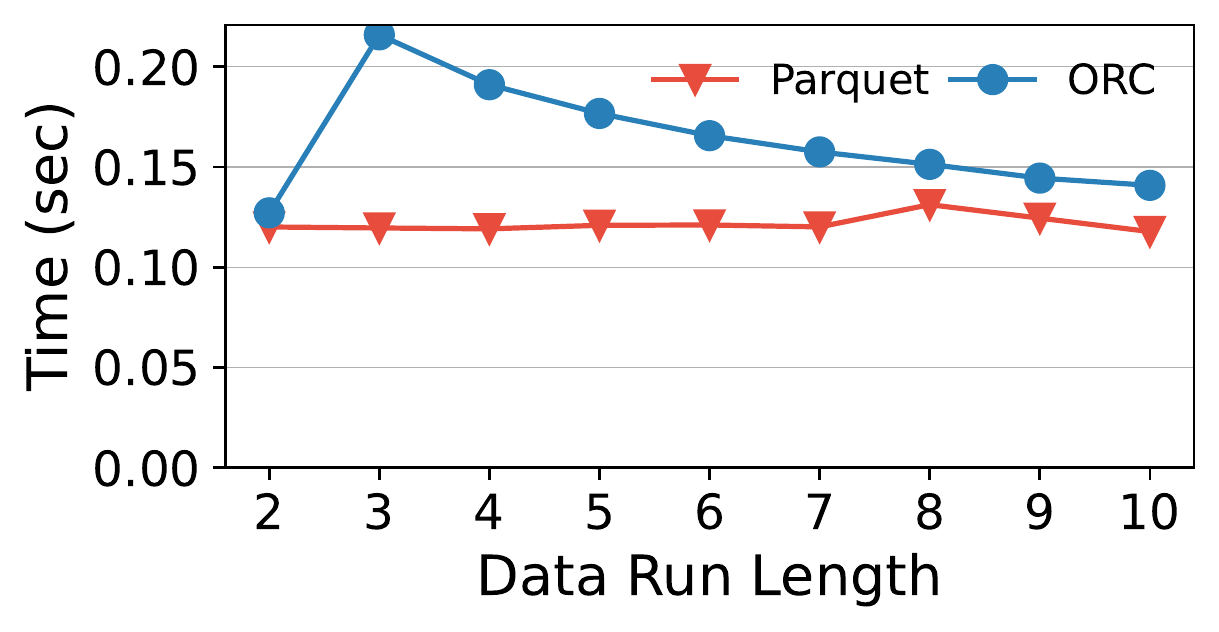}
    \vspace{-.2in}
    \caption{Scan Time}
    \label{fig:decode_run_time}
  \end{subfigure}
  \begin{subfigure}[t]{0.48\linewidth}%
    \center
    \includegraphics[width=\linewidth]{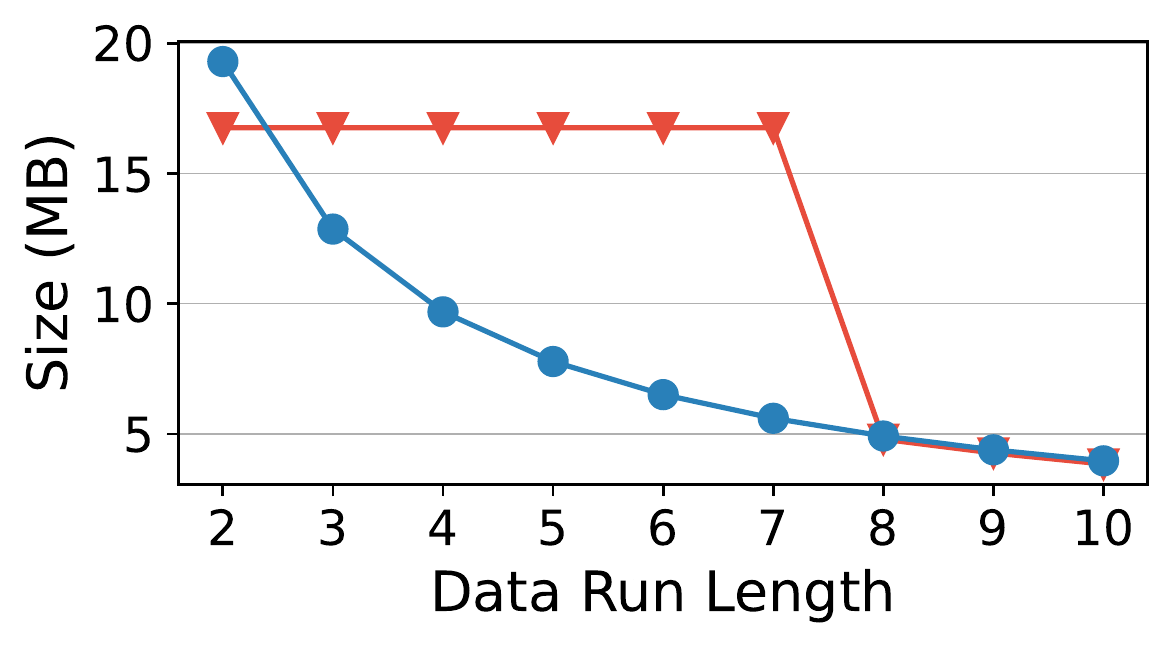}
    \caption{File Size}
    \label{fig:decode_run_size}
  \end{subfigure}
  \caption{\textbf{Varying run length on string, w/o compression.}}
  \label{fig:decode_run}
\end{figure}


\subsubsection{Decoding Speed} \label{sssec:decode_speed}

We next benchmark the decoding speed of \pq and \orc.
We use the data sets in \cref{sssec:encoding_size} that follow the default \texttt{core} workload.
Block compression is still disabled in the experiments in this section.
We perform a full table scan on each file and measure the I/O time
and the decoding time separately.

\begin{table}[ht]
  \begin{minipage}{0.4\columnwidth}
    \centering
    \small{\begin{tabular}{p{0.25in}>{\raggedleft\arraybackslash}p{0.2in}p{0.2in}p{0.2in}X}
    \toprule
    \multicolumn{1}{c}{} & \multicolumn{3}{c}{Workload}                                    \\
    Format               & \textbf{int}                 & \textbf{string} & \textbf{float} \\
    \midrule
    ORC                  & 2.9M                         & 3.1M            & 0.3M           \\
    Parquet              & 0.9M                         & 1.9M            & 0.6M           \\
    \bottomrule
\end{tabular}
}
    \caption{\edit{\textbf{Branch Mispredictions of \cref{fig:scan_none}.}}}
    \label{tab:branch-misses}
  \end{minipage}
  \hfill
  \begin{minipage}{0.55\columnwidth}
    \centering
    \small{
\begin{tabular}{p{0.3in}>{\raggedleft\arraybackslash}p{0.3in}p{0.3in}p{0.3in}X}
    \toprule
    \multicolumn{4}{c}{Encoding}                                     \\
    \textbf{RLE} & \textbf{Bitpack} & \textbf{Delta} & \textbf{PFOR} \\
    \midrule
    .7M(16\%)    & .7M(32\%)        & .2M(49\%)      & .01M(3\%)     \\
    .2M(46\%)    & .2M(54\%)        & 0              & 0             \\
    \bottomrule
\end{tabular}
}
    \caption{\edit{\textbf{Subsequences count and data percentage for integer in \cref{tab:branch-misses}.}}}
    \label{tab:enc-breakdown}
  \end{minipage}
\end{table}

\cref{fig:scan_none} shows that \pq has faster decoding than \orc for integer and
string columns.
As explained in \cref{ssec:general}, there are two main reasons behind this:
(1) \pq relies more on the fast \bitpack and applies RLE less aggressively than \orc,
and (2) \pq has a simpler integer encoding scheme that involves fewer algorithm options.
\edit{
  \marginpar{\weakpt{1}{3}\&}\marginpar{\detailedev{1}{1(b)}}
  As shown in \cref{tab:branch-misses}, switching between the four integer encoding algorithms in
  \orc generates 3$\times$ more branch mispredictions than \pq during the decoding process (done on a similar physical machine to collect the performance counters).
  According to the breakdown in \cref{tab:enc-breakdown}, \orc has 4$\times$ more subsequences to decode than \pq, and the encoding algorithm distribution among the subsequences is unfriendly to branch prediction.
}
\pq's decoding-speed advantage over \orc shrinks for integers
compared to strings, indicating a (slight) decoding overhead due to its additional
dictionary layer for integer columns.
\pq also optimizes the bit-unpacking procedure using SIMD
instructions and code generation to avoid unnecessary branches.

To further illustrate the performance and space trade-off between \bitpack and RLE,
we construct a string column with a pre-configured parameter $r$
where each string value repeats $r$ times consecutively in the column.
Recall that \orc applies RLE when $r \ge 3$,
while the RLE threshold for \pq is $r \ge 8$.
\cref{fig:decode_run} shows the decoding speed and file sizes of \pq and \orc
with different $r$'s.
We observe that RLE takes longer to decode compared to \bitpack for short repetitions.
As $r$ increases, this performance gap shrinks quickly.
The file sizes show the opposite trend (\cref{fig:decode_run_size})
as RLE achieves a much better compression ratio than \bitpack.

For float columns, \orc achieves a better decoding performance than \pq
because \orc does not apply any encoding algorithms on floating-point values.
Although the float columns in \orc occupy much larger space than the
dictionary-encoded ones in \pq (as shown in \cref{fig:encoding_size}),
the saving in computation outweighs the I/O overhead with modern NVMe SSDs.

\textbf{Discussion:} Although more advanced encoding algorithms,
such as FSST~\cite{boncz2020fsst}, HOPE~\cite{zhang2020hope}, Chimp~\cite{liakos2022chimp} and LeCo~\cite{liu2023leco},
have been proposed recently,
it is important to keep the encoding scheme in an open format simple
to guarantee a fast decoding speed.
Selecting from multiple encoding algorithms at run time 
imposes noticeable performance overhead on decoding.
Future format designs should be cautious about including encoding algorithms
that only excel at specific situations in the decoding critical path.

In addition, as the storage device gets faster,
the local I/O time could be negligible during query processing.
According to the float results in \cref{fig:scan_none},
even a scheme as lightweight as Dictionary Encoding adds significant
computational overhead for a sequential scan,
and this overhead cannot get covered by the I/O time savings.
This indicates that most encoding algorithms
still make trade-offs between storage efficiency and decoding speed
with modern hardware (instead of a Pareto improvement as in the past).
Future formats may not want to make any lightweight encoding algorithms
``mandatory'' (e.g., leave raw data as an option).
Also, the ability to operate on compressed data is important with today's hardware.



\subsection{Block Compression} \label{ssec:compression_exp}

\begin{figure}[t]
  \begin{minipage}[t]{0.515\linewidth}
    \begin{subfigure}[t]{\linewidth}%
      \center
      \includegraphics[width=\linewidth]{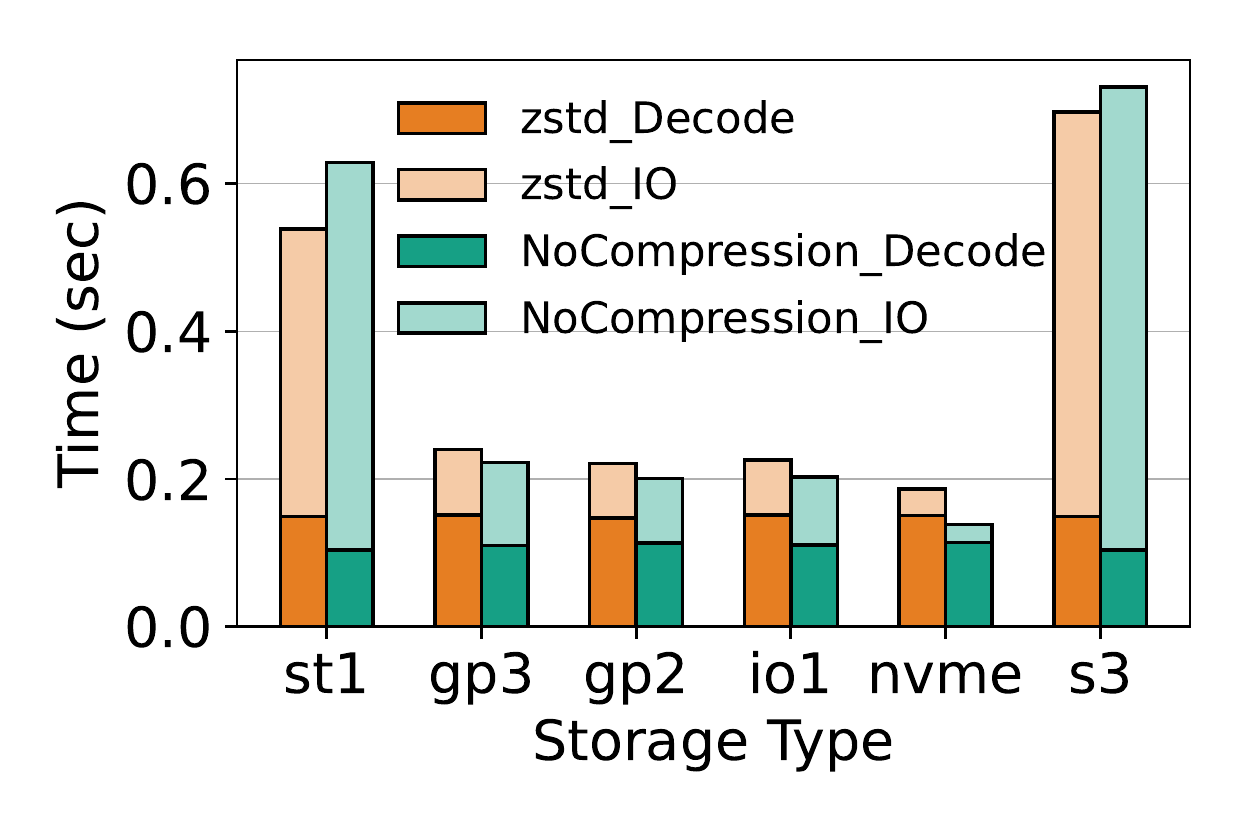}
    \end{subfigure}%
    \caption{\textbf{Block Compression}
    }
    \label{fig:compression_device}
  \end{minipage}
  \hfill
  \begin{minipage}[t]{0.475\linewidth}
    \begin{subfigure}[t]{\linewidth}%
      \center
      \includegraphics[width=\linewidth]{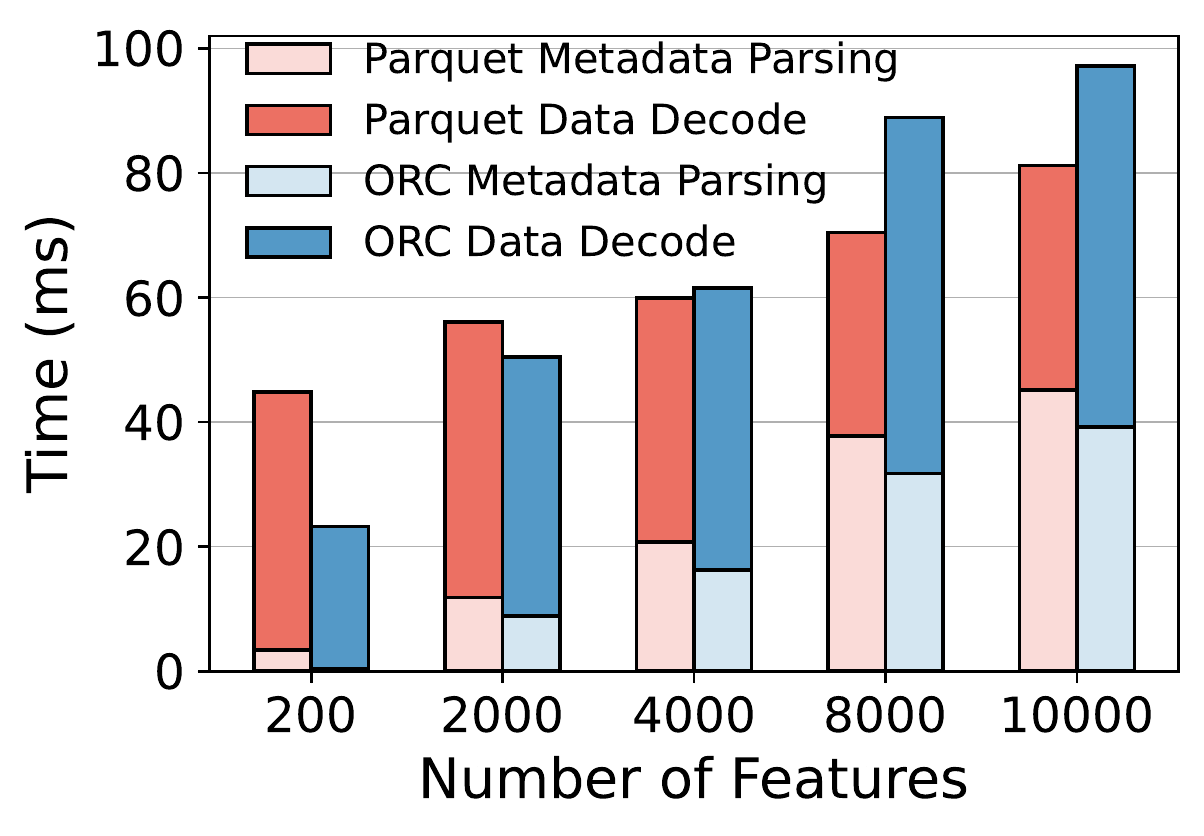}
    \end{subfigure}%
    \caption{\edit{\textbf{Wide-Table Projection}}}
    \label{fig:proj}
  \end{minipage}

\end{figure}

\begin{figure}
  \begin{minipage}{0.4\linewidth}
    \centering
    \begin{subfigure}[b]{.8\linewidth}%
      \center
      \includegraphics[width=\linewidth]{encoding/legend.pdf}
      \vspace{-.171in}
    \end{subfigure}%
  \end{minipage}
  \begin{subfigure}[t]{0.51\linewidth}%
    \center
    \includegraphics[width=\linewidth]{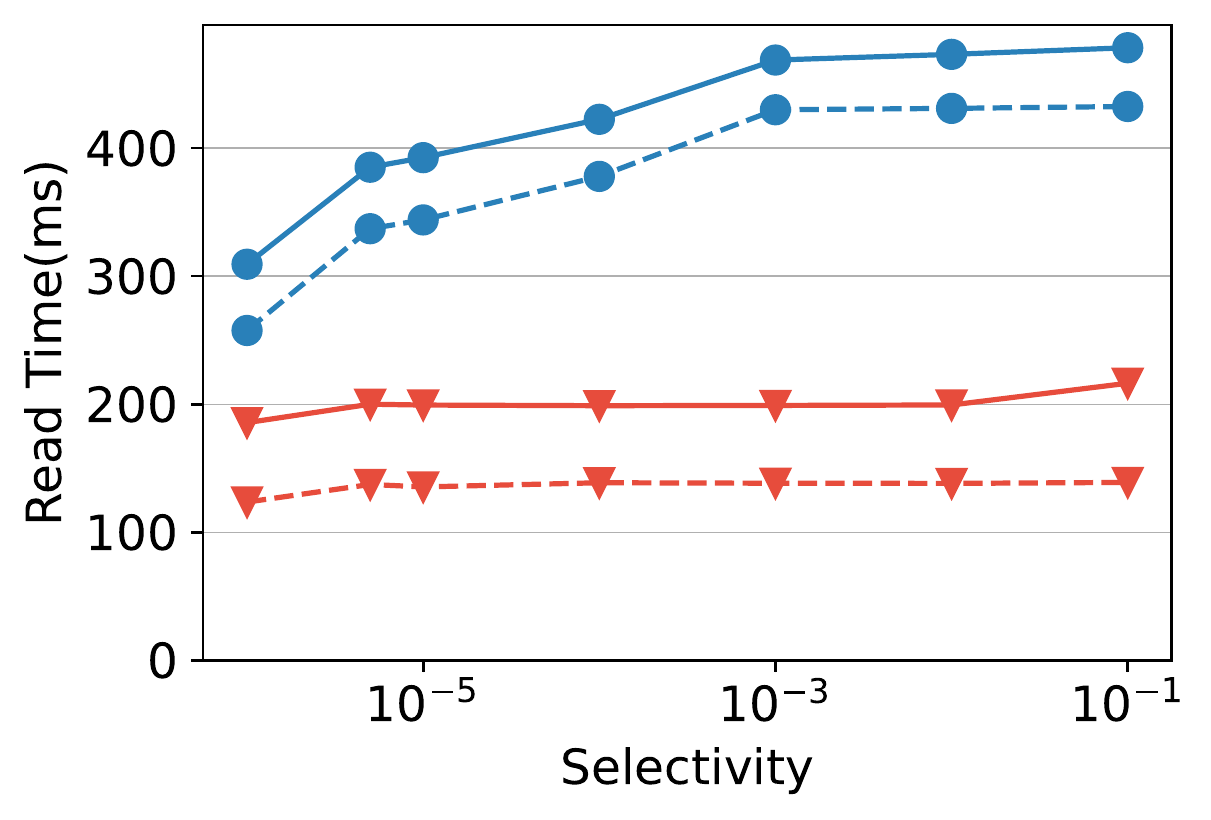}
    \caption{Low NDV ratio Column}\label{fig:filterscan_lowcar}
  \end{subfigure}%
  \begin{subfigure}[t]{0.48\linewidth}%
    \center
    \includegraphics[width=\linewidth]{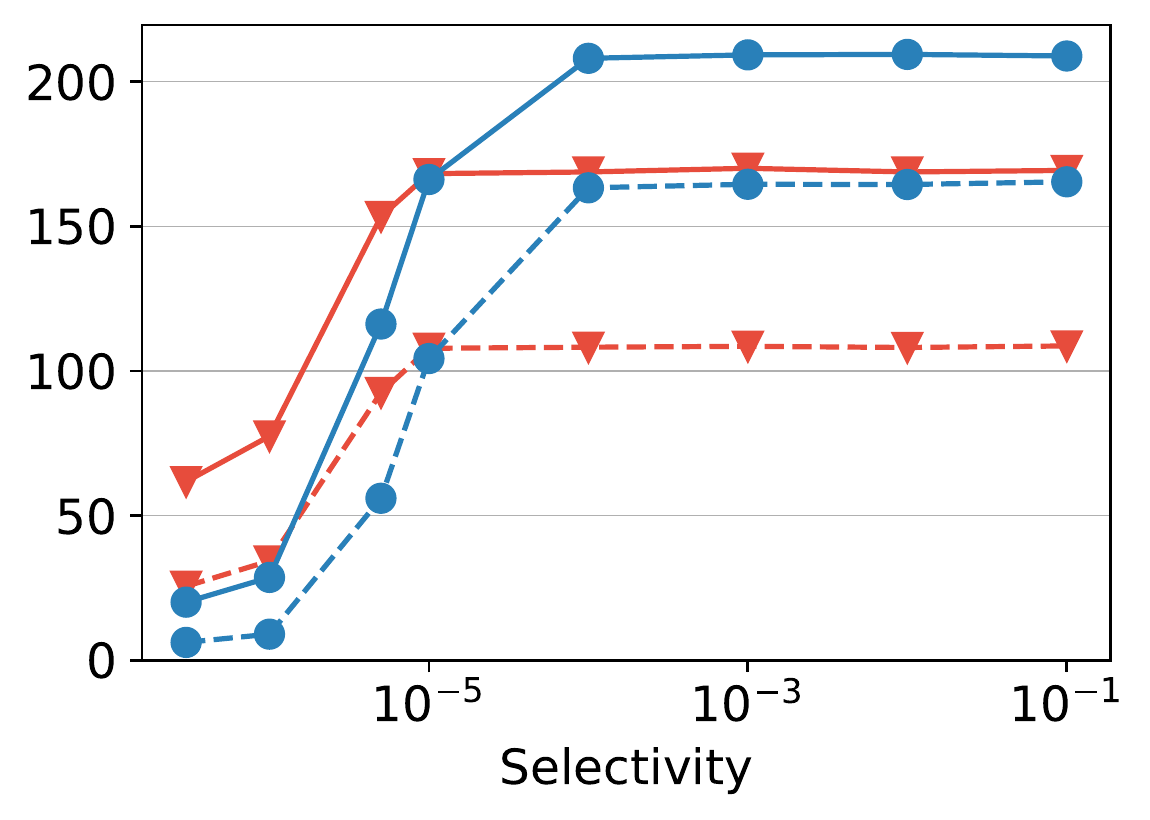}
    \caption{High NDV ratio Column}\label{fig:filterscan_highcar}
  \end{subfigure}%
  \caption{\textbf{Select performance with range predicate on float data.}}
  \label{fig:filterscan}
\end{figure}
We study the performance-space trade-offs of block compression on the formats in this section.
We first repeat the decoding-speed experiments in \cref{sssec:decode_speed}
with different algorithms (i.e., Snappy~\cite{snappy}, Zstd~\cite{zstd}).
As shown in \cref{fig:size_none,fig:size_snappy,fig:size_zstd},
Zstd achieves a better compression ratio than Snappy for all data types.
The results also show that block compression is effective on float columns in \orc
because they contain raw values.
For the rest of the columns in both \pq and \orc, however, the space savings of such compression
is limited because they are already compressed via lightweight
encoding algorithms.
\cref{fig:scan_none,fig:scan_snappy,fig:scan_zstd} also shows that block compression imposes
up 4.2$\times$ performance overhead to scanning.

We further investigate the I/O benefit and the computational overhead of
block compression on \pq across different storage-device tiers available in AWS.
The x-axis labels in \cref{fig:compression_device} show the storage tiers,
where \texttt{st1}, \texttt{gp3}, \texttt{gp2}, and \texttt{io1} are from Amazon EBS,
while \texttt{nvme} is from an AWS i3 instance.
These storage tiers are ordered by an increasing read bandwidth.
We generate a table with 1m rows and 20 columns according to the \texttt{core} workload
and store the data in \pq.
The file sizes are 34~MB and 25~MB with Zstd disabled and enabled, respectively.
We then perform scans on the \pq files stored in each storage tier
using a single thread.

As shown in \cref{fig:compression_device},
applying Zstd to \pq only speeds up scans on slow storage tiers
(e.g., \texttt{st1}) where I/O dominates the execution time.
For faster storage devices, especially NVMe SSDs,
the I/O time is negligible compared to the computation time.
In this case, the decompression overhead of Zstd hinders scan performance.
The situation is different with S3 because of its high access latency~\cite{armbrust2020delta}.
Reading a \pq file requires several round trips,
including fetching the footer length, the footer, and lastly the column chunks.
Therefore, even with multi-threaded optimization to fully utilize S3's bandwidth,
the I/O cost of reading a medium-sized (e.g., 10s-100s~MB) \pq file is still noticeable.

\textbf{Discussion:} As storage gets faster and cheaper,
the computational overhead of block compression dominates the I/O and storage savings
for a storage format.
Unless the application is constrained by storage space,
such compression should not be used in future formats.
Moreover, as more data is located on cloud-resident object stores (e.g., S3),
it is necessary to design a columnar format specifically for this operating environment (e.g.,
high bandwidth and high latency).
Potential optimizations include storing all the metadata continuously in the format
to avoid multiple round trips, appropriately sizing the row groups (or files) to hide the access latency,
and coalescing small-range requests to better utilize the cloud storage bandwidth~\cite{arrow-dataset,bian2022pixels}.

\subsection{Wide-Table Projection}\label{ssec:projection}

\edit{\marginpar{\detailedev{2}{1}}According to our discussion with Meta's Alpha~\cite{meta-alpha} team, it is common to store a large number of features (thousands of key-value pairs) for ML \textbf{training} in \orc files using the ``flat map'' data type where the keys and values are stored in separate columns. Because each ML training process often fetches a subset of the features, the columnar format must support wide-table projection efficiently. In this experiment, we generate a table of 10K rows with a varying number of float attributes. We store the table in \pq and \orc and randomly select 10 attributes to project. \cref{fig:proj} shows the breakdown of the average latency of the projection queries. }

\edit{As the number of attributes (i.e., features) in the table grows, the metadata parsing overhead increases almost linearly even though the number of projection columns stays fixed. This is because the footer structures in \pq and \orc do not support efficient random access. The schema information is serialized in Thrift (\pq) or Protocol Buffer (\orc), which only supports sequential decoding. We also notice that \orc's performance declines as the table gets wider because there are fewer entries in each row group whose size has a physical limit (64~MB).}

\edit{\textbf{Discussion: }
    Wide tables are common, especially when storing features for ML training. Future formats must organize the metadata to support efficient random access to the per-column schema.}

\subsection{Scan with Predicates} \label{ssec:index_exp}

This section evaluates how the index and filter can boost scans with predicates.
A scan with predicates contains the following two steps: 1. \textit{select}: Scan the predicate column(s) and output a bitvector containing the matched value positions. 2. \textit{bitvector project}: Use the bitvector to fetch the matched values of projection column(s). Because Parquet C++ does not implement Page Index and \bloomf, we use \pq's Rust implementation in this section.

\subsubsection{Select} \label{sssec:filterscan}

\begin{figure*}[t]
  \begin{minipage}{\linewidth}
    \centering
    \begin{subfigure}[b]{.3\linewidth}%
      \center
      \includegraphics[width=\linewidth]{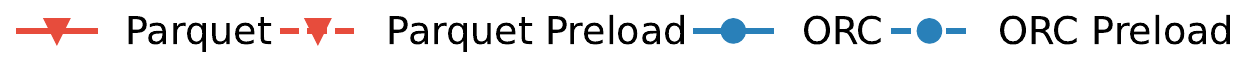}
      \vspace{-.16in}
    \end{subfigure}%
  \end{minipage}
  \begin{minipage}[t]{0.32\linewidth}
    \begin{subfigure}[t]{\linewidth}%
      \center
      \includegraphics[width=\linewidth]{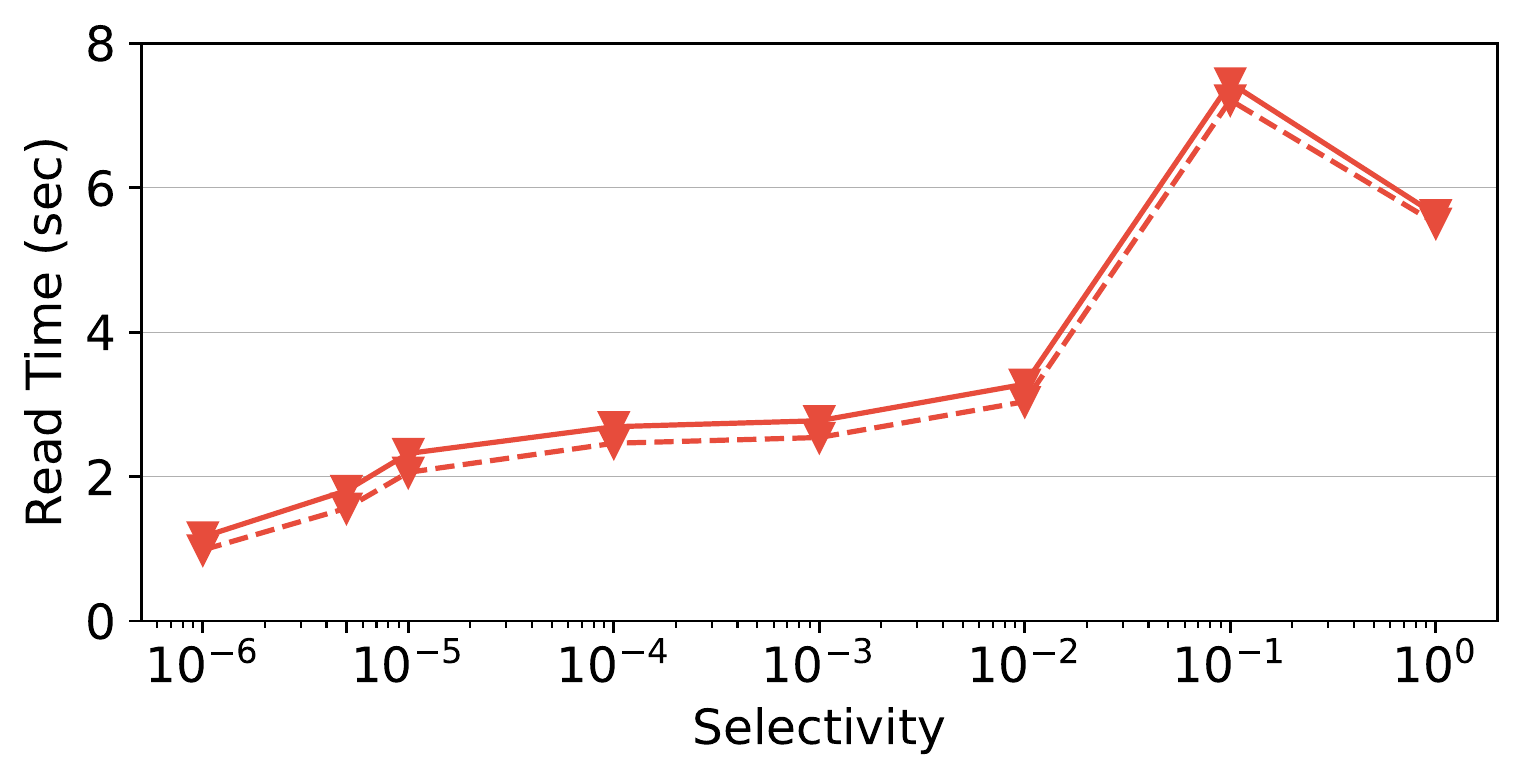}
    \end{subfigure}%
    \vspace{.25in}
    \caption{\pq-selection scan}\label{fig:parquet_e2e}
  \end{minipage}
  \begin{minipage}[t]{0.67\linewidth}
    \begin{subfigure}[t]{.5\linewidth}%
      \center
      \includegraphics[width=\linewidth]{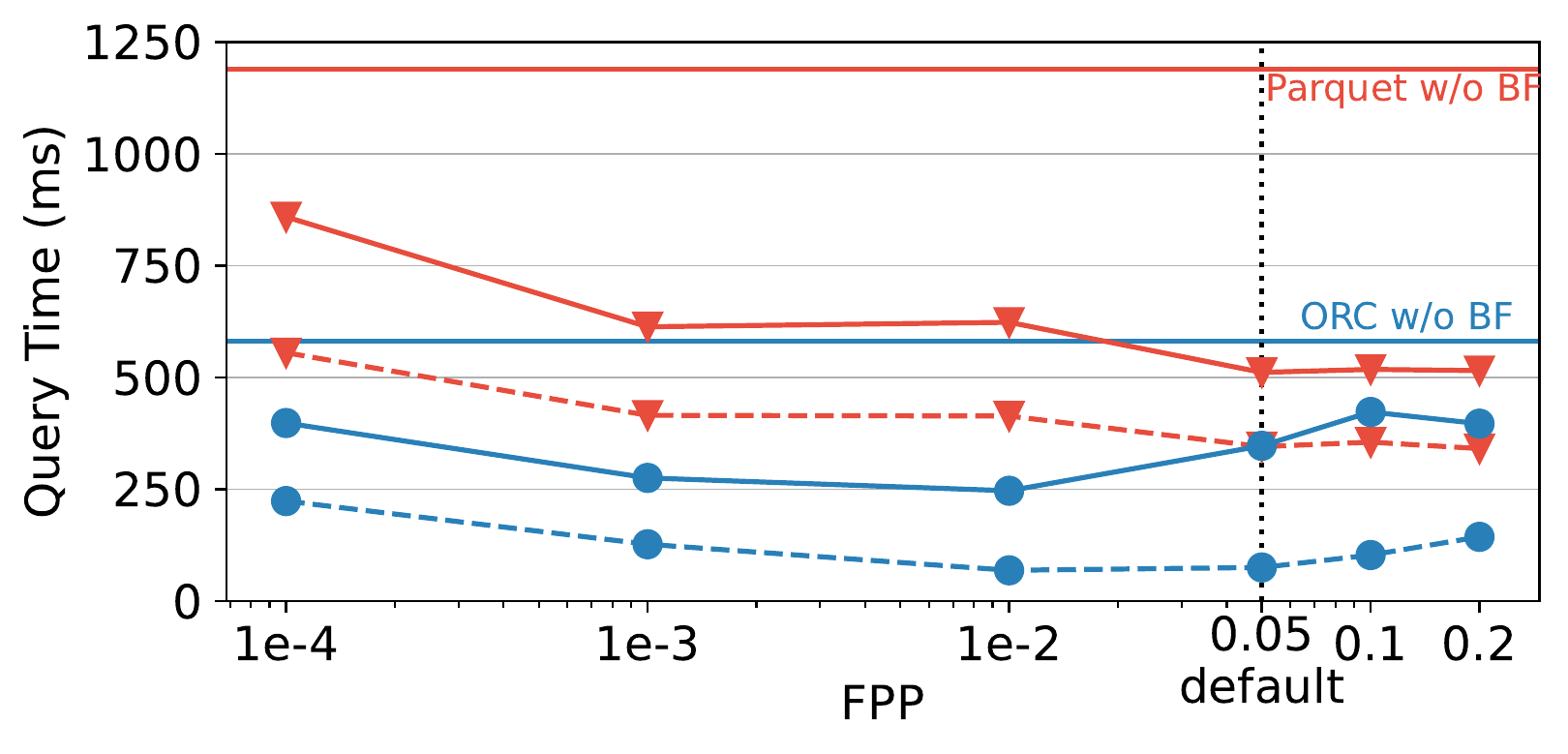}
      \caption{Query Time}\label{fig:bf_time}
    \end{subfigure}%
    \begin{subfigure}[t]{.5\linewidth}%
      \center
      \includegraphics[width=\linewidth]{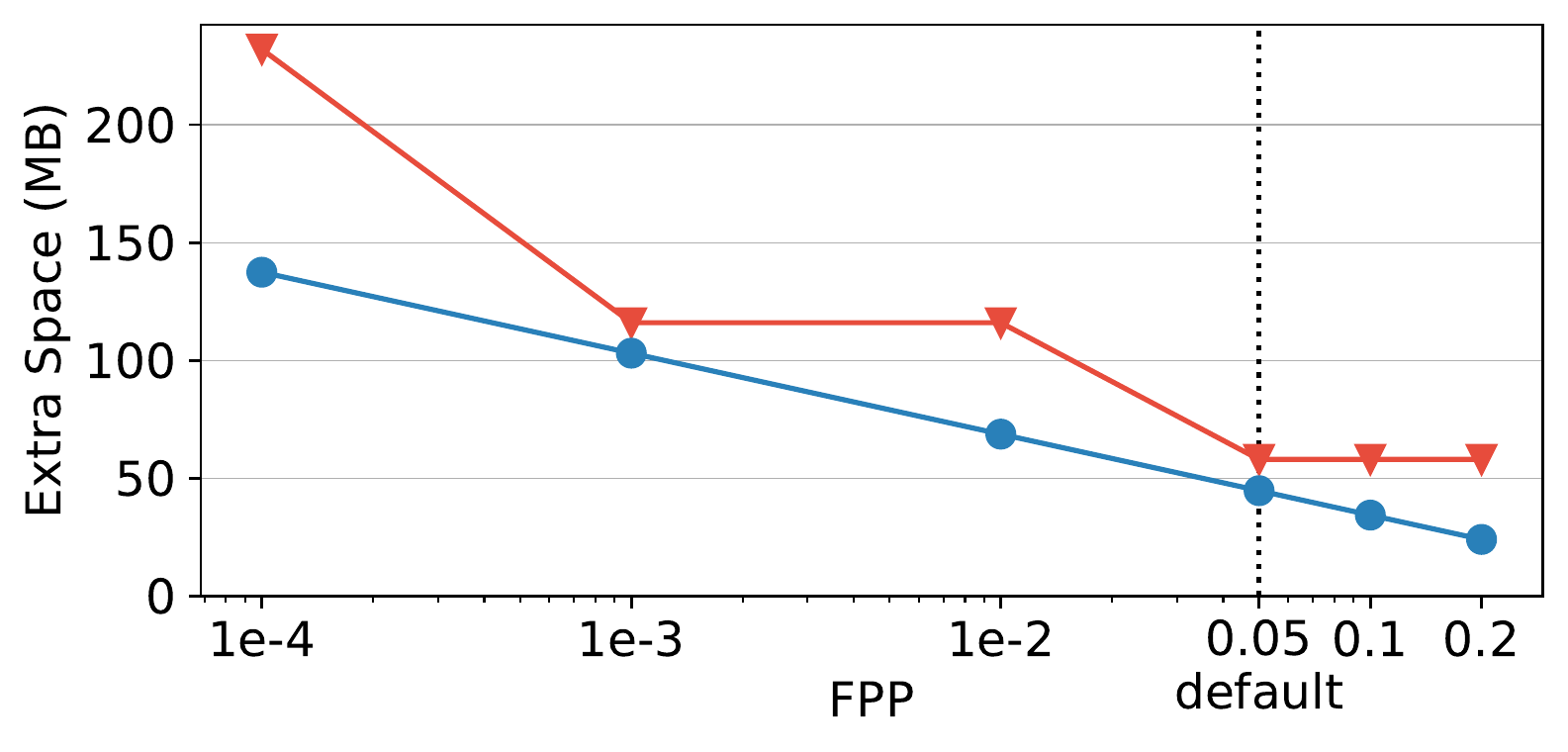}
      \caption{Extra Space}
      \label{fig:bf_space}
    \end{subfigure}%
    \caption{Select performance on varying FPP of \bloomf.}
    \label{fig:bf}
  \end{minipage}
\end{figure*}

We first evaluate the performance of select.
We generate a table with 10 million rows and 20 columns using the \texttt{core} workload and then pick two float columns with NDV ratios of 0.03 and 0.75, respectively.
Then we generate range predicates with different target selectivities on the two columns.

The results are in \cref{fig:filterscan}. As the selectivity goes down, \orc's reading time decreases faster than \pq's for both columns. In \cref{fig:filterscan_highcar}, \orc outperforms \pq when selectivity is lower than $10^{-5}$ even though \orc's full scan performance is worse than \pq's on float columns. The reason is that \orc's \zm granularity is smaller than \pq's.
Recall that \pq's smallest \zm granularity is 1~MB page while \orc is 10k values. This means for 8-byte numerical data, \pq's \zm contains 128k values without encoding and compression. Therefore \orc has more opportunities to skip values using \zms.

\textbf{Discussion: }
Zone maps are only useful on columns where the values are well-clustered.
Deciding when to have a zone map (in what granularity) is important
for future formats to improve the selection performance with minimal
space overhead.


\subsubsection{Case Study: \pq select + bitvector project}
Next, we conduct a case study on the performance of a full scan with predicates in \pq.
We did not include \orc because it does not implement bitvector project.
We assume the projection columns are all the columns and only one predicate column exists. We use the same table, predicate column, and predicates in \cref{fig:filterscan_lowcar}.

The results in \cref{fig:parquet_e2e} indicate that when conducting a complete end-to-end selection scan, the Page Index design of \pq demonstrates efficient data-skipping capabilities through its employment of late materialization (i.e., only the pages containing records indicated by the bitvector produced by select are decoded). Nevertheless, if the selectivity of the query is high (i.e., 0.1 in \cref{fig:parquet_e2e}), the computation required for determining the pages to skip surpasses the benefits of skipping. Furthermore, in case of extremely low selectivity, the achievable improvement is bounded by the time to decode a single compression unit.

\textbf{Discussion: }The implementation of the formats should decide wisely on whether to do late materialization (i.e., bitvector project) or a full table scan, depending on predicate selectivity and data distribution. Also, to enable faster low selectivity query, future formats can consider breaking the access granularity limit of a compression unit to enable finer-grained access~\cite{lance}.

\subsubsection{Point Query with \bloomf}
We further evaluate \pq and \orc's \bloomf performance and overhead. False Positive Probability (FPP) of the \bloomf is adjustable in both formats. The smaller the FPP, the more space the \bloomf takes. We aim to study how the FPP affects each format's space and query time.
Because we want the target scenario to be where \bloomf can take effect, we use a uniformly distributed 8-byte integer column as the predicate column, with value range [1, 2000000]. We build \bloomf on the column with varying FPP in the two formats.
Then we run a point query to each file where only 22 rows out of 60 million rows are selected and we record the extra file size compared with the file without building \bloomf.

The results are in \cref{fig:bf}. \cref{fig:bf_time} shows that the \bloomf can boost low selectivity point query speed. Interestingly, because \orc's \bloomf granularity (10k rows, smallest \zm level) is much finer than \pq's (1Mi rows, row group level), decreasing FPP to 0.01 helps \orc skip more data. At the same time, there is no benefit to \pq.
Instead, \pq's performance decreases as FPP is lower, which is also the same for ORC as FPP is below 0.01, because of the more space and decoding overhead.

\cref{fig:bf_space} further shows the extra storage cost introduced by the \bloomf. It shows that \orc's \bloomf design is more space-efficient than \pq's. Therefore, although \pq and \orc's performance and space overhead are close under the default FPP level (0.05), \orc offers better options to choose from.

\textbf{Discussion: }
The design of \bloomf accelerates point query on uniform data, and finer granularity of \bloomf brings more benefits to low selectivity queries.

Overall, \zms and Bloom Filters can boost the performance of low-selectivity queries.
However, \zms are effective only for a smaller number of well-clustered columns,
while Bloom Filters are only useful for point queries.
When designing future columnar formats,
we should consider more advanced indexing and filtering structures,
such as column indexes~\cite{li2013bitweaving,hentschel2018column,sidirourgos2013column} and range filters~\cite{zhang2018surf,vaidya2022snarf},
and study their performance-space trade-offs.

\subsection{Nested Data Model} \label{ssec:nested_exp}

\begin{figure}[t]

    \begin{subfigure}[t]{0.5\linewidth}%
        \center
        \raisebox{0.15\height}{\includegraphics[width=\linewidth]{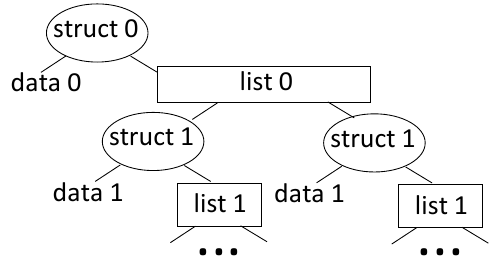}}
        \caption{File Schema. Number=nested depth}\label{fig:nested_schema}
    \end{subfigure}%
    \begin{subfigure}[t]{0.5\linewidth}%
        \center
        \includegraphics[width=\linewidth]{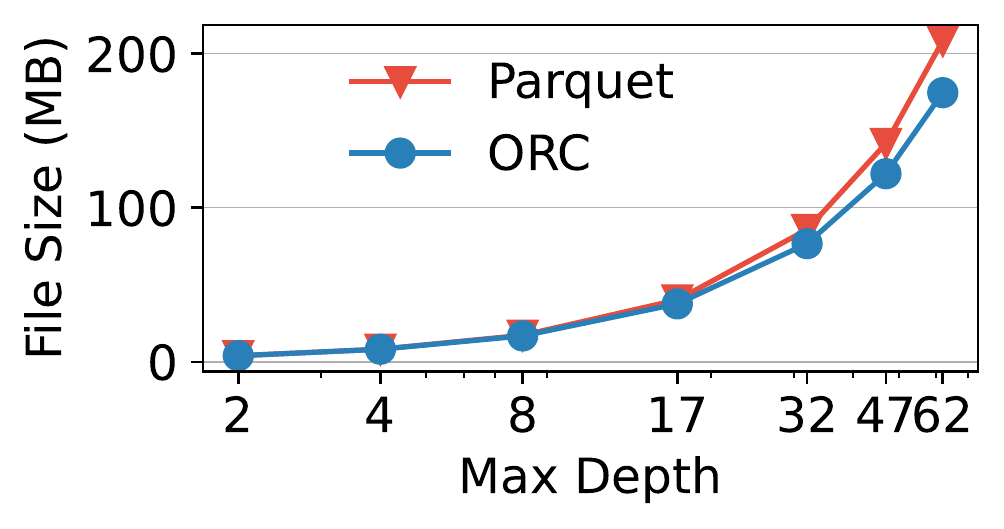}
        \caption{File Size}\label{fig:nested_size}
    \end{subfigure}%
    \\
    \begin{subfigure}[t]{0.5\linewidth}%
        \center
        \includegraphics[width=\linewidth]{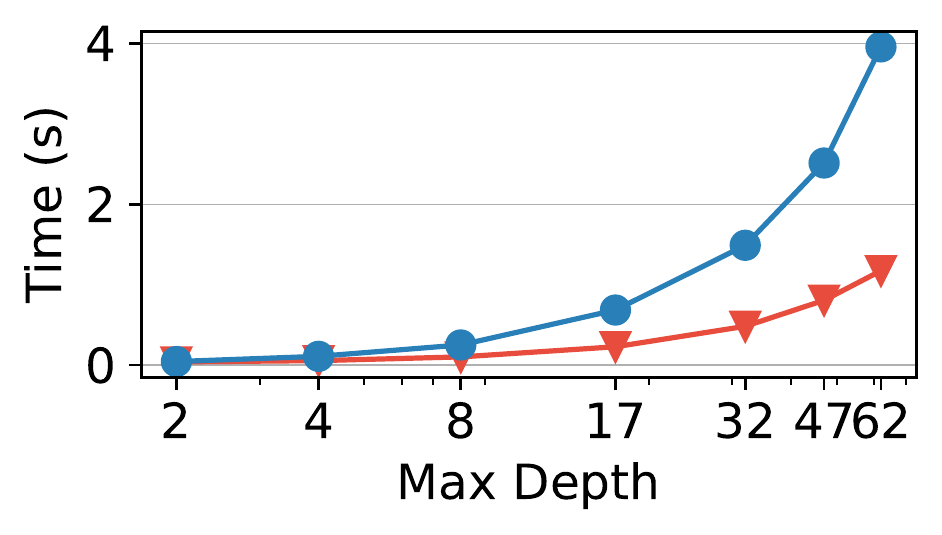}
        \caption{Time of Scanning to Arrow}\label{fig:nested_time_arrow}
    \end{subfigure}%
    \begin{subfigure}[t]{0.5\linewidth}%
        \center
        \includegraphics[width=\linewidth]{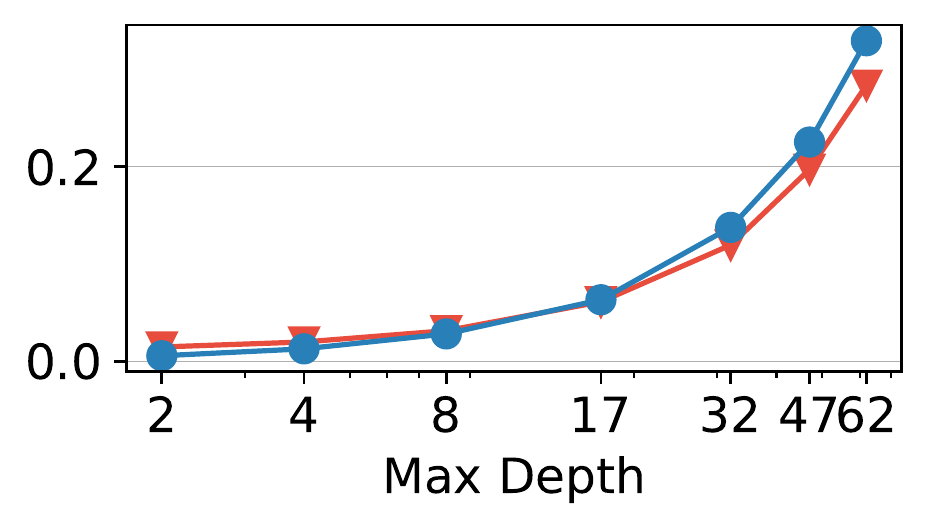}
        \caption{Nested Info Decode Overhead}\label{fig:nested_time}
    \end{subfigure}%
    \caption{\textbf{Nested Data Model}
        --
        Varying max depth in the data.
    }
    \label{fig:nested}
\end{figure}

In this section, we quantitatively evaluate the trade-off on the nested data model between \pq and \orc. To only test the nested model and isolate other noise, we use float data so we can disable encoding and compression on both formats. We test against a synthetic nested schema tree which we design as follows (as shown in \cref{fig:nested_schema}): The root node is a struct containing a float field and a list field. The list recursively contains 0-2 structs with the same schema as the root.
97\% of the lists contain one struct, and 1\% contains no struct.
We generate a series of Arrow tables with 256k rows on different max depths of the schema tree and write them into \pq and \orc. During table generation, the tree of a record stops growing when the depth of the tree reaches the desired max depth. Then we record the file size, the time to read the file into an Arrow table, and the time to decode the nested structure during the table scan.

As shown in \cref{fig:nested_size}, as the depth of the schema tree gets larger, the \pq file size grows faster than \orc. On the other hand, \orc spends much more time transforming to Arrow (\cref{fig:nested_time_arrow}). The reason is that \orc needs to be read into its in-memory data structure and then transformed to Arrow. And the transformation is not optimized.
Therefore, we further profile the time decoding the nested information during the scan. The result in \cref{fig:nested_time} shows that \orc's overhead to decode the nested structure information is getting larger than \pq's as the schema gets deeper. The reason is that \orc needs to decode structure information of struct and list while \pq only needs to decode leaf fields along with their levels. This result is consistent with Dremel's retrospective work~\cite{melnik2020dremel}.

\textbf{Discussion: } 
The trade-offs between the two nested data models only manifest when the depth is large.
Future formats should pay more attention to avoiding extra overhead during the translation between the on-disk and in-memory nested models.

\subsection{Machine Learning Workloads} \label{ssec:ml_exp}

\edit{\marginpar{\weakpt{2}{2}}\marginpar{\weakpt{3}{1-2}}
    We next investigate how well the columnar formats support common ML workloads.
    Besides raw data (e.g., image URLs, text) and the associated metadata (e.g., image dimensions,
    tags), an ML data set often contains the vector embeddings of the raw data, which is a vector of
    floating-point numbers to enable similarity search in applications such as text-image matching and
    ad recommendation. It is common to store the entire ML data set in \pq
    files~\cite{huggingface-parquet}, where the vector embeddings are stored as lists in \pq's nested
    model. Additionally, ML applications often build separate vector indexes directly from \pq to speed up similarity
    search~\cite{autofaiss-build-index}.
}

\subsubsection{Compression Ratio and Deserialization Performance with Vector Embeddings}
\label{sssec:embedding}

\begin{figure}[t]
    \begin{subfigure}[t]{0.5\linewidth}%
        \center
        \includegraphics[width=\linewidth]{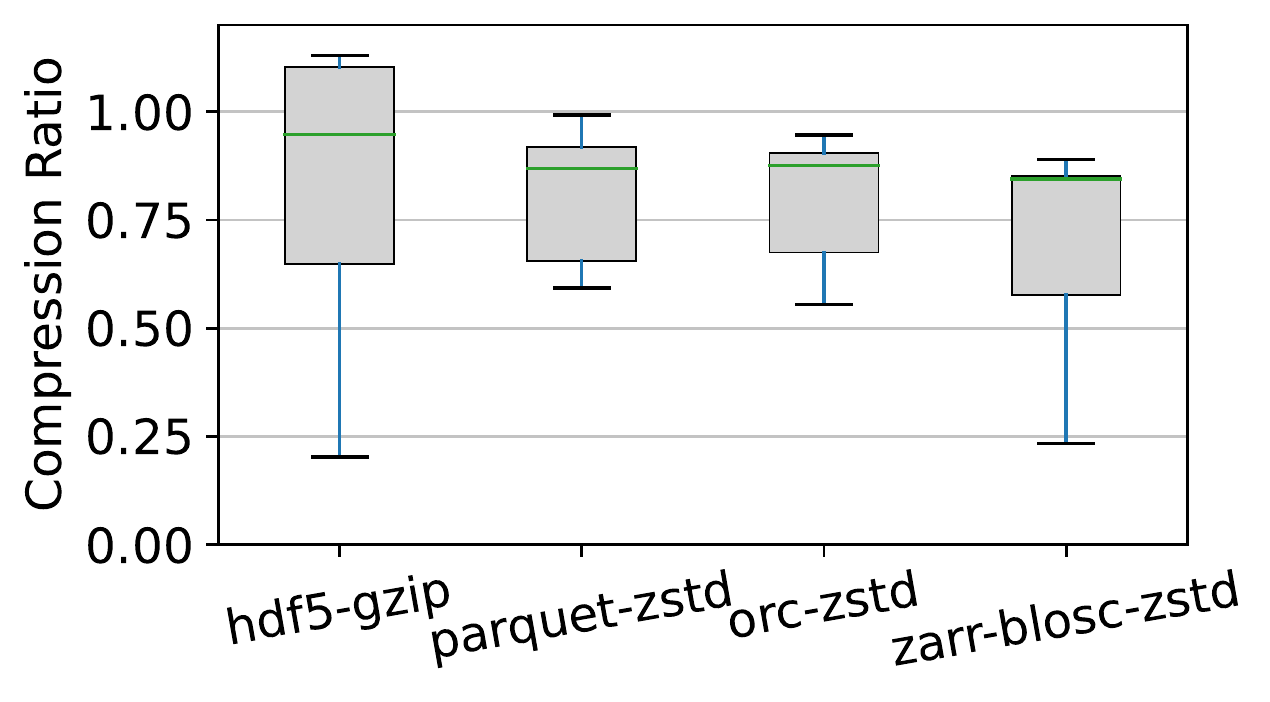}
        \caption{\edit{Compression Ratio}}\label{fig:vector_cr}
    \end{subfigure}%
    \begin{subfigure}[t]{0.495\linewidth}%
        \center
        \includegraphics[width=\linewidth]{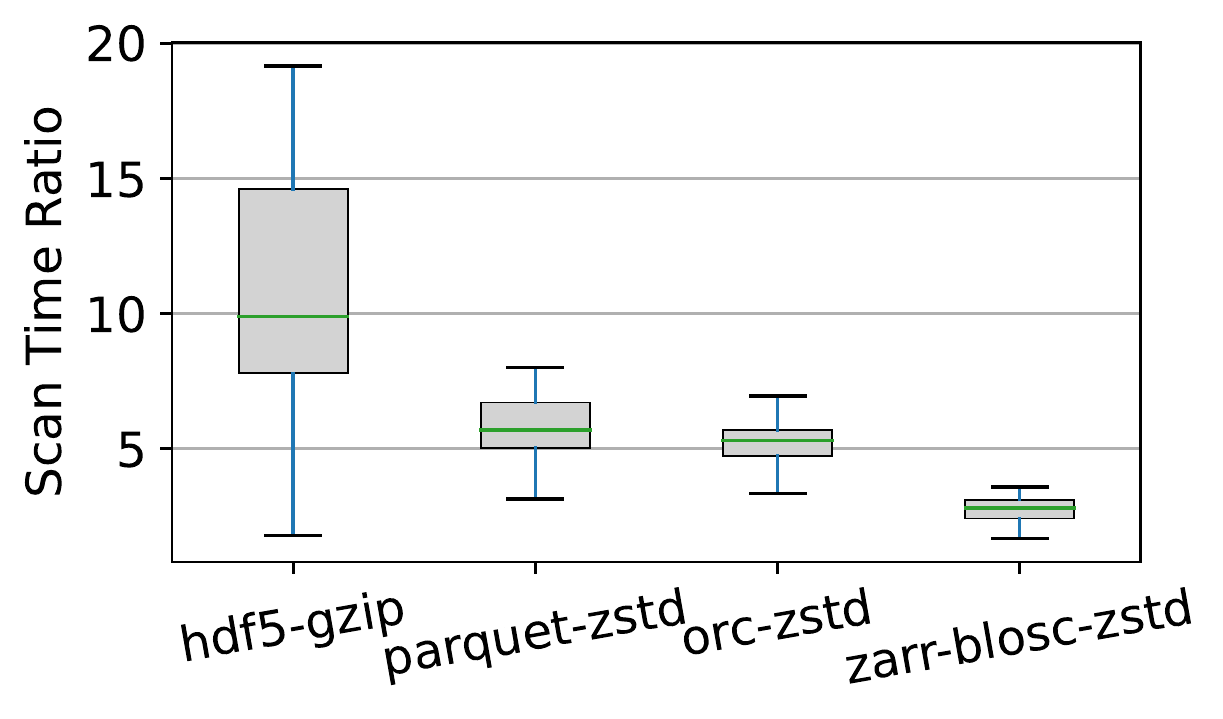}
        \caption{\edit{Scan Time w.r.t Numpy}}
        \label{fig:vector_scan}
    \end{subfigure}%
    \caption{\edit{\textbf{Efficiency of storing and scanning embeddings}}}
    \label{fig:vector}
\end{figure}

\edit{In this experiment, we collect 30 data sets
    with vector embeddings from the top downloaded
    and top trending lists on Hugging Face and store the embeddings in four different formats: \pq,
    \orc, HDF5, and Zarr. We then scan those files into in-memory Numpy arrays and record the scan time
    for each file. We report the median, 25/75\%, and min/max of the compression ratio (format\_size /
    Numpy\_size) and the scan slowdown (format\_scan\_time / disk\_Numpy\_scan\_time) in
    \cref{fig:vector}.
}

\edit{\cref{fig:vector_cr} shows that none of the four formats achieves good compression with
    vector embeddings, although Zarr is optimized for storing numerical arrays. Zarr, however, incurs a
    smaller scanning overhead compared to \pq and \orc, as shown in \cref{fig:vector_scan}. This is
    because Zarr divides a list of (fixed-length) vector embeddings into grid chunks to facilitate
    parallel scanning/decoding of the vectors. On the other hand, \pq and \orc only support sequential
    decoding within a row group.
}

\edit{\textbf{Discussion: }Existing columnar formats are less optimized to store and deserialize
    vector embeddings, which are prevailing in ML data sets. Future format designs should include
    specialized data types/structures to allow better floating point compression~\cite{liakos2022chimp,pelkonen2015gorilla,btrblocks} and better parallelism.}

\subsubsection{Integration with Vector Search Pipeline}
\edit{Despite the emerging vector databases~\cite{wang2021milvus,chroma,pinecone}, it is still
    common to perform the vector search directly in the data lake to avoid the expensive ETL process.
    Databricks recently announced their vision of Vector Data Lakes to support querying vector
    embeddings stored in \pq inside Delta Lake~\cite{vector-data-lake}. In this experiment, we evaluate
    the performance of \pq and \orc in top-k similarity search queries.}

\edit{We use the image-text LAION-5B data set~\cite{DBLP:conf/nips/SchuhmannBVGWCC22} with the
    corresponding vector embeddings. We store the first 100M entries in \pq/\orc and then use the
    embeddings from the rest of the data set to perform top-k similarity search queries (k = 10). We
    maintain an in-memory vector index auto-tuned using the FAISS library~\cite{johnson2019billion,
        autofaiss}. Each query first searches the vector index to get the row IDs of the top 10 most similar
    entries. The query then uses those row IDs to fetch the corresponding image URLs and text from the
    underlying columnar storage. We batch the queries to amortize the read amplification to the files.
}

\begin{figure}[t]
    \begin{minipage}{0.9\linewidth}
        \begin{subfigure}[b]{\linewidth}%
            \includegraphics[width=\linewidth]{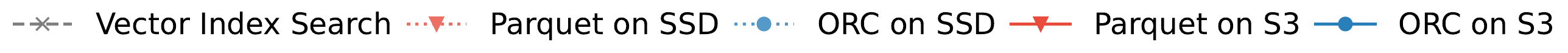}
            \vspace{-.2in}
        \end{subfigure}%
    \end{minipage}
    \begin{subfigure}[t]{0.5\linewidth}%
        \center
        \includegraphics[width=\linewidth]{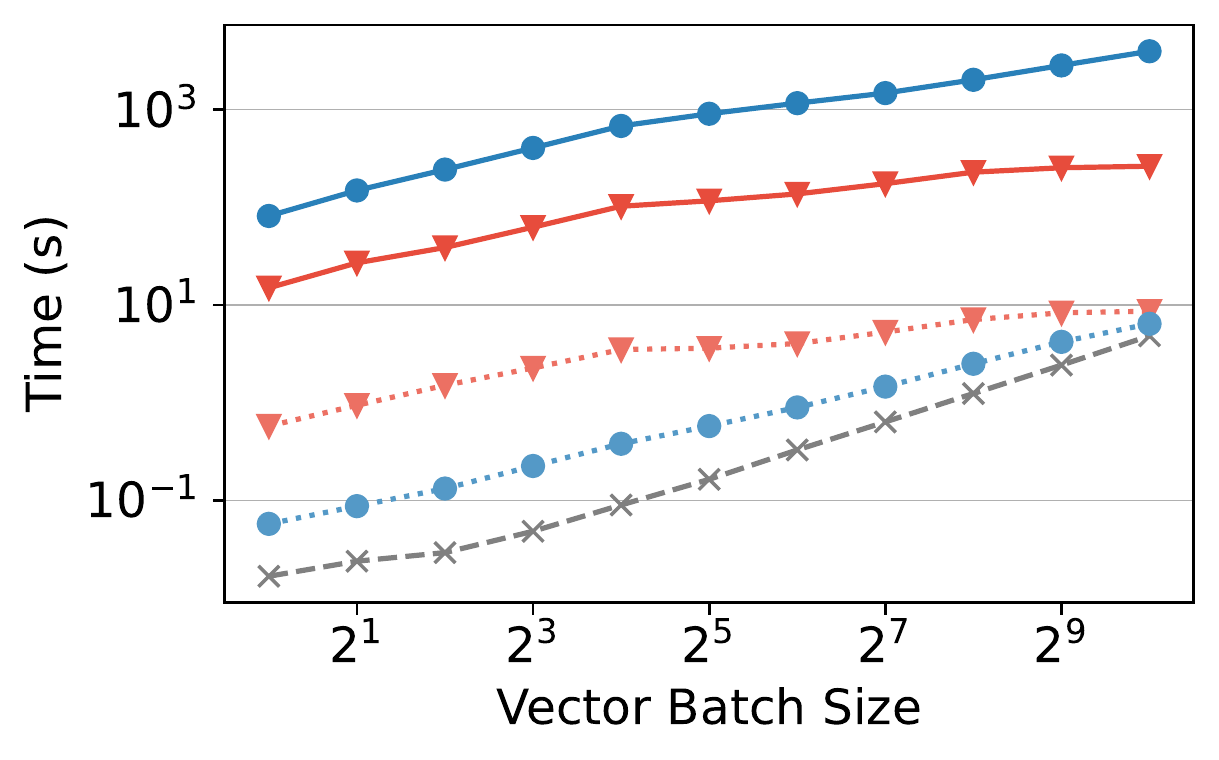}
        \caption{\edit{Time of vector index search vs. selection on files using resulting row IDs}}\label{fig:vs_local}
    \end{subfigure}%
    \begin{subfigure}[t]{0.48\linewidth}%
        \center
        \includegraphics[width=\linewidth]{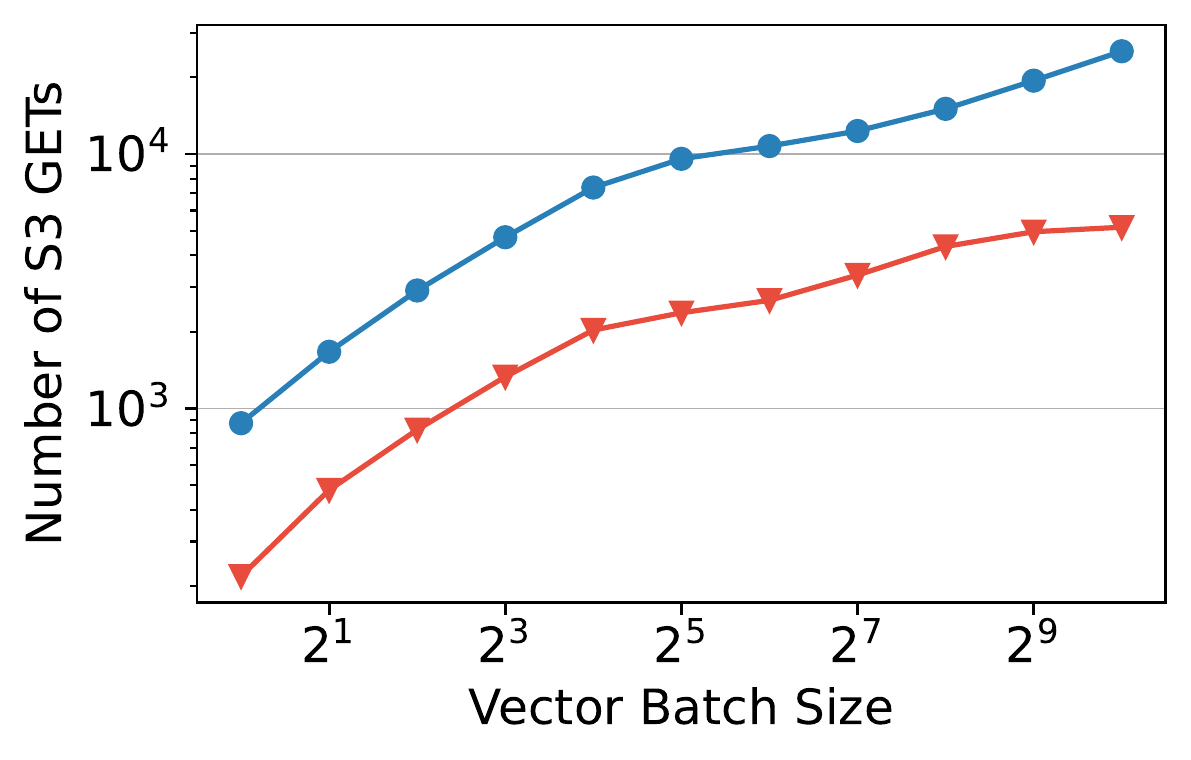}
        \caption{\edit{S3 GET requests issued}}
        \label{fig:vs_s3}
    \end{subfigure}%
    \caption{\edit{\textbf{Top-k Search Workflow Breakdown (k = 10)}}}
    \label{fig:vs}
\end{figure}

\edit{\cref{fig:vs_local} shows the average time (over 20 trials) of performing the top-k
    queries with a varying batch size on the x-axis. We repeated the queries using local NVMe SSDs and
    AWS S3 for storage. We observe that the selection operations in \orc are faster than those in \pq on
    local SSDs because \orc includes fine-grained \zms to reduce the read
    amplification. As the query batch size gets larger, the performance gap between \orc and \pq
    shrinks because the query batch fetches a significant portion of the file. The result is different
    when the files are stored in S3. Fetching records is much slower in \orc because it issues
    $\approx4\times$ S3 GET than \pq during the process, as shown in \cref{fig:vs_s3}. The reason is
    that the \zms in \orc are scattered in the row-group footers while those in \pq are centralized in
    the file footer.}

\edit{\textbf{Discussion: } ML workloads often involve low-selectivity vector search queries. Although aggressive query batching could amortize the read amplification, fine-grained indexes (e.g., \zms) are necessary to guarantee the search latency. Also, as more and more large-scale ML data sets reside in data lakes, it is critical for future formats to reduce the number of small reads (e.g., zone map fetches in \orc) to the high-latency cloud object stores.}

\subsubsection{Storage of Unstructured Data}
\edit{Besides tabular data, deep learning data sets often include unstructured data such as images, audio, and videos. One approach for storing them in the columnar format is to use their external URLs, as done in the LAION-5B data set above. This approach, however, could suffer from massive http-get requests and invalid URLs over time. Therefore, it is beneficial to store the unstructured data within the same file~\cite{image-parquet}.}

\begin{figure}[t]
    \begin{minipage}{0.6\linewidth}
        \begin{subfigure}[b]{\linewidth}%
            \includegraphics[width=\linewidth]{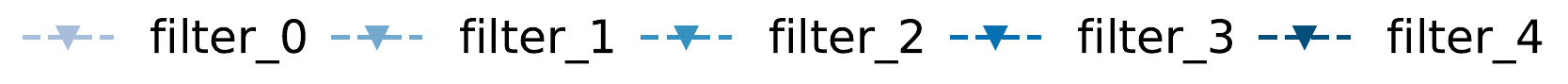}
            \vspace{-.2in}
        \end{subfigure}%
    \end{minipage}
    \begin{subfigure}[t]{0.49\linewidth}%
        \center
        \includegraphics[width=\linewidth]{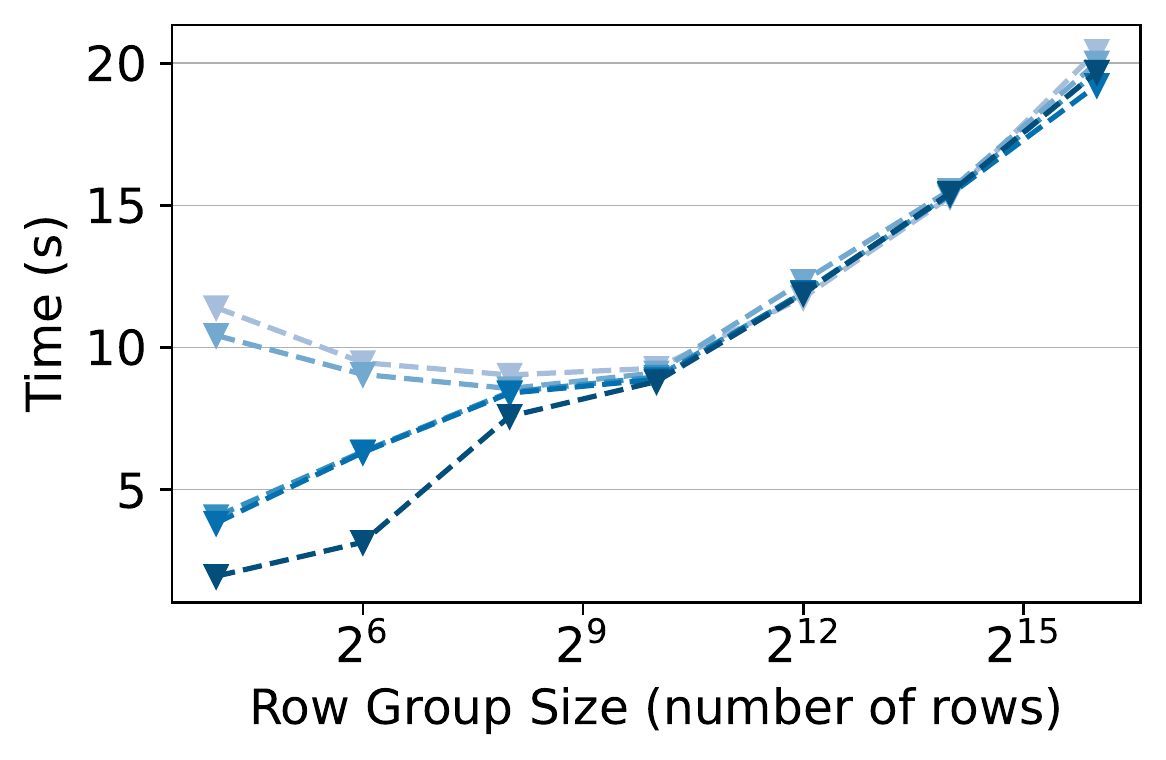}
        \caption{\edit{With images in projection}}\label{fig:image_filterscan}
    \end{subfigure}%
    \hfill
    \begin{subfigure}[t]{0.48\linewidth}%
        \center
        \includegraphics[width=\linewidth]{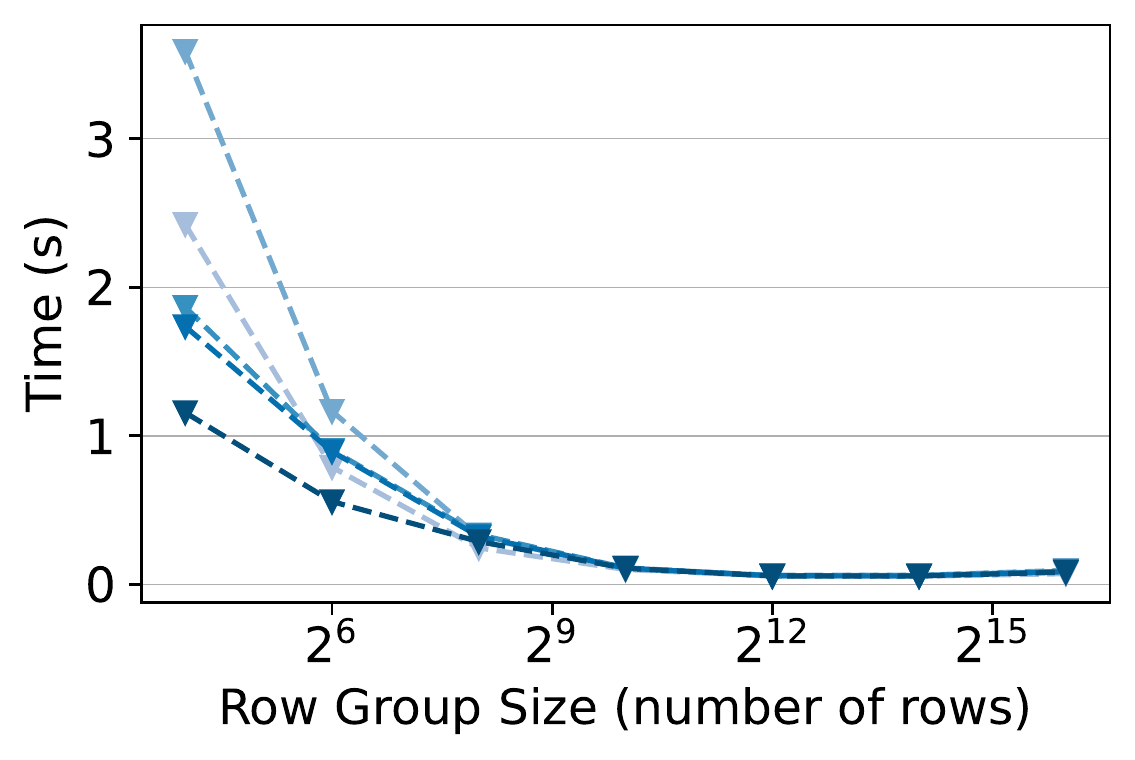}
        \caption{\edit{Without images in projection}}
        \label{fig:tabular_filterscan}
    \end{subfigure}
    \caption{\edit{\textbf{Filterscan on Image Data in \pq} -- Filters 0-4 correspond to low
            to high selectivities. Filters are applied on tabular data.}}
    \label{fig:image}
\end{figure}

\edit{We evaluate this on \pq using the LAION-5B data set with the image URLs replaced by the original binaries. The result \pq file is 13~GB with 219K rows and is stored on NVMe SSD. We perform scans with five different filters (filter\_0 - filter\_4) whose selectivities are $1$, $0.1$, $0.01$, $0.001$, and $0.0001$, respectively. We enable parallel read and pre-buffer of column chunks via Arrow's API. \cref{fig:image_filterscan} shows the query times when the image column is projected, while \cref{fig:tabular_filterscan} presents the query times with only the tabular columns projected. We vary the row-group size of the \pq file on the x-axis. We observe that a smaller row-group size works better when fetching the images because more row groups allow better parallel read of the large binaries with asynchronous I/Os. A smaller row group, however, compromises the compression of the structured data, and the increased I/O time dominates the latency of queries that only project structured data.}

\edit{\textbf{Discussion:} It is inefficient to store large binaries with structured data in the same PAX format with a default row-group size. Future designs should separate them in the physical layout of the format while providing a unified query interface logically.}

\subsection{GPU Decoding} \label{ssec:gpu_exp}

\edit{\marginpar{\weakpt{2}{2}}\marginpar{\weakpt{3}{1-2}}
    Besides machine learning, GPUs are used to speed up data analytics~\cite{yogatama2022orchestrating,shanbhag2020study} and decompression~\cite{Shanbhag-gpucompression}. In this section, we investigate the decoding efficiency of \pq and \orc with GPUs. We use state-of-the-art GPU readers for \pq and \orc in RAPIDS cuDF 23.10a~\cite{RAPIDS}. The machine for the experiments is equipped with NVIDIA GeForce RTX 3090, AMD EPYC 7H12 with 128 cores, 512GB DRAM, and Intel P5530 NVMe SSD. We generate the data set using the \texttt{core} workload with a table of 32 columns and a varying number of rows.}

\begin{figure}[t]
    \begin{minipage}{0.8\linewidth}
        \centering
        \begin{subfigure}[b]{.9\linewidth}%
            \includegraphics[width=\linewidth]{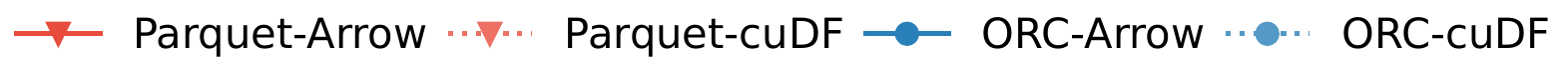}
            \vspace{-.2in}
        \end{subfigure}%
    \end{minipage}
    \begin{subfigure}[t]{0.49\linewidth}%
        \center
        \includegraphics[width=\linewidth]{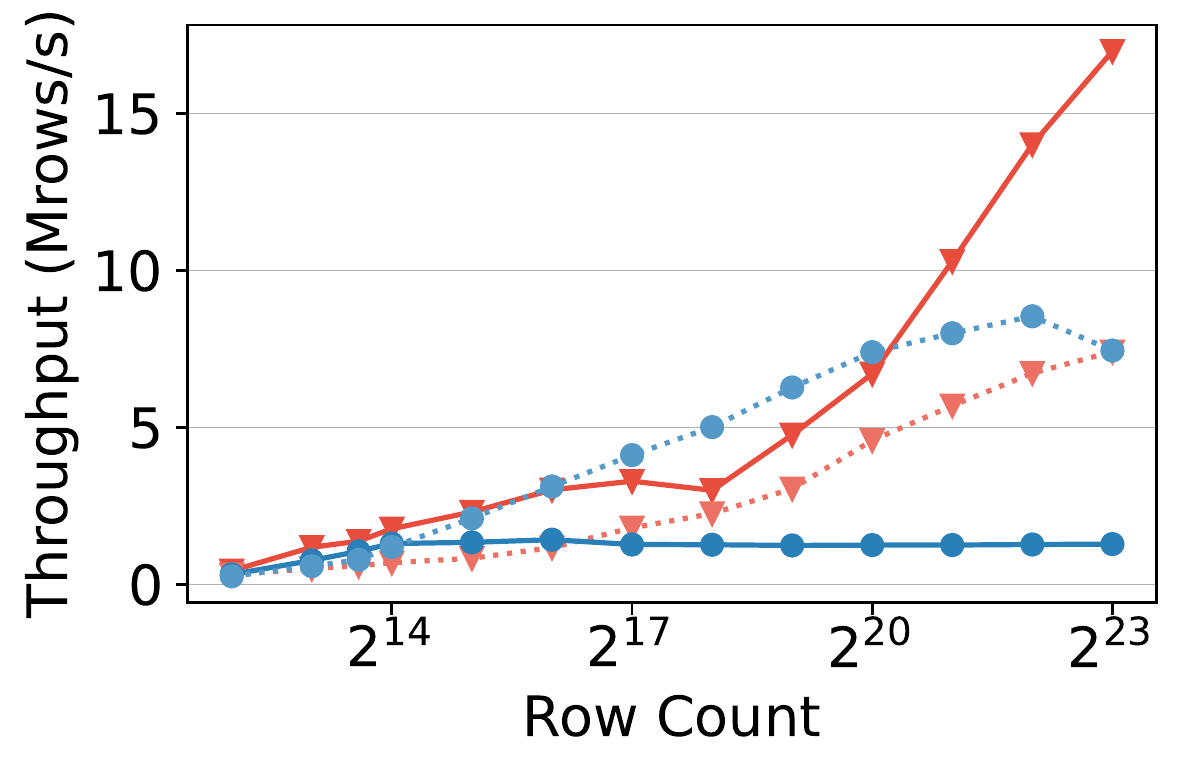}
        \caption{\texttt{core} workload}\label{fig:gpu_core}
    \end{subfigure}%
    \begin{subfigure}[t]{0.51\linewidth}%
        \center
        \includegraphics[width=\linewidth]{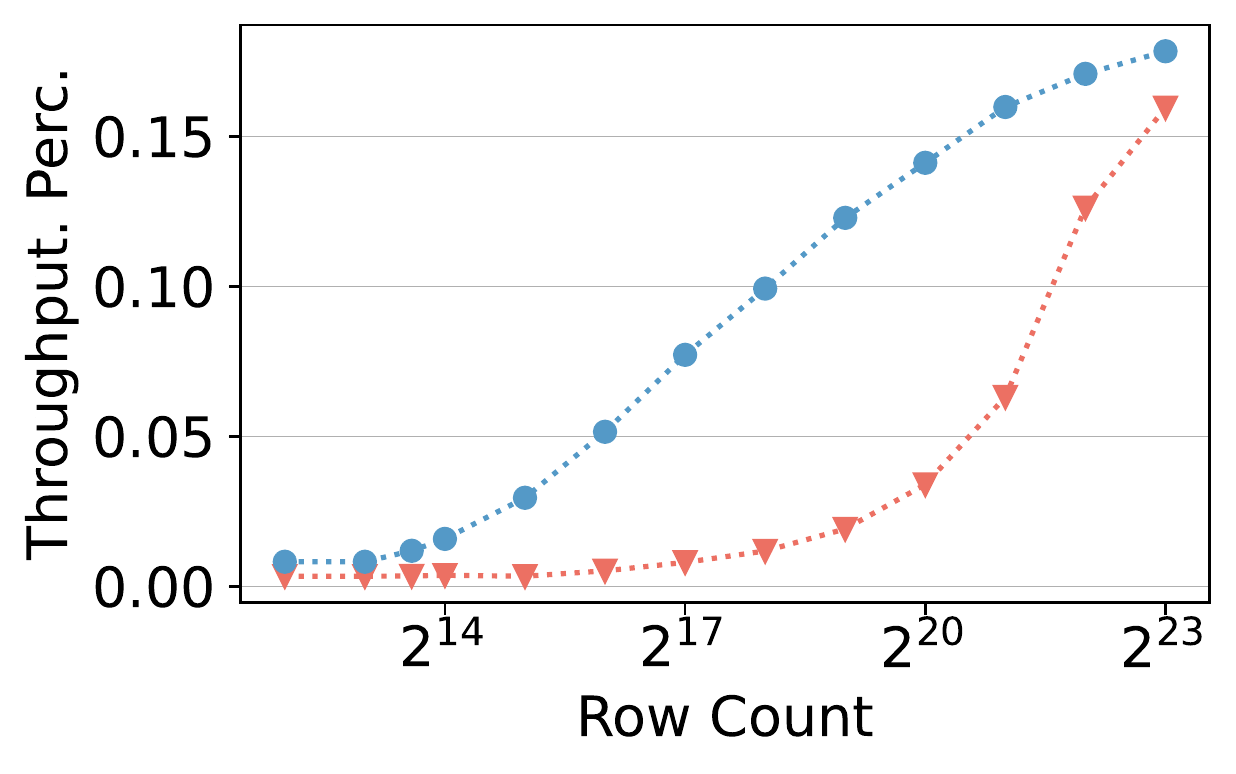}
        \caption{Peak GPU Throughput Percentage}\label{fig:gpu_sm}
    \end{subfigure}%
    \\
    \begin{subfigure}[t]{0.48\linewidth}%
        \center
        \includegraphics[width=\linewidth]{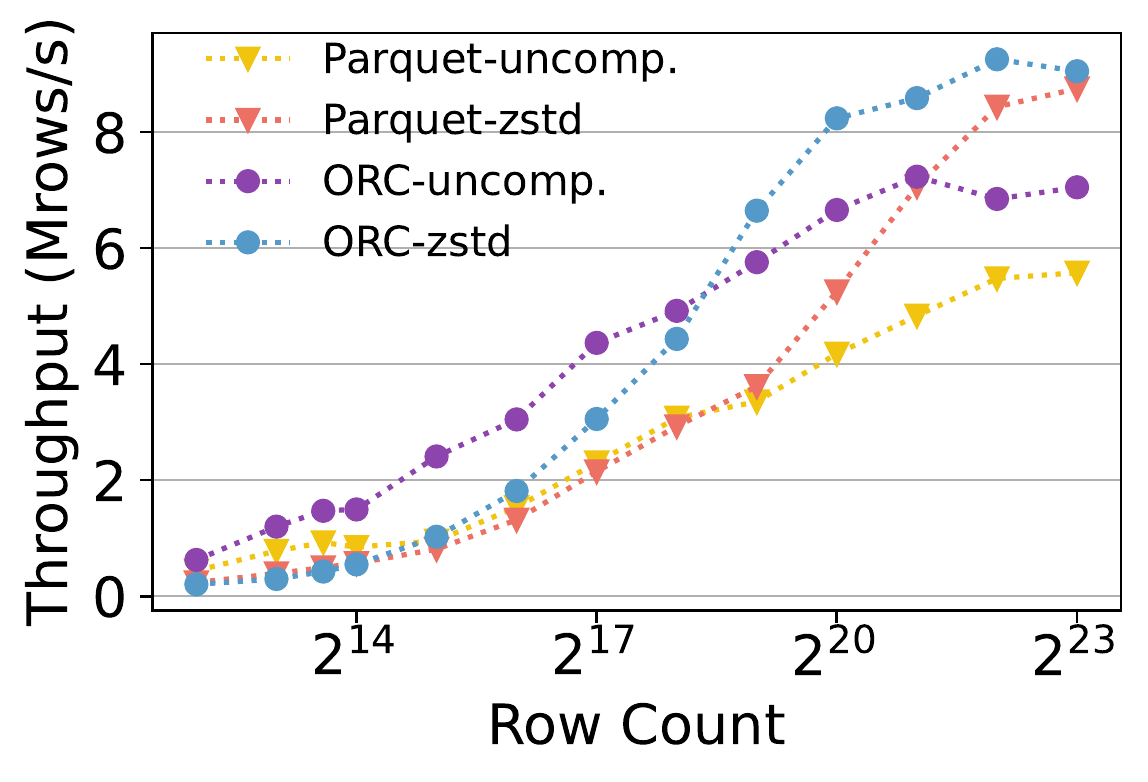}
        \caption{cuDF varying compression}\label{fig:gpu_compress}
    \end{subfigure}%
    \begin{subfigure}[t]{0.51\linewidth}%
        \center
        \includegraphics[width=\linewidth]{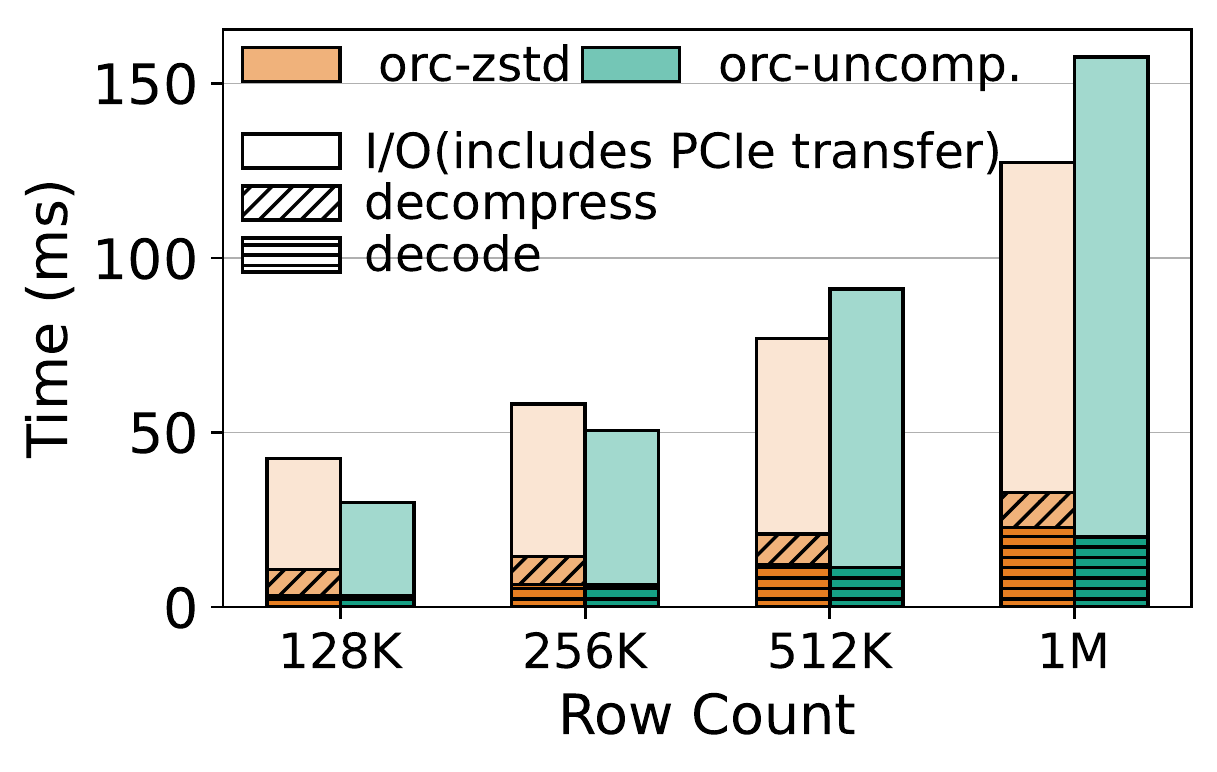}
        \caption{Time breakdown of ORC in (c)}\label{fig:gpu_comp_breakdown}
    \end{subfigure}%
    \caption{\edit{\textbf{GPU Decoding}}
    }\label{fig:gpu}
\end{figure}

\edit{In the first experiment, we scan and decode the files using Arrow (with multithread and I/O prefetching enabled) and cuDF, respectively. As shown in \cref{fig:gpu_core}, \orc-cuDF exhibits higher decoding throughput than \pq-cuDF because \orc has more independent blocks to better utilize the massive parallelism provided by the GPU: the smallest \zm in \orc maps to fewer rows than \pq's, and each GPU thread block is assigned to each smallest \zm region in cuDF. As the number of rows increases in the files, the decoding throughput of \pq-Arrow scales because there are more row groups to leverage for multi-core parallel decoding with asynchronous I/O. On the contrary, the Arrow implementation for \orc does not support parallel read.}

\edit{We further profile the GPU's peak throughput in the above experiment over its theoretical maximum throughput using Nsight Compute~\cite{nsight-compute}. As shown in \cref{fig:gpu_sm}, the overall compute utilization is low (although the GPU occupancy is full when row count reaches 8M). This is because the integer encoding algorithms used in \pq and \orc (e.g., hybrid RLE + \bitpack) are not designed for parallel processing: all threads must wait for the first thread to scan the entire data block to obtain their offsets in the input and output buffers. Moreover, because cuDF assigns a warp to each encoded run, a short run (e.g., a length-3 RLE run in \orc) would cause the threads within a warp underutilized.}

\edit{We next performed a controlled experiment under the same setting as above to evaluate how block compression affects GPU decoding. The results in \cref{fig:gpu_compress} show that applying zstd improves the scan throughput for both \pq and \orc when there are enough rows in the files (i.e., enough data to leverage GPU parallelism). \cref{fig:gpu_comp_breakdown} shows the scan time breakdown. We observe that the I/O time (including the PCIe transfer between GPU and CPU) dominates the scan performance, making aggressive block compression pay off.}

\edit{\textbf{Discussion: }Existing columnar formats are not designed to be GPU-friendly. The integer encoding algorithms operate on variable-length subsequences, making decoding hard to parallelize efficiently. Future formats should favor encodings with better parallel processing potentials. Besides, aggressive block compression is beneficial to alleviate the dominating I/O overheads (unlike with CPUs).}

\section{Lessons and Future Directions}\label{sec:lessons}

We summarize the lessons learned from our evaluation of \pq and \orc
to guide future innovations in columnar storage formats.

\textbf{Lesson 1.}
Dictionary Encoding is effective across data types (even for floating-point values)
because most real-world data have low NDV ratios.
\edit{\marginpar{\detailedev{3}{1}\&}\marginpar{\revision{3}{2}}Future formats should continue to apply the technique aggressively, as in \pq.}

\textbf{Lesson 2.}
It is important to keep the encoding scheme simple in a columnar format
to guarantee a competitive decoding performance.
\edit{Future format designers should pay attention to the performance cost
of selecting from many codec algorithms during decoding.}

\textbf{Lesson 3.}
The bottleneck of query processing is shifting from storage to
(CPU) computation on modern hardware.
\edit{Future formats should limit the use of block compression and
other heavyweight encodings unless the benefits are justified in specific cases.}

\edit{\textbf{Lesson 4.}
The metadata layout in future formats should be centralized and friendly to random access
to better support wide (feature) tables common in ML training.
The size of the basic I/O block should be optimized for the high-latency cloud storage.}

\textbf{Lesson 5.}
As storage is getting cheaper,
\edit{future formats could afford to store more sophisticated indexing and filtering structures
to speed up query processing.}

\textbf{Lesson 6.}
\edit{Future formats should design their nested data models with an affinity to modern in-memory formats
to reduce the translation overhead.}

\edit{\textbf{Lesson 7.}
The characteristics of common machine learning workloads require future formats to support both
wide-table projections and low-selectivity selections efficiently. This calls for better metadata
organization and more effective indexing. Besides, future formats should allocate separate regions
for large binary objects and incorporate compression techniques specifically designed for floats
in vector embeddings.}

\edit{\textbf{Lesson 8.}
Future formats should consider the decoding efficiency with GPUs.
This requires not only sufficient parallel data blocks at the file level
but also encoding algorithms that are parallelizable to fully utilize the computation within a GPU thread block.}

\section{Conclusion} \label{sec:Conclusion}
In this paper, we comprehensively evaluate the common columnar formats, including \pq and \orc. We build a taxonomy of the two widely-used formats to summarize the design of their format internals. To better test the formats' trade-offs, we analyze real-world data sets and design a benchmark that can sweep data distribution to demonstrate the differences in encoding algorithms. Using the benchmark, we conduct experiments on various metrics of the formats. Our results highlight essential design considerations that are advantageous for modern hardware and emerging machine learning workloads.


\balance
\bibliographystyle{ACM-Reference-Format}
\bibliography{main.bib,db.bib,format.bib}
\end{document}